\documentclass[]{aa}

\usepackage{graphicx}
\usepackage{multirow}
\usepackage{amsmath}
\usepackage{mathtools}
\usepackage{siunitx}
\usepackage{lipsum}
\usepackage{placeins}
\usepackage{hyperref}
\usepackage{threeparttable}
\usepackage{tablefootnote}
\usepackage{mathtools}
\usepackage[thinc]{esdiff}
\usepackage{rotating,tabularx}
\usepackage{caption}
\usepackage{arydshln}
\usepackage{stfloats}

\DeclareRobustCommand{\rchi}{{\mathpalette\irchi\relax}}
\newcommand{\irchi}[2]{\raisebox{\depth}{$#1\chi$}} 

\usepackage[varg]{txfonts}

\def \bp{$\beta$\,Pic}
\def \betapictoris{$\beta$\,Pictoris}

\def\Hi{H\,{\sc i}}

\def\fei{Fe\,{\sc i}}
\def\feii{Fe\,{\sc ii}}

\def\Oi{O\,{\sc i}}

\def\Ni{N\,{\sc i}}

\def\mnii{Mn\,{\sc ii}}

\def\znii{Zn\,{\sc ii}}

\def\mgi{Mg\,{\sc i}}
\def\mgii{Mg\,{\sc ii}}

\def\niii{Ni\,{\sc ii}}

\def\nai{Na\,{\sc i}}

\def\caii{Ca\,{\sc ii}}
\def\caiii{Ca\,{\sc iii}}

\def\crii{Cr\,{\sc ii}}

\def\coii{Co\,{\sc ii}}

\def\siii{Si\,{\sc ii}}

\def\siiv{Si\,{\sc iv}}
\def\Si{S\,{\sc i}}

\def\alii{Al\,{\sc ii}}
\def\aliii{Al\,{\sc iii}}

\def\Ci{C\,{\sc i}}
\def\Cii{C\,{\sc ii}}

\def\Civ{C\,{\sc iv}}
\def\srii{Sr\,{\sc ii}}
\def\Xi{X\,{\sc i}}

\def \A{$\si{\angstrom}$}
\def\feplus{Fe$^+$}

\def\caplus{Ca$^+$}

\begin{document}

\title{Transit distances and composition of low-velocity exocomets\newline in the \bp\ system}
\titlerunning{Inner structure of the \bp\ disc}

\author{
T.\ Vrignaud\inst{1}
\and
A. Lecavelier des Etangs\inst{1}
}

\institute{
Institut d'Astrophysique de Paris, CNRS, UMR 7095, Sorbonne Université, 98$^{\rm bis}$ boulevard Arago, 75014 Paris, France 
}

\abstract{\betapictoris\ is a young nearby A5V star, about 20 Myr old, embedded in a prominent debris disc. For the past 40 years, variable absorption features have been observed in the stellar spectrum, produced by the gaseous tails of exocomets transiting the star. Yet, despite the large number of observations available, the origin and dynamical evolution of the exocomets remain poorly understood. Here we present new spectroscopic observations of \bp\, obtained on April 29, 2025, with the \emph{Hubble} Space Telescope (HST) and the High Accuracy Radial Velocity Planet Searcher (HARPS). We report the detection of three strong exocomet signatures at low radial velocities ($-7.5$, $+2.5$ and $+10$ km/s), in a large set of lines from various species and excitation levels. We show that the three exocometary tails have different excitation states, indicating that they are located at different distances from the star. Using a detailed modelling of the excitation state of the transiting gas, which includes both radiative and collisional excitation, we derive the transit distance of the three exocometary gaseous tails to be $0.88 \pm 0.08$,\, $4.7 \pm 0.3$\, and $1.52 \pm 0.15$\,au. These values are much larger than previous estimates, which generally placed the transient features within $0.2$\,au. This reveals that gaseous tails produced by exocomets sublimating close to the star can expand and migrate over large distances, while still remaining detectable in absorption spectroscopy. Our study provides a new method to measure the transit distance of exocomets, based on excitation modelling, complementing the acceleration method only applicable for high-velocity objects.

}


\maketitle
\section{Introduction}
\label{Sect. Introduction}

\betapictoris\ is a young \citep[20 Myr,][]{Miret-Roig_2020}, nearby \citep[19.3 pc,][]{Crifo1997} A5V star, which hosts a complex extrasolar system. The star is surrounded a prominent debris disc seen edge-on, expanding over hundreds of astronomical units \citep{Smith_1984, Lecavelier_1993}, and composed of dust \citep[][]{Apai_2015, Rebollido2024} and volatile-rich gas \citep[][]{Roberge2000, Dent2014, Brandeker_2016}. The system also hosts two massive planets, \bp\ b \citep{Lagrange_2010} and \bp\ c \citep{Lagrange_2019, Nowak2020}, orbiting at 2.7 and 9\,au, whose presence were initially predicted from the disc morphology \citep[e.g.,][]{Heap2000, Okamoto2004}. Recent observations with JWST have shown evidence of a recent ($\sim$ 100 yr) collision between large planetesimals at $\sim 100$ au from \bp\ \citep{Chen2024, Rebollido2024}, hinting that the system is still rapidly evolving. 

Besides the disc and the planets, \bp\ is widely known for harbouring a unique system of extrasolar comets, or exocomets, which are analogues of comets in our own solar system. These objects are detected when their tails, composed of dust and gas, transit the star and produce absorption signatures in the stellar spectrum \citep[e.g.,][]{Ferlet_1987, Vidal-Madjar_1994, Lagrange_1995} or the star's light curve \citep[][]{Zieba_2019, Pavlenko_2022, Lecavelier_2022}. While signatures of such objects have been detected in other systems, such as HD\,172555 \citep{Kiefer_2014b} or 49\,Cet \citep{Miles_2016}, \bp\ stands out by the intensity of its cometary activity. On average, each optical \bp\ spectrum in the \caii\ H\&K lines shows six signatures of individual exocomets and, as of now, thousands of such objects have been detected, mostly using the \caii\ doublet \citep{Kiefer_2014}. The complexity of the system, combined with the close proximity of the star and the favourable orientation of the disc, make of \bp\ a unique laboratory for studying the last stages of planetary formation.

Many efforts have been devoted to study the orbital dynamics of \bp\ exocomets. Using the extensive set of observations obtained with the HARPS spectrograph, \cite{Kiefer_2014} showed that the comets can be divided into two broad dynamical families: the S family (for \emph{shallow}), producing faint and wide absorption features at high velocities (typically from –50 to more than +100 km/s), and the D family (\emph{deep}), producing narrow and deep features at lower velocities (–10 to +30 km/s). \cite{Kiefer_2014} showed that the S and D families can be distinguished by the depth of their absorption in the \caii\ K line, with D family comets producing absorption depth deeper than 40\,\%. Based on the hydrodynamical model of \cite{Beust_1993}, \cite{Kiefer_2014} also inferred the typical transit distances of exocomets from the two families, finding $10 \pm 3\,R_\star$ ($0.07 \pm 0.02$ au) for the S family and $19 \pm 4\,R_\star$ ($0.14 \pm 0.03$ au) for the D family. These distances are much smaller than the typical periastron distances of the brightest comets in the solar system (e.g., 0.91\,au for Hale-Bopp). They are however consistent with the large amounts of refractory ions (\feplus, \caplus, etc.) found in \bp\ exocomets gaseous tails \citep[e.g.][]{Vrignaud24}, which are released by the sublimation of dust grains only at stellocentric distances below a few tens of stellar radii (for instance, \cite{Beust1996} provides a sublimation distance of 35\,$R_\star\,=\,0.25$\,au for tektite glass).

Directly measuring the transit distance (and, more generally, the orbital elements) of exocomets is a difficult task. Based on the acceleration of exocomet signatures in \caii\ lines,  \cite{Kennedy_2018} was able to measure the transit distance of a few \bp\ exocomets. All the studied objects were found to transit below $20\,R_\star$ (0.14\,au), in agreement with \cite{Kiefer_2014}. More recently, \cite{Vrignaud24b} showed that the excitation state of \feii\ in \bp\ exocomets is primarily set by the stellar radiation, and is therefore a direct probe of the distance to the star. Applying this principle to an exocomet with a RV of $\sim$30\,km/s observed with the \emph{Hubble} Space Telescope (HST), \cite{Vrignaud24b} showed that the comet transited at less than $60\,R_\star$ (0.4 au), similar to the objects studied in \cite{Kennedy_2018}.

However, as of now, all the exocomets which had their transit distance directly measured \citep[][]{Kennedy_2018, Vrignaud24b} belong to the S family. This is because S-family comets signatures are generally more isolated and exhibit larger accelerations, owing to their closer proximity to \bp. In contrast, the properties of D family comets remain poorly constrained. Their absorption signatures are frequently blended and overlap with the stable circumstellar disc absorption \citep[see, e.g.,][]{Kiefer_2014}, making them much more difficult to analyse. As a result, no direct constraints currently exist on the dynamical properties of the D family. Their transit distances ($19 \pm 4,R_\star$) were inferred indirectly by \citet{Kiefer_2014} using the dynamical model of \citet{Beust_1993}. More recently, \citet{Beust2024} showed that the low radial velocities of D family comets are consistent with an origin close to \bp\ (0.5–1.2 au), where bodies can fall in high-order mean-motion resonances with \bp\ c.

\begin{table*}[!b]
    \centering 
    \begin{threeparttable}
    \caption{Log of the \bp\ spectra obtained on 2025, April 29.}
    \renewcommand{\arraystretch}{1.2} \begin{tabular}{ c c c c c c c c c}                    
    \cline{1-8}      
    \noalign{\smallskip}
    \cline{1-8}      
     
     Instrument & Aperture & Filter/Grating & Wavelength & Start time & Exposure time & Spectral resolution   & \\
                &          &                &    (\A)    &    (UT)    &     (s)      &   (km/s) &            &  \\
                  
    \cline{1-8}      
    \noalign{\smallskip}

    \multirow{6}{*}{STIS} & 0.1x0.03 & E230H & 1880 - 2140 &   15:08 & 382 & 1.9    \\ 
    & 31X0.05NDA & E230H  & 2670 - 2930 & 15:19 & 500 &  2.3 &  \\ 
    & 0.2X0.2 & E140H  & 1500 - 1680 & 16:16 & 2474  & 1.5  \\ 
    & 0.2X0.2 & E230H  & 1630 - 1900 & 17:48 & 855 & 2.3 \\ 
    & 0.1x0.03 & E230H  & 2130 - 2400 & 18:09 & 555 & 1.9 \\ 
    & 0.1x0.03 & E230H  & 2380 - 2650 & 18:23 & 555 & 1.9 \\ 

    \noalign{\smallskip}
    \cline{1-8}      
    \noalign{\smallskip}

    COS & PSA &  G130M & 1140 - 1430 & 19:28 & 2144  & $15-20$ \\ 
    
    \noalign{\smallskip}
    \cline{1-8}      
    \noalign{\smallskip}

    HARPS & & &  3780 - 6910 & 22:53 & 8100\tnote{a} & $2.7$ \\ 
    
    \noalign{\smallskip}
    \cline{1-8}      
\end{tabular}

    \begin{tablenotes}
      \item[a] To avoid saturation, the full HARPS observation consists in a sequence of 27 exposures of 300 s each.
    \end{tablenotes}
    
\label{Tab. data}

\end{threeparttable}
\end{table*}

\FloatBarrier

\begin{figure*}[h!]
\centering
    \includegraphics[scale = 0.44,     trim = 55 10 0 50,clip]{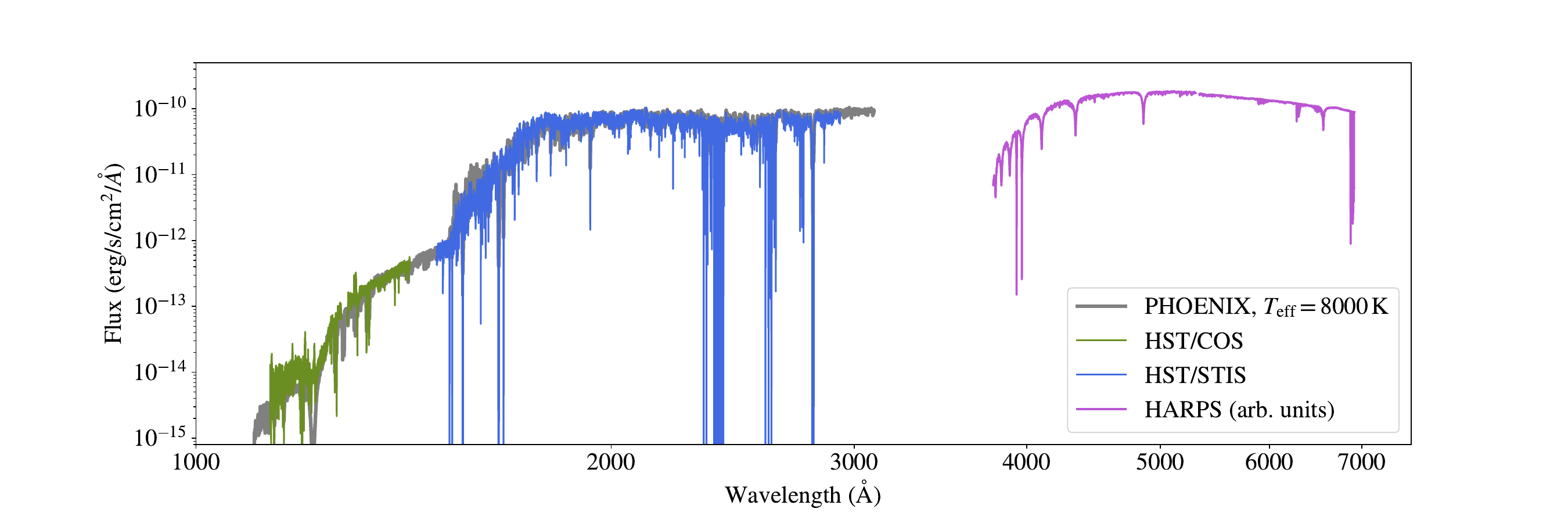}    
    \caption{Spectroscopic observations of \bp\ obtained on April 29, 2025. The HARPS spectrum is not flux-calibrated and was simply renormalized to a flux level similar to that of the STIS data. Numerous saturated absorption lines are visible in the STIS and HARPS spectra, attributed to \siii, \Ci, \alii, \feii, \mgii\ and \caii. The HST spectra are consistent with a stellar model from the PHOENIX library \citep[thick grey line,][]{PHOENIX} at $T_{\rm eff} = 8000$\,K (see Sect. \ref{Sect. Input stellar spectrum}). }
    \label{Fig. full spectrum}
\end{figure*}

The goal of the present study is to directly constrain the orbital properties of D family comets, using spectroscopic observations obtained with the HST in 2025. These observations cover numerous UV \feii\ lines spanning a wide range of excitation levels, which can be used as a probe of the transit distance of exocometary tails \citep{Vrignaud24b}. The present paper is organised as follows. In Sect.~\ref{Sect. Studied observations}, we present our spectroscopic observations and the three exocomets under study. Then, in Sect.~\ref{Sect. Modelling the circumstellar absorption} and~\ref{Sect. Fit}, we describe our approach to model the absorption spectra of the transiting exocomets and fit our spectroscopic data. Finally, in Sect.~\ref{Sect. Discussion}, we report our measurements of the transit distances and compositions of the studied exocomets, and we discuss the implication for the dynamical evolution of exocometary tails in the \bp\ system.

\section{Description of the observations}
\label{Sect. Studied observations}

\subsection{Studied spectrum}
\label{Sect. Raw data}

Spectroscopic observations of \bp\ were obtained on April 29, 2025 from three instruments: the Space Telescope Imaging Spectrograph (STIS) and the Cosmic Origin Spectrograph (COS) onboard the HST (program \#115.282N, PI T. Vrignaud), and the High Accuracy Radial Velocity Planet Searcher (HARPS) mounted on the 3.6m telescope of the European Southern Observatory (program \#115.282N PI T. Vrignaud). A summary of the observations used in the present study is provided in Table~\ref{Tab. data}, and the full spectrum is shown in Fig.~\ref{Fig. full spectrum}.

The studied dataset covers almost the whole wavelength range from 1000 to 7000 \A. Specifically, the COS spectrum covers the 1100-1500 \A\ range, the six STIS spectra cover the 1500-2900~\A\ range, and the HARPS spectrum covers the 3800-6900~\A\ range. Altogether, they provide access to a large range of spectral lines (e.g., \fei, \feii, \caii, \aliii, \Ci, or \Cii), allowing for a detailed analysis of the physical and chemical properties of the circumstellar gas occulting the star. Simultaneous observations of such a large number of chemical species in the \bp\ system are unprecedented.

The present analysis is based on the 1D extracted orders from the six STIS spectra (25 to 65 orders per spectrum), and the two spectral segments from the COS observations. For HARPS, we use the merged 1D spectrum (in which all spectral orders were resampled on a unique wavelength grid), summed over the full 2.25-hour observing sequence (27 exposures of 300\,s each). The wavelengths table of the studied spectra were shifted to the \bp\ rest frame, assuming a RV of 20.5\,km/s \citep{Brandeker_2011}. In the following, all RVs will be expressed in \bp\ rest frame.

Little reduction was applied to our dataset. The six STIS spectra, which partly overlap,  were all renormalised to a common flux level, using a procedure similar to that described in Sect. 2 of \cite{Vrignaud24}. This method consists in interpolating each individual spectrum with a cubic spline, calculated on spectral regions free of circumstellar and interstellar absorption. Each spectrum is then divided by its corresponding spline, which mitigates the effects of flux calibration variations from one exposure to another.

\subsection{Overview of the detected signatures}
\label{Sect. Recovery of a reference spectrum}

\begin{figure}[h]
\centering
    \includegraphics[scale = 0.385,     trim = 30 40 0 22,clip]{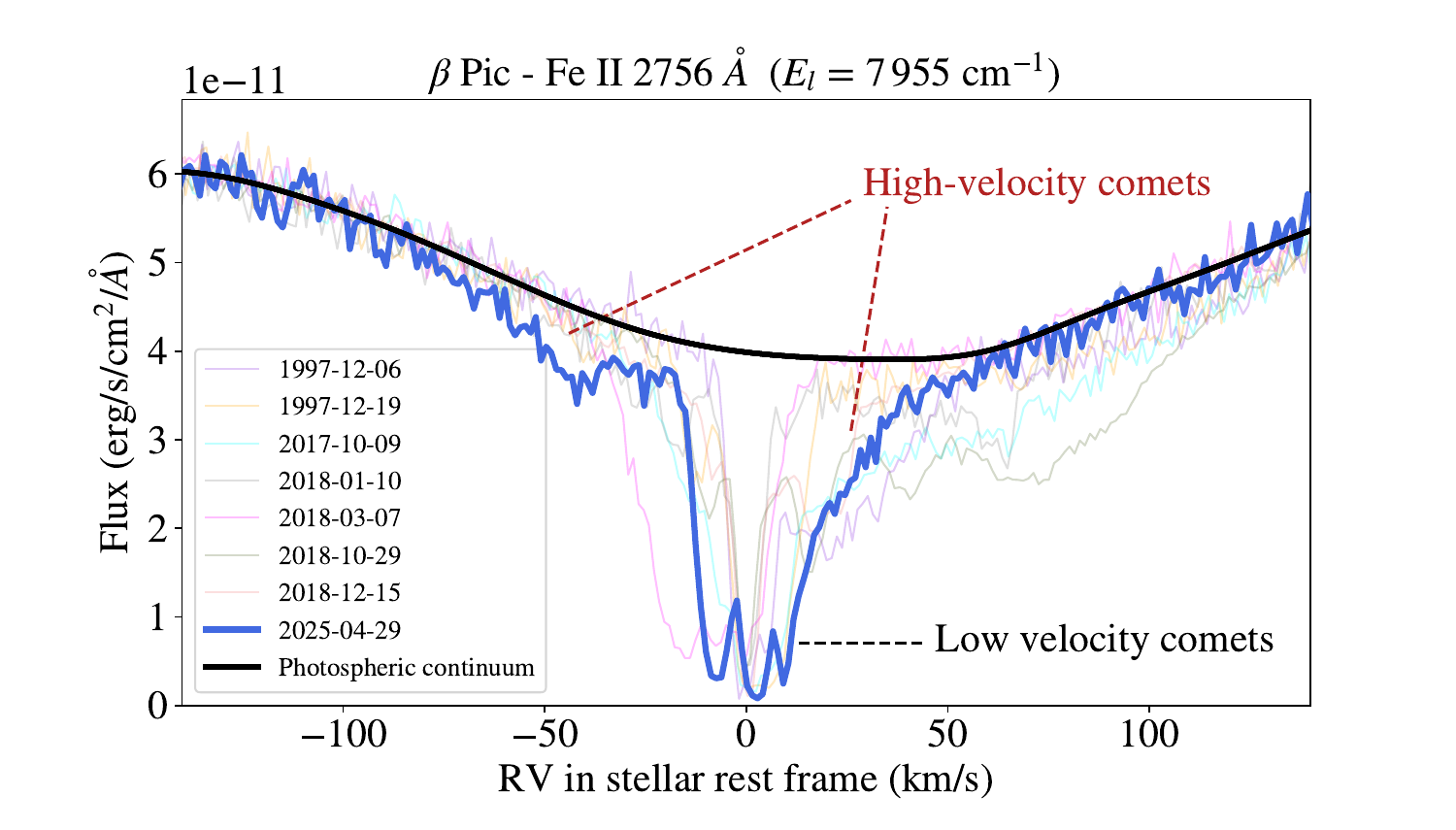}    
    \includegraphics[scale = 0.385,     trim = 30 10 0 12,clip]{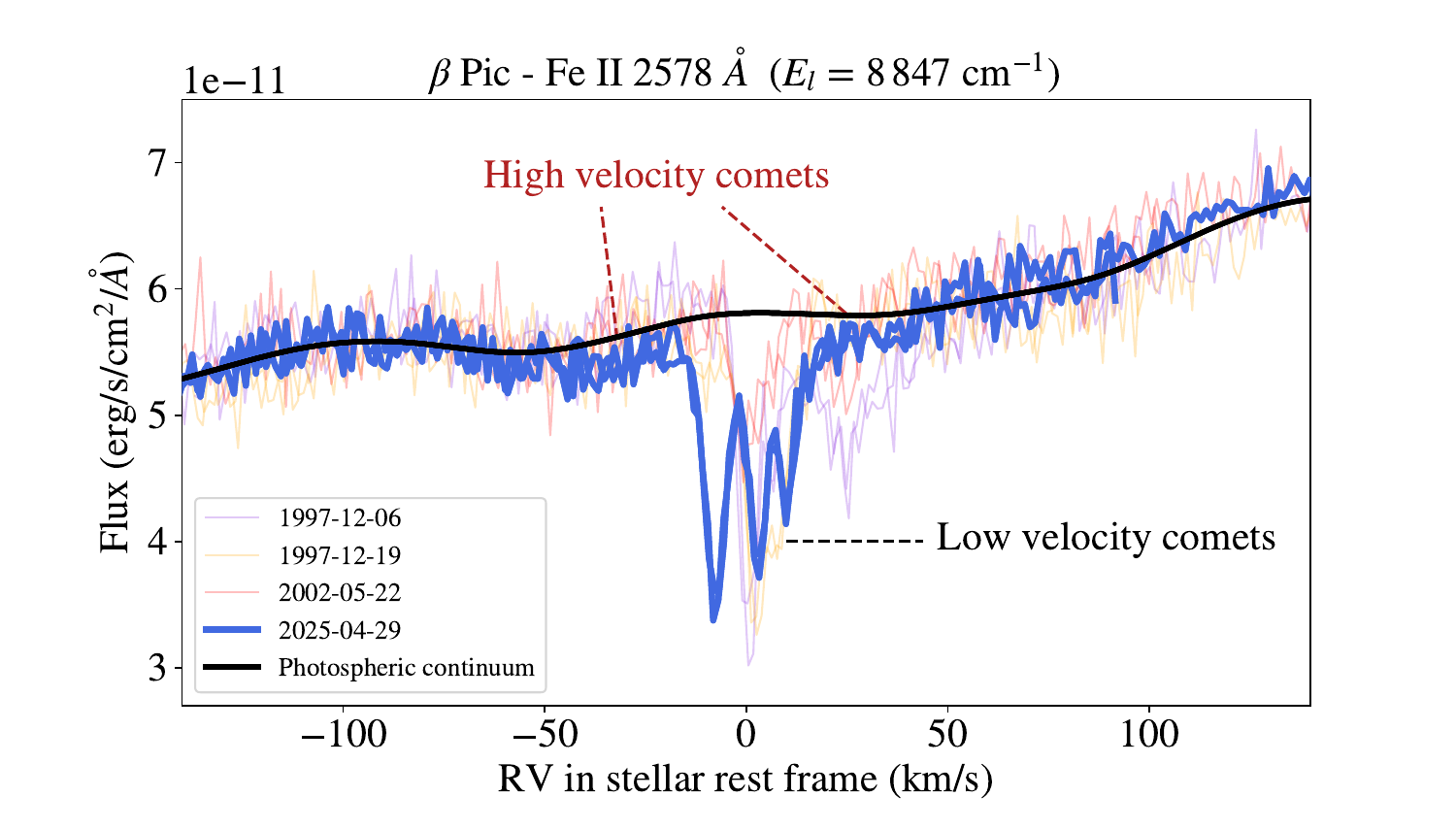}    
    \caption{Zoomed-in view of the STIS spectra of \bp\ of April 29, 2025, around two \feii\ lines. \textbf{Top}: View of the strong \feii\ line at 2756.5 \A. Archival STIS spectra of \bp\ covering the same line are shown in transparent font. The solid black line represents an estimate of the stellar photospheric continuum, free of circumstellar absorption. \textbf{Bottom}: Same view of the shallower \feii\ line at 2578.7 \A. 
    This wavelength region is covered by two orders of the STIS echelle spectrograph. Since these two lines rise from excited states of \feplus, they don't show any signature from the circumstellar disc (see Sect. \ref{Sect. Closer look at the central components}).}
    \label{Fig. Comparison 2025 and before}
\end{figure}

A sample of the reduced STIS spectra obtained on April 29, 2025 is provided in Fig.~\ref{Fig. Comparison 2025 and before} around two \feii\ lines. For comparison, we also show archival STIS spectra of \bp\ obtained between 1997 and 2018, as well as an estimate of the \bp\ spectrum free of circumstellar absorption (black), reconstructed using all archival observations. The April 29, 2025 spectrum displays many absorption features, which we associate with three types of objects:

\begin{itemize}
    \item Low-velocity comets (LVCs), which produce deep and narrow absorption lines at RVs below 15 km/s. The April 29, 2025 spectrum shows three signatures from LVCs, around -7.5, +2.5 and +10 km/s (see Fig.~\ref{Fig. Comparison 2025 and before}). We associate these objects with the D family of \cite{Kiefer_2014}.
    \item High-velocity comets (HVCs), which produce shallow and broad absorption features at higher velocities (typically from about $-100$\,km/s to more than +100\,km/s). Several HVCs are detected in the April 29, 2025 spectrum, particularly around $-50$\,km/s and +30\,km/s (see Fig.~\ref{Fig. Comparison 2025 and before}). These objects have the typical characteristics of the S family \citep{Kiefer_2014}. 
    \item The circumstellar disc, which produces a narrow and stable signature at 0\,km/s. Since the disc is located further out, it is primarily detected in the lines from the ground state level (see Sect.~\ref{Sect. Closer look at the central components}).
\end{itemize}

The HVCs seen in the April 29, 2025 spectrum resemble the exocomet signature observed on December 6, 1997 and studied in \cite{Vrignaud24b}. The transit distance of this previous comet was found to be below 0.4 au, based on the excitation state of \feii\ in the gaseous tail. Given the similarity between this previously studied signature and the HVCs observed in the April 29, 2025 spectrum, it is likely that the latter are also produced by gaseous tails at short stellar distances, typically well below 1 au. 

LVCs, however, are more difficult to interpret. Previous studies \citep[e.g.][]{Beust1998, Tobin2019, Hoeijmakers_2025} showed that these features are fairly different from HVCs: they are much less time-variable (typically lasting for several days), don't show any acceleration \citep[e.g.][]{Kennedy_2018} and have much narrower absorption lines, with typical width of 1 to 5\,km/s. The goal of the present study is to investigate the physical properties of these transiting components, using the spectrum obtained on April 29, 2025.

\subsection{The low velocity comets of April 29, 2025}
\label{Sect. Closer look at the central components}

The April 29, 2025 spectrum shows absorption signatures from three strong LVCs (see Fig.~\ref{Fig. Comparison 2025 and before}). These LVCs are consistently detected in lines of \feii, \Si, \niii, \crii, \mnii, \siii\ (STIS and COS data), and \caii\ (HARPS data). Their signatures also seem to be present in \Ci, \alii\ and \mgii\ (STIS), also these lines are much harder to interpret due to their extreme saturation. Faint signatures may also be present in \znii, \coii\ and \srii.

To isolate the absorption spectra of these LVCs, we constructed a reference spectrum free of their absorption. This reference spectrum was computed using a cubic spline interpolation across the full STIS, COS, and HARPS datasets, excluding all wavelength bins located within 20~km/s of any line where absorption from the LVCs is expected. The list of lines considered in this calculation was obtained through the following procedure: for all the species listed above (\feii, \niii, ...), we identified a set of metastable states with clear signatures from the LVCs (for instance, the 8 lowest levels of \niii). For each of these metastable states, we then selected all lines with large oscillator strengths (typically > 0.01). The 20~km/s threshold was chosen because signatures from the three LVCs are confined within the [-20,+20] km/s range for all the detected lines.

\begin{figure}[h!]
\centering
    \includegraphics[scale = 0.385,     trim = 25 30 0 15,clip]{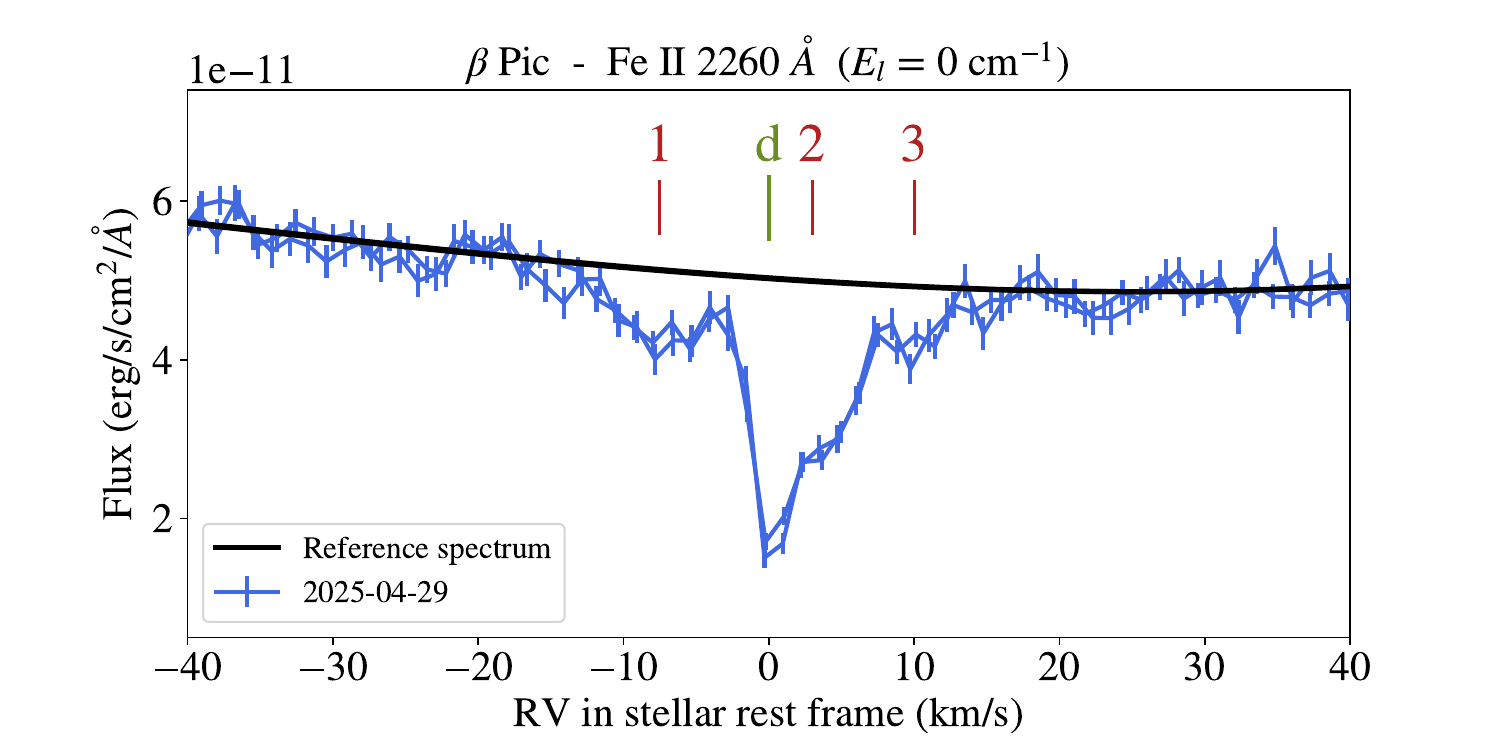}    
    \includegraphics[scale = 0.385,     trim = 25 30 0 0,clip]{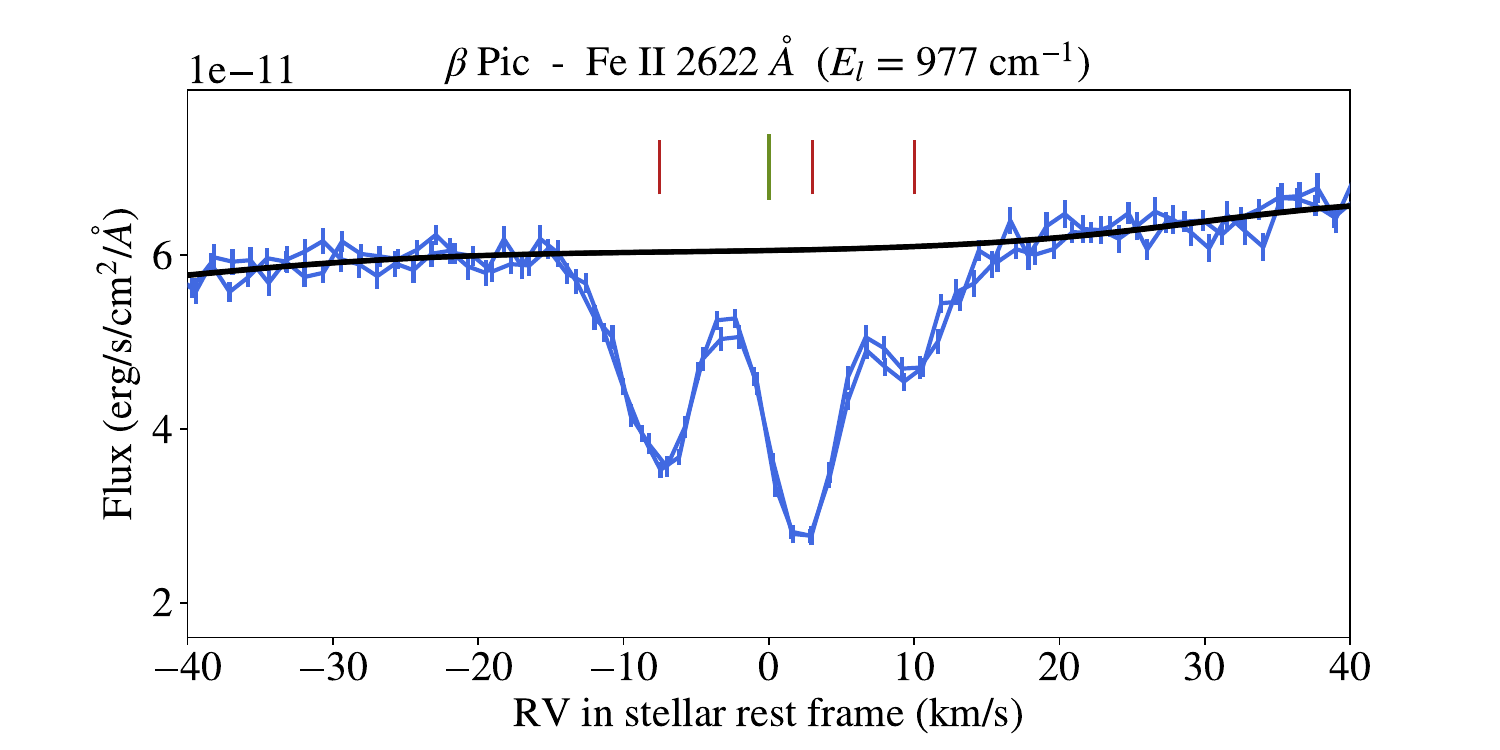}    
    \includegraphics[scale = 0.385,     trim = 25 30 0 0,clip]{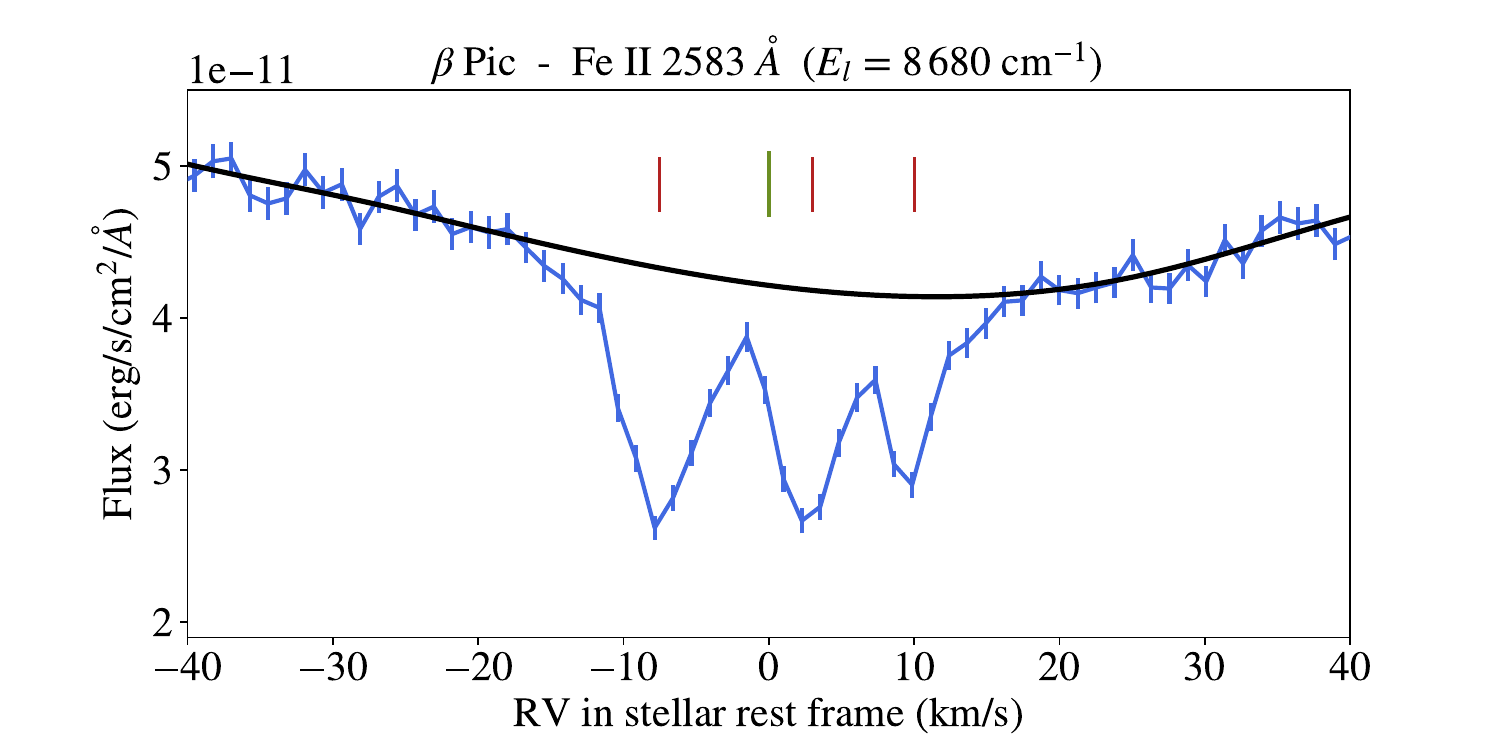}    
    \includegraphics[scale = 0.385,     trim = 25 10 0 0,clip]{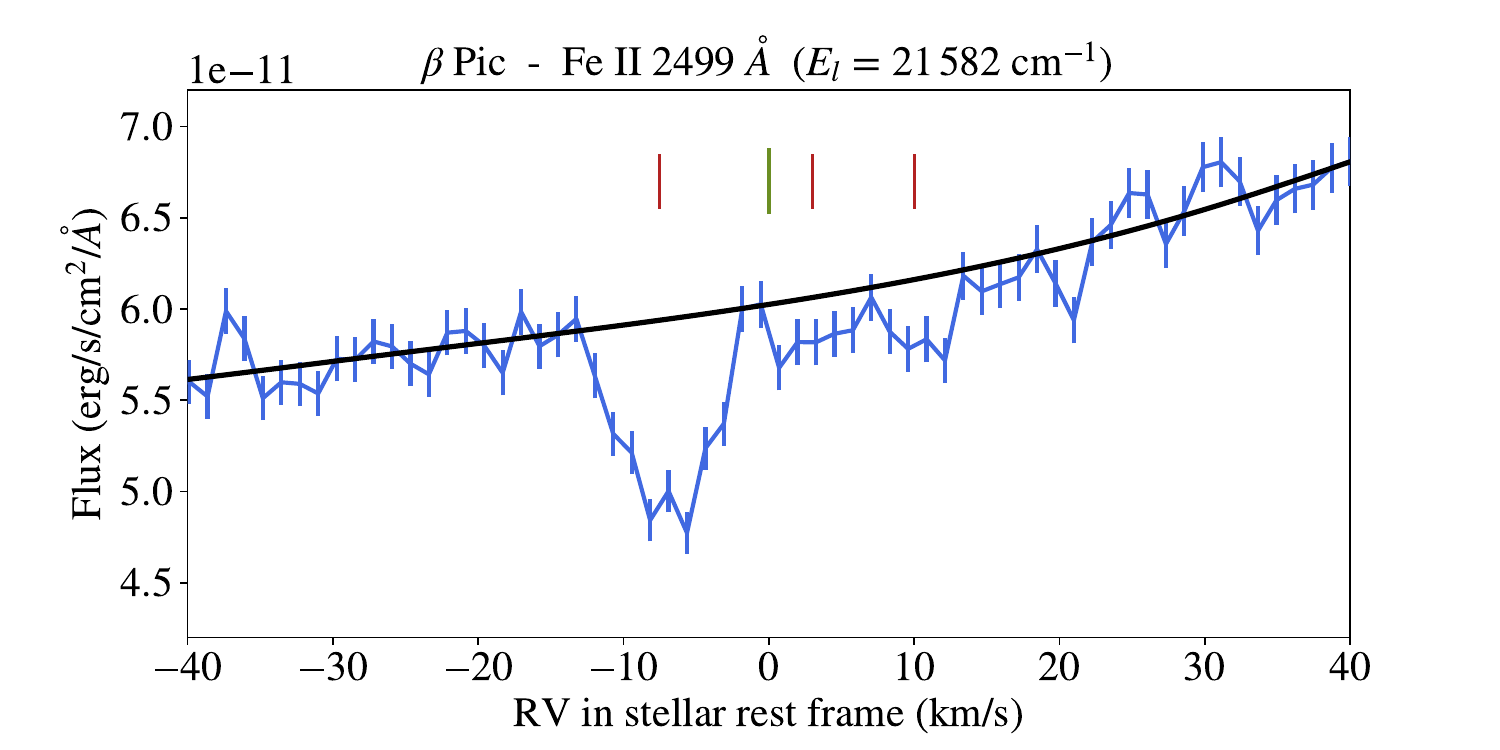}    
    \caption{Zoomed-in view of the STIS spectra of \bp\ obtained on April 29, 2025, around four \feii\ lines rising from different excitation levels. Errors bars were tabulated from the STIS data reduction pipeline. Red ticks indicate the velocities of the three main LVCs. The circumstellar disc (green tick, 0 km/s) is detected in the ground state of \feii. Note that the $\lambda \lambda$2260 and 2622 \A\ lines are covered by two spectral orders.
    }
    \label{Fig. Fe II lines}
\end{figure}

\begin{figure}[h!]
\centering
    \includegraphics[scale = 0.385,     trim = 25 30 0 15,clip]{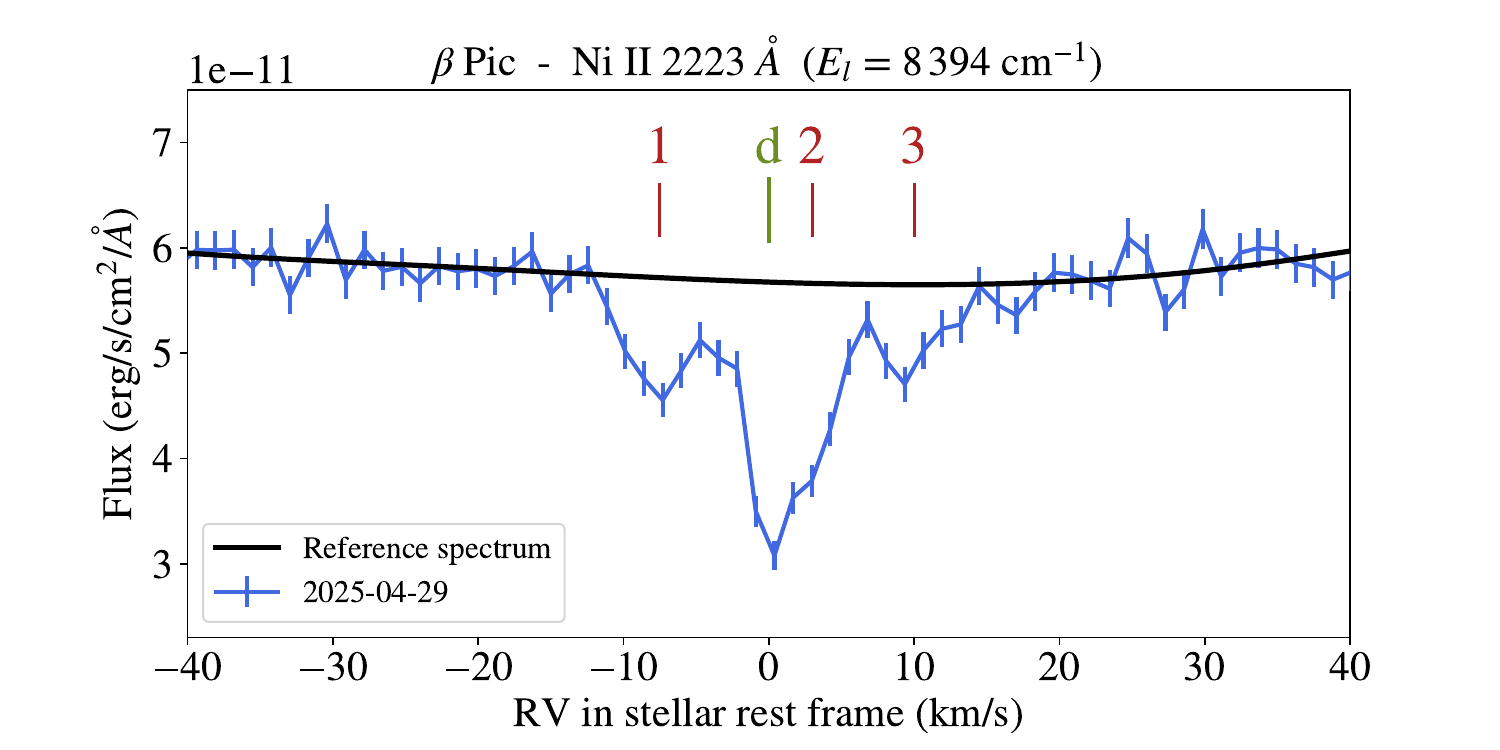}    
    \includegraphics[scale = 0.385,     trim = 25 30 0 0,clip]{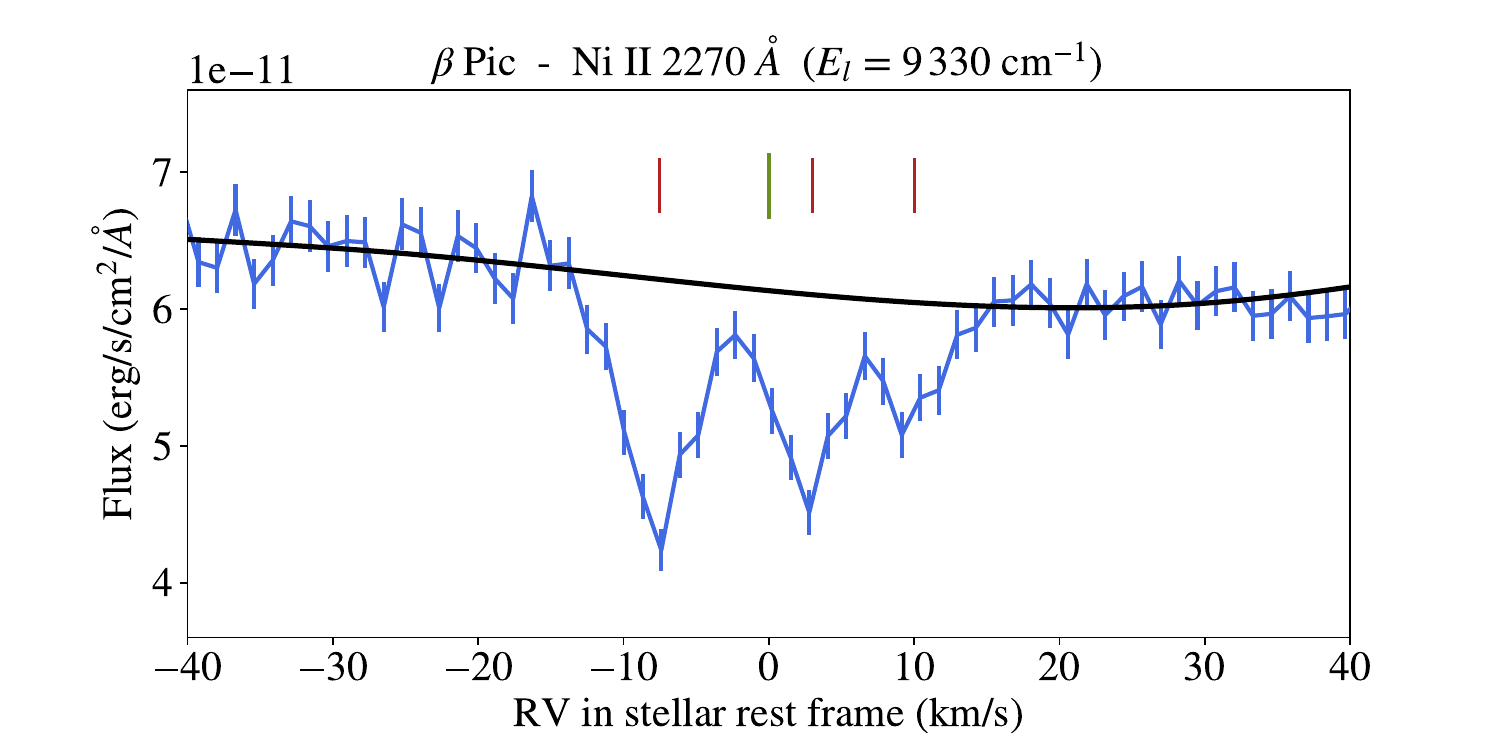}   
    \includegraphics[scale = 0.385,     trim = 25 30 0 0,clip]{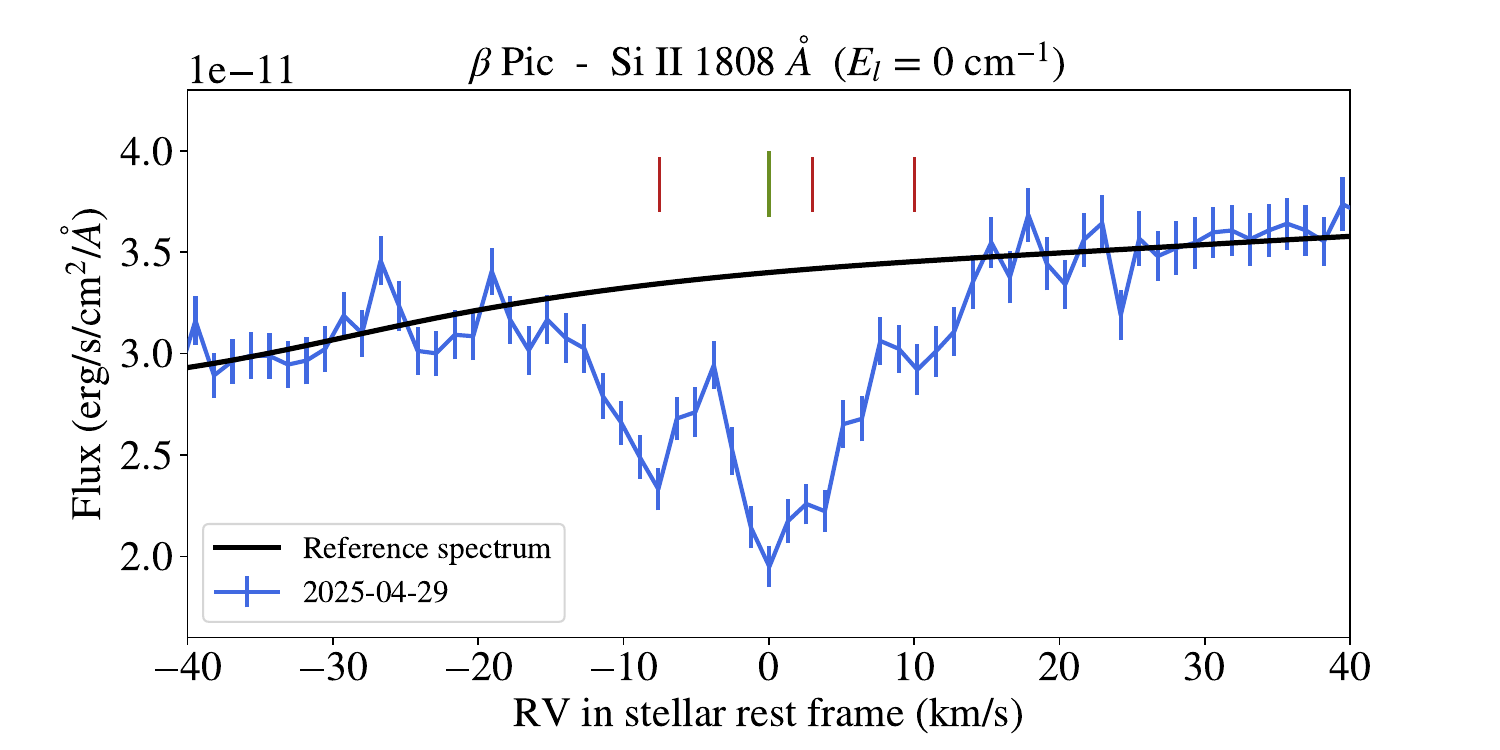}    
    \includegraphics[scale = 0.385,     trim = 25 10 0 0,clip]{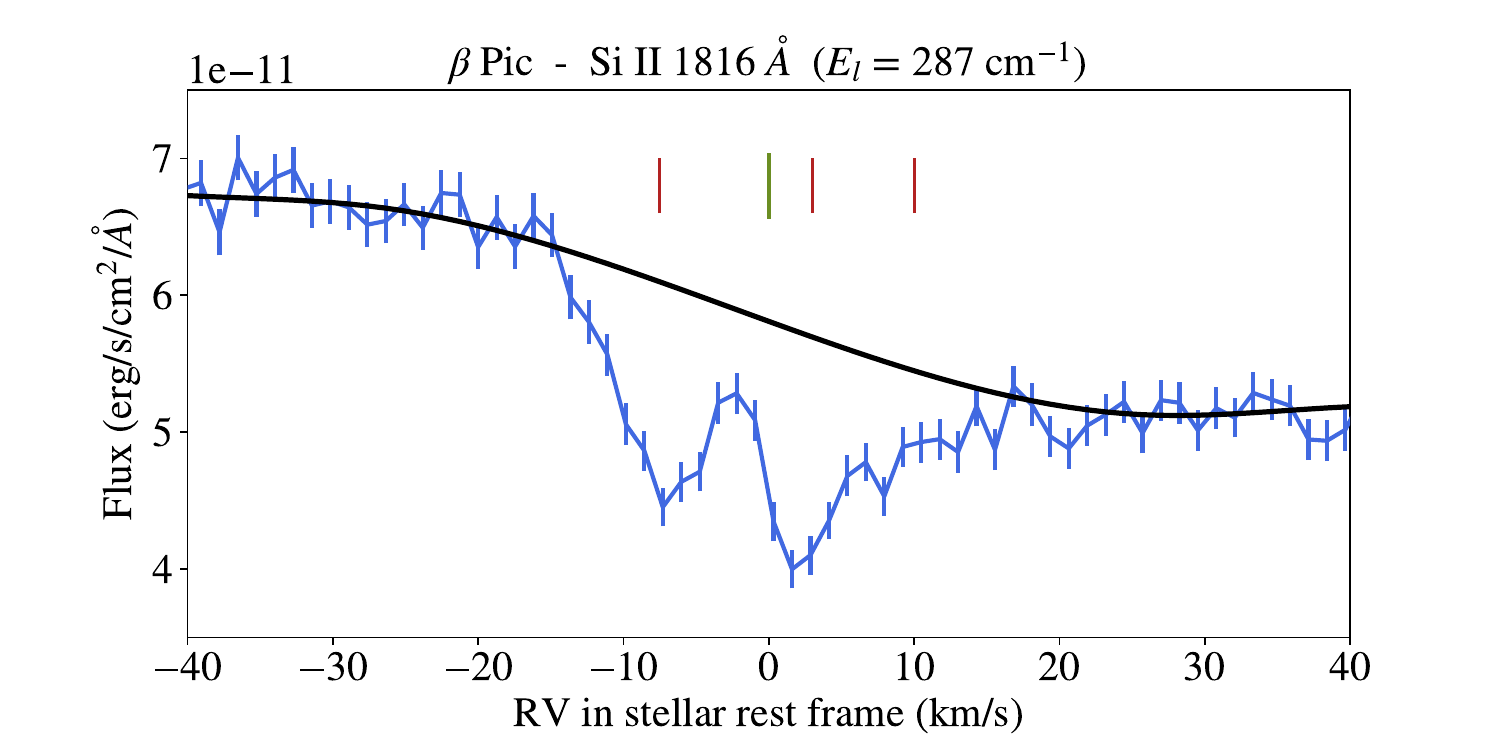}   
    \caption{Same as Fig.~\ref{Fig. Fe II lines} for two \niii\ lines and two \siii\ lines. The disc is detected in the metastable state of \niii\ at 8393 cm$^{-1}$, and in the ground state of \siii.}
    \label{Fig. Ni II + Si II lines}
\end{figure}

This reference spectrum is shown on Figs.~\ref{Fig. Fe II lines} and \ref{Fig. Ni II + Si II lines} around several \feii, \niii\ and \siii\ lines. We emphasize that it is not intended to represent the stellar photospheric continuum, as it remains polluted by broad absorption features from HVCs. Instead, the goal of this reference spectrum is to estimate what the April 29, 2025 \bp\ spectrum would look like without the three LVCs. The absorption spectrum of the three LVCs alone is obtained by taking the ratio of the observed spectrum to this reference.

The velocities of the three LVCs are identified with red ticks on Figs. \ref{Fig. Fe II lines} and \ref{Fig. Ni II + Si II lines}. Following the nomenclature introduced in \cite{Lecavelier_2026}, we designate these comets as C20250429a, C20250429b, and C20250429c. For brevity, we will refer to them hereafter as LVCs \#1, 2 and 3; LVC~\#1 is centred around -7.5 km/s, LVC~\#2 around 2.5 km/s, and LVC~\#3 around 10 km/s. A first look at the \feii\ and \niii\ signatures of the three LVCs reveals differences in their excitation states. Lines arising from low-excitation levels show a strong contribution from LVC~\#2 (centred around $\simeq$ 2.5 km/s) compared to LVCs~\#1 and 3. In contrast, lines from highly excited levels show a very strong signature from LVC~\#1 (see the bottom panel of Fig.~\ref{Fig. Fe II lines}). Since the excitation state of exocometary tails is primarily controlled by the local radiation flux, and thus by the distance to the host star \citep[][]{Vrignaud24b}, this suggests that the three LVCs are located at different distances from \bp. Specifically, LVC \#1 seems to be rather close to the star (given the presence of excited \feii\ levels), while LVC \#2 appears to be further out. The goal of the remainder of this work is to quantify this general picture.

In addition to the three LVCs, Figs. \ref{Fig. Fe II lines} and \ref{Fig. Ni II + Si II lines} show the presence of an additional narrow absorption feature centred around 0\,km/s. Among the lines shown in these figures, this central component is only detected in the \feii\ $\lambda$2260\,\A\ and \siii\ $\lambda$1808\,\A\ lines rising from the ground states of \feii\ and \siii\ ($E_l$=0\,cm$^{-1}$), and in the \niii\ $\lambda$2223\,\A\ line rising from the metastable state at $E_l$=8394\,cm$^{-1}$. This signature is naturally associated with the circumstellar disc surrounding \bp. It is detected in the ground state of all atomic and ionised species (e.g. \Si, \feii, \niii, \crii, \siii...), as well as in the $E_l$=8394\,cm$^{-1}$ metastable state of \niii, which has a long radiative lifetime. As discussed below, the weaker excitation of the disc is explained by its large distance to \bp\ and to its low density, which limit the rates of radiative and collision excitation.

\section{Modelling the low velocity comets' and disc's absorptions}
\label{Sect. Modelling the circumstellar absorption}

\subsection{Principle}

The various absorption features detected in the April 29, 2025 observations offer a favourable configuration for analysis: the three LVCs are spectrally resolved, and since none of them peaks around 0 km/s, the contribution from the circumstellar disc can also be isolated. The goal of the following sections is to build a model of the absorption spectra of these four components (the three LVCs and the disc), and to fit this model to the spectra obtained on April 29, 2025 in order to infer their physical and chemical properties. To achieve this, we follow and extend the approach of \citet{Vrignaud24b}, who studied the excitation state of \feii\ in a \bp\ exocomet. Contrary to this previous study, we will here directly fit the observed spectra of the transiting components, rather than their excitation diagrams. This will allow us to rigorously account for the blending between the absorption signatures of the four components, the instrumental spectral dispersion, and the mixing of closely spaced multiplets.

We propose to model each transiting component with the simplest parameterisation possible. We assume that each component is located at a unique distance $d$ to \bp, and has a unique electronic temperature $T_e$ and electronic density $n_e$ (since the tails are ionised, electrons are considered to be the main collision partner.). The line profile of each component is described by a Voigt profile, with a central velocity $v$ and a Doppler broadening parameter $b$. We assume that the studied species are well mixed within each component, due to efficient momentum transfer between ions through Coulomb interactions \citep[][]{Beust1989, Vrignaud2025}. This assumption is supported by the similarity of the line profile observed in different species, and by the quasi-solar composition found in all transiting components (see Sect. \ref{Sect. ionisation state of the gas}).

\subsection{Self-absorption inside the gaseous clouds}
\label{Sect : Self-absorption inside the gaseous clouds}

As many lines in our spectrum are found to be highly saturated (most notably \feii, \niii\ and \siii), we also take into account the self-absorption of the transiting objects. To do so, each component is divided into a finite number of gas bins (typically $\sim$10–30), each containing an equal fraction of the total column density. These bins are arranged sequentially along the line of sight, such that the first bin is fully exposed to the stellar flux, while each subsequent bin receives a progressively attenuated flux due to absorption from the preceding bins. The transmitted spectrum is calculated as follows:

\begin{itemize}
    \item We start by considering the first bin of the innermost component, directly exposed to the  stellar flux. For all the studied species, we compute the level populations by solving the statistical equilibrium, taking into account both radiative and collisional excitation.
    \item Using a comprehensive list of spectral lines from all the studied species and the computed level populations, we calculate the flux transmitted by the first bin. This flux is then passed to the second bin, located just behind the first, which thus receives a slightly weaker radiation field.  
    \item The statistical equilibrium of the second bin is then solved, using the absorbed stellar spectrum. We compute its transmitted flux, which in turn becomes the input for the third bin. This process is repeated iteratively across all bins of all gaseous components, which are processed by increasing distance from the star.
    \item After all components have been considered, the last transmitted flux can be compared to our observations to constrain the physical parameters of the transiting gas clouds. 
\end{itemize}

In the following, each transiting component is discretized into a number of gas bins chosen such that each bin has a \feii\ column density of $2\times10^{13}$ cm$^{-2}$, except for the last bin. Since the \feii\ column densities in our components range from 1.5 to $6 \times 10^{14}$ cm$^{-2}$ (see Sect. \ref{Sect. Discussion} below), this results in a reasonable number of 10-30 gas bins per component. The full model is computed in $\sim 1$ s on a single CPU.

\subsection{Statistical equilibrium}
\label{Sect. Statistical equilibrium}

To calculate the model spectrum, the statistical equilibrium needs to be computed in each bin of the transiting clouds. 

Let us consider a gaseous component located at a distance $d$ from the star, characterised by an electronic density $n_e$ and an electronic temperature $T_e$. For any excitation level $i$ of any atom or ion $X$, the statistical equilibrium equation is expressed as: 
$$\diff{n_i}{t} = 0,$$
where $n_i$ is the population of level $i$. Taking into account all radiative and collisional exchanges with other levels $j \neq i$, this reads \citep{Vrignaud24b}:

\begin{flalign}
\label{Eq. Statistical equilibrium 1}
  & \sum_{j<i} \left( C_{ji} n_e n_j - C_{ij} n_e n_i + B_{ji} \overline{J}_{ij} n_j - B_{ij} \overline{J}_{ij} n_i - A_{ij} n_i \right) & \notag \\
+ & \sum_{j>i} \left( C_{ji} n_e n_j - C_{ij} n_e n_i + B_{ji} \overline{J}_{ij} n_j - B_{ij} \overline{J}_{ij} n_i + A_{ji} n_j \right) = 0, &
\end{flalign}
where $C_{ij}$ and $C_{ji}$ are the collisional excitation rates for the $i\rightarrow j$ and $j\rightarrow i$ transitions (which depend on the electronic density and temperature), $A_{ji}$, $B_{ij}$ and $B_{ji}$ the Einstein coefficients for spontaneous decay, absorption and stimulated emission, and $\overline{J}_{ij}$ the line-averaged flux in the $i\leftrightarrow j$ transition. $\overline{J}_{ij}$ is computed as:
\begin{align}
\label{Eq. J_nu}
    \hspace{2.9 cm} \overline{J}_{ij} = \int_0^{+\infty} J_\nu \cdot \phi_{ij}(\nu) \ \mathrm{d} \nu,
\end{align}
where $\phi_{i,j}$ is the line profile of the $i\leftrightarrow j$ transition in the considered component and $J_\nu$ the solid angle-averaged specific intensity at frequency $\nu$: 
$$J_\nu = \frac{1}{4 \pi} \int_{4 \pi} I_\nu\, \mathrm{d}\Omega.$$

In the case of an optically thin medium, $J_\nu$ is dominated by the stellar flux, and is thus given by $J_\nu= W \, I_\nu^\star$, where $W = \frac{1}{2} \ \Big (1 - \sqrt{1 - (R_\star/d)^2 } \ \Big ) \approx (\pi R_\star^2) / (4\pi d^2)$ is the dilution factor and $I_\nu^\star$ the average specific intensity of the stellar disc (unit: erg/s/cm$^2$/Hz/sr). This approximation was used in \cite{Vrignaud24b}. In the present case, however, we choose to take self-absorption into account, by progressively reducing the radiation field going through the gas after it is absorbed (Sect.~\ref{Sect : Self-absorption inside the gaseous clouds}). 
The stellar flux $J_\nu$ received by a gas bin is thus given by: 
$$J_\nu =    W \, I_\nu^\star \, e^{-\tau_\nu},$$
where $\tau_\nu$ is the optical thickness of the material occulting the current bin from the star. The corresponding line-integrated flux $\overline{J}_{ij}$ can be written as:
$$
\overline{J_{i,j}} = \overline{J}_{ij, \, \text{abs}}^\star :=  W \int_0^{+\infty}  I_\nu^\star \ e^{-\tau_\nu} \cdot \phi_{ij}(\nu) \ \mathrm{d} \nu.
$$

In addition, Eq. \ref{Eq. Statistical equilibrium 1} neglects the possibility that photons produced by spontaneous emission are immediately reabsorbed by the surrounding gas when the cometary tail is optically thick. In this case, the excitation energy of a given ion is simply transferred to another particle, and the global level populations remain  unchanged. This process is likely significant for the exocomets analysed in this work, whose signatures in \feii\ or \siii\ lines are often strongly saturated. To account for this, we make use of the escape probability formalism, which consists in weighting the spontaneous decay rates $A_{ij}n_j$ and $A_{ji}n_i$, appearing in Eq. \ref{Eq. Statistical equilibrium 1}, by the probability that the emitted photon escapes the cloud in a single flight. To calculate this probability, we assume that the components have symmetric geometries relative to the line of sight, and that the transverse column densities $N_{\rm trans}(X)$, taken from the cloud centre to their edge, are much smaller than the radial column densities $N(X)$. For each component, we introduce a new parameter $f$, defined as the ratio between the transverse and radial column density:
$$
N_\text{trans}(X) = f \cdot N(X).
$$
For the three LVCs, the escape probability of photons $P_{ij}$ in a given transition $i \leftrightarrow j$ of species $X$ is calculated assuming that the cometary tails have cylindrical, tube-like geometries. Integrating over all possible directions and frequencies, one finds: 
$$
P_{ij} = \frac{1}{4\pi}\displaystyle \int_{0}^{\pi} 2\pi \sin(\theta) \, \mathrm{d}\theta \int_0^{+\infty} \phi_{ij}(\nu) \, \mathrm{d}\nu \, \exp\left(-\frac{\tau_{\rm trans}(\nu)}{\sin(\theta)}\right),
$$
where $\tau_{\rm trans}(\nu)$ is the optical depth of the cloud at frequency $\nu$ in the transverse direction, and $\theta$ is the angle between the line of sight symmetry axis and the escape direction. The optical depth $\tau_{\rm trans}(\nu)$ is directly computed from the transverse column density $N_\text{trans}(X)$ and the level populations in the cloud, and thus depends on $f$.

For the disc, we adopt a slab geometry. The escape probability takes a slightly different form: 
$$
P_{ij} = \frac{1}{4\pi} \displaystyle \int_{0}^{\pi} 2\pi \sin(\theta) \, \mathrm{d}\theta \int_0^{+\infty} \phi_{ij}(\nu) \, \mathrm{d}\nu \, \exp\left(-\frac{\tau_{\rm trans}(\nu)}{\cos(\theta)}\right),
$$
where $\tau_{\rm trans}(\nu)$ is here the optical depth of the disc at frequency $\nu$ in the direction perpendicular to the disc plane, and is also computed from $N_\text{trans}(X)$.

Ultimately, the full statistical equilibrium equation used to calculate the level populations $n_i$ is given by:
\begin{flalign}
\label{Eq. Statistical equilibrium 2}
  & \sum_{j<i} \left( C_{ji} n_e n_j - C_{ij} n_e n_i + B_{ji} \overline{J}_{ij, \, \text{abs}}^\star n_j - B_{ij} \overline{J}_{ij, \, \text{abs}}^\star n_i - P_{ij} A_{ij} n_i \right) & \notag \\
+ & \sum_{j>i} \left( C_{ji} n_e n_j - C_{ij} n_e n_i + B_{ji} \overline{J}_{ij, \, \text{abs}}^\star n_j - B_{ij} \overline{J}_{ij, \, \text{abs}}^\star n_i + P_{ij} A_{ji} n_j \right) & \notag \\
 & = 0.
\end{flalign}
In the model, each component is modelled with six global parameters, plus one column density per chemical species. The global parameters are the distance $d$ from the star; the electronic density $n_e$ and electronic temperature $T_e$; the component velocity $v$ and Doppler broadening parameter $b$; and the ratio $f$ between the transverse and radial column densities.

\section{Fit procedure}
\label{Sect. Fit}

\subsection{Input stellar spectrum}
\label{Sect. Input stellar spectrum}

\begin{figure}[h!]
\centering
    \includegraphics[scale = 0.385,     trim = 20 0 0 25,clip]{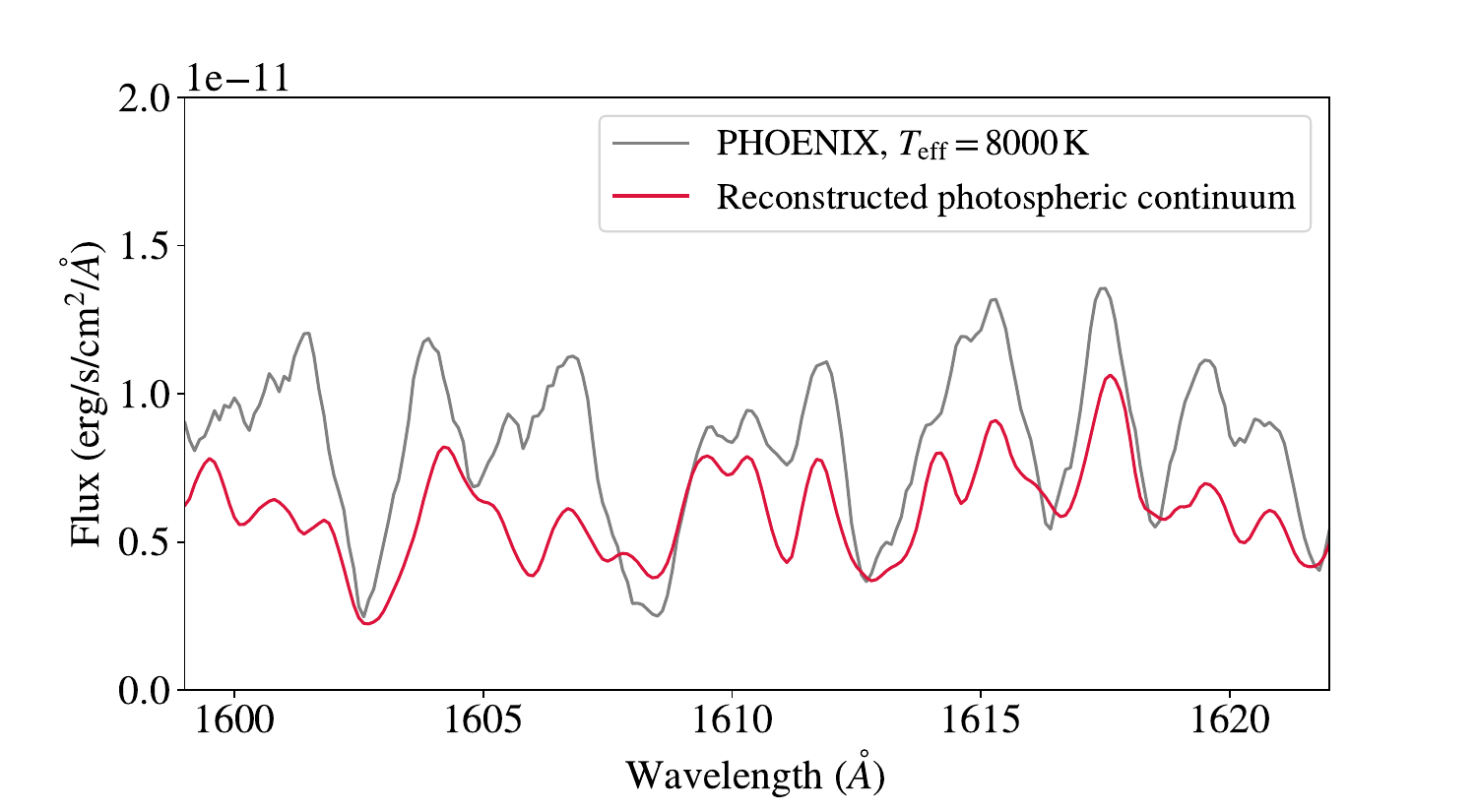}    
    \caption{Comparison between the PHOENIX model and the observed photospheric spectrum of \bp\ (also shown on Fig. \ref{Fig. Comparison 2025 and before}), reconstructed from archival observations. Fluxes are provided as observed from Earth (19.3 pc). The spectral region displayed includes several strong \feii\ transitions at 1608, 1612 and 1618~\A.}
    \label{Fig. model spectrum}
\end{figure}

To model the statistical equilibrium of the gas clouds transiting \bp, a model of the stellar spectrum $I_\nu^\star$ is needed (Eq.~\ref{Eq. Statistical equilibrium 2}). Following \cite{Vrignaud24b}, we use a synthetic stellar spectrum from the PHOENIX library \citep{PHOENIX}, with photospheric parameters close to that of \bp\ ($T_{\rm eff} = 8000 \ \si{K}$, $\log(g) = 4.5$, [Fe/H] = 0, [$\alpha$/M] = 0). This model matches well the overall shape of the observed \bp\ spectrum (see Fig. \ref{Fig. full spectrum}). However, we noted significant discrepancies in the depth of photospheric lines. To mitigate this, the PHOENIX spectrum was replaced with the observed photospheric continuum of \bp\ between 1100 and 3000~\AA, reconstructed from archival STIS and COS observations (see Fig.~\ref{Fig. Comparison 2025 and before}). A comparison between the original PHOENIX spectrum and the reconstructed photospheric spectrum is shown in Fig.~\ref{Fig. model spectrum}. This refined stellar spectrum provides more reliable estimates of the stellar flux in the transitions of the studied species ($\overline{J}_{i,j}$, see Eq. \ref{Eq. Statistical equilibrium 1}).

\subsection{Modelled species}

We choose to limit our study to abundant, neutral or singly ionised refractory species, with ionisation threshold above 10\,eV. Seven chemical species are considered: \feii, \niii, \caii, \crii, \mnii, \siii, and \Si. All these species exhibit strong optically-allowed transitions in the spectral range covered by our observations, and show very similar absorption patterns, with clear signatures from both the three LVCs and the circumstellar disc. Other species showing circumstellar absorption were not considered in the analysis, for the following reasons: 

\begin{itemize}
    \item Highly ionised species, such as \aliii, \Civ\ or \siiv, were not analysed, as they are dominated by broad absorption signatures from HVCs. In these lines, the signatures of the three LVCs appear to be weak, but are difficult to disentangle from HVCs;
    \item Neutral species with low ionisation potential, such as \fei, \nai\ or \mgi\ (5 - 8 eV), were also not analysed, because they are not detected in the LVCs;
    \item \mgii\ and \alii\ were not included in the fit, due to the extreme saturation of the available lines (1670 \A, 2800 \A).
    \item Finally, volatile tracer species, such as \Oi, \Hi, \Ci\ and \Ni, are left for a future study. These lines are particularly difficult to analyse, because they are affected of geocoronal emission (e.g. in \Hi\ Ly-$\alpha$) and appear to be very saturated (e.g. \Ci).
\end{itemize}

\subsection{Atomic data}

In order to compute the statistical equilibrium of each studied species, extensive datasets of transition probabilities, line wavelengths and effective collision strengths were collected. In most cases, the transition probabilities and line wavelengths were obtained from the NIST database \citep{NIST_ASD}, and the effective collision strength from CHIANTI \citep{CHIANTI_Dere_1997}. However, for species with complex spectra, such as \feii\ or \niii, the NIST and CHIANTI datasets were completed with more exhaustive studies (see Table \ref{Tab. fitted species}). The hyperfine structure of \mnii\ (which has a non-zero nuclear spin of $I = 5/2$) was taken into account, using hyperfine structure constants from \cite{Blackwell_2005}.

\subsection{Interstellar medium}

In spectral lines from the ground states of singly ionised species, the signature of LVC \#1 overlaps with that of the interstellar medium (ISM). Following \cite{Wilson2017} and \cite{Wilson2019}, we model the ISM using a temperature of 7000\,K, a turbulent broadening parameter of 1.5 km/s, and a RV of -10 km/s in \bp\ rest frame. The column density of \feii\ in the ISM was measured to be $1.3 \pm 0.1 \times10^{13}$ cm$^{-2}$ using the HST spectrum obtained on December 6, 1997, which do not show any exocomet absorption at velocities below $-5$\,km/s. The ISM column densities of other species were obtained by scaling the column density of \feii\ with the gas-phase element fractions of \cite{Jenkins2009}, and assuming a depletion strength factor $F_\star = 0$ in the Local Bubble. All the gas from the ISM is assumed to be in the ground state.

\subsection{Line spread functions}

Once calculated, the flux transmitted by the three LVCs, the circumstellar disc and the ISM needs to be convolved with the instrumental line spread function (LSF). The LSF of STIS and COS were retrieved from the STScI website\footnote{See the tabulated line spread functions for STIS an COS on {\tt www.stsci.edu/hst/instrumentation/}.}. For STIS, tabulated LSFs are available for only a limited number of apertures; in particular, no LSF is provided for the 31$\times$0.05NDA aperture. In this case, we adopted the LSF of the 0.1$\times$0.2 aperture, which has a similar width and therefore provides the closest available approximation to the 31$\times$0.05NDA aperture. This choice has no significant impact on our fits, as the studied absorption features are broader than the instrumental LSF. For the HARPS spectrum, we used the LSF calculated by \cite{Brandeker_2011} using a two-gaussian profile.

\subsection{Fit to the data}

Each of the 4 studied components (the three LVCs and the disc) is described by 13 parameters: the seven column densities of the studied species, and the six physical parameters presented in Sect. \ref{Sect. Modelling the circumstellar absorption} (the distance to the star $d$, the central velocity $v$ and Doppler parameter $b$, the electronic density $n_e$ and temperature $T_e$, and the ratio between the transverse and radial column densities of the cloud $f$). In total, the four components are thus described by 52 parameters. However, the RV of the disc was fixed to 0 km/s in \bp\ rest frame, resulting in 51 free parameters.

Our model was fitted to all the optically allowed transitions rising from the main stable and metastable states of the studied species (340 transitions per component, yielding a total of 1360 absorption lines). Each line was fitted over the full RV range from –15 to +15 km,s$^{-1}$, resulting in a total of $10,100$ degrees of freedom. Most of the studied transitions show clear signatures from the LVCs and the disc, although several non-detected lines were also included in the fit to better constrain some parameters. The fit was performed using a Markov chains-Monte Carlo (MCMC) algorithm from the {\tt emcee} Python package \citep{emcee2013}. 150 independent chains were launched over 30\,000 steps, discarding the first 20,000 steps as burn-in.

\subsection{Wavelength shifts}

Our initial fit matched the observed spectrum reasonably well, yielding a reduced $\rchi^2$ of 1.78. This value, however, exceeds unity, indicating that the observed data is not fully reproduced. 

\begin{figure}[h!]
\centering
    \includegraphics[scale = 0.385,     trim = 25 10 0 15,clip]{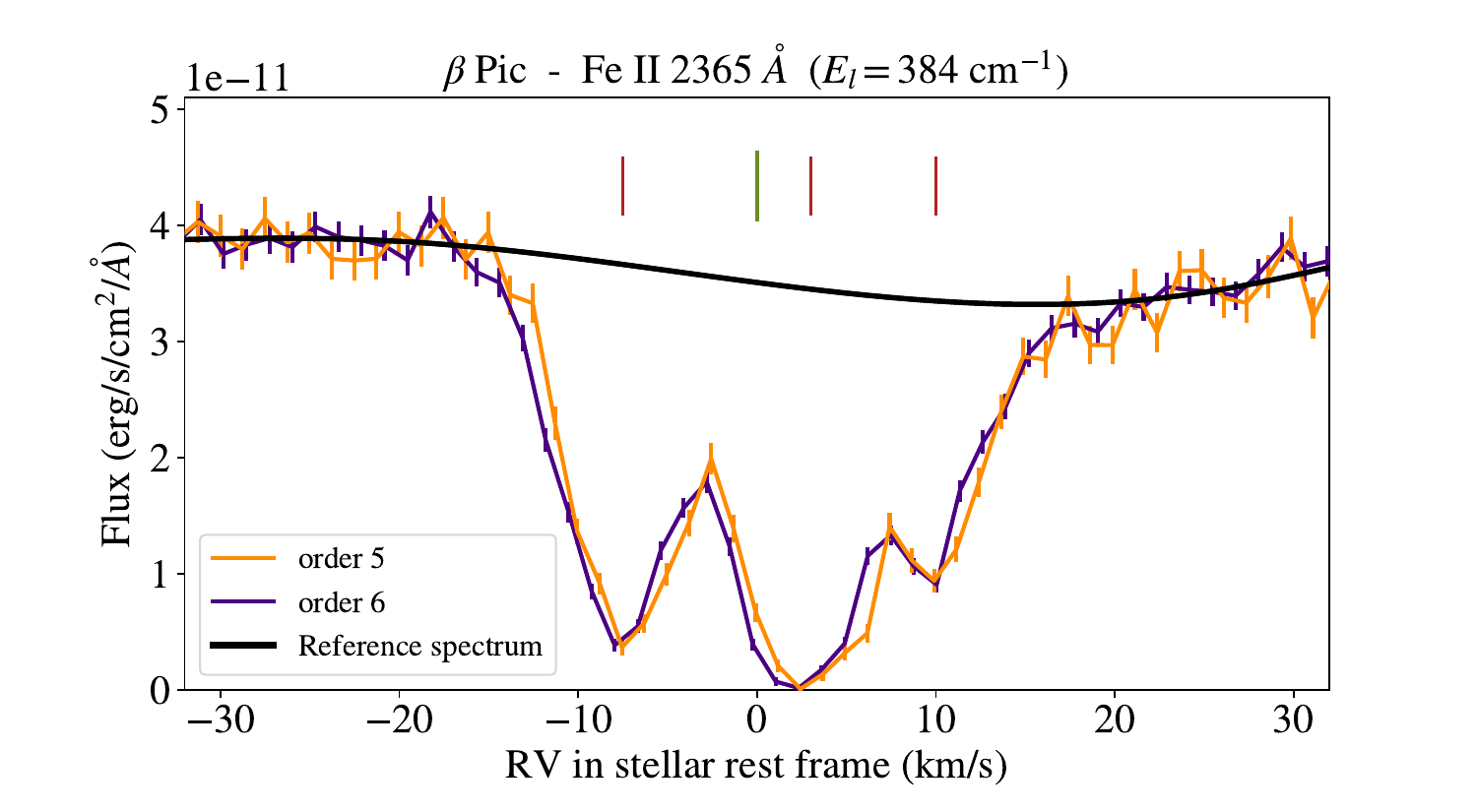}        \caption{View of the \feii\ line at 2365 \A, covered by two orders of the same STIS spectrum. A clear wavelength offset is visible between the two orders; here order \#5 is found to be redshifted by 0.5 km/s compared to order \#6. The red and green ticks mark respectively the RV of the three LVCs and the disc.}
    \label{Fig. shift Fe II}
\end{figure}

An important source of discrepancy appeared to rise from the imperfect wavelength calibration of the STIS data. Many lines are indeed detected at a slightly lower or higher wavelength than expected from NIST, with typical shifts of 0.5--1 km/s. This bias is instrumental, as lines covered by several spectral orders often exhibit different shifts from one order to another (see Fig.~\ref{Fig. shift Fe II}). To correct for this, we applied small RV adjustments to the spectral regions were wavelength offsets were noted. These corrections were restricted to the strongest lines, for which the RV shift could be reliably identified and measured (through a $\rchi^2$ minimisation). This allowed us to significantly improve the fit, reaching a reduced $\rchi^2$ of 1.51.

\vspace{-0.2 cm}
\subsection{Systematic uncertainties}
\label{Sect. systematic uncertainties}

The value of the final reduced $\rchi^2$ (1.51) remains well above unity. The remaining mismatch between our model and the observed spectrum can be explained by several factors: 

\begin{itemize}
    \item First, our model is highly simplified. In reality, exocometary tails likely have complex geometries, non-gaussian line-profiles, and non-homogeneous temperature and density profiles, resulting in a much more complicated absorption spectrum than produced by our model.
    \item Our transmission model also relies on extensive datasets of oscillator strengths and effective collisions strengths, which are generally accurate within 10\%, at best. For instance, most \feii\ and \niii\ lines listed in NIST are classified as B or B+, corresponding to uncertainties of 10\% and 7\% on the transition probabilities, respectively. Similarly, effective collision strengths for \feii, taken from \cite{Tayal2018}, are accurate to within about 10 \% for the ground multiplets. 
    \item Finally, only four transiting components are modelled (the three LVCs and the disc). In practice, however, the star could be occulted by additional, minor components, not clearly detected in optically thin lines but affecting our fit in the strongest lines.
\end{itemize}

\begin{figure}[h!]
\centering
    \includegraphics[scale = 0.385,     trim = 20 5 0 35,clip]{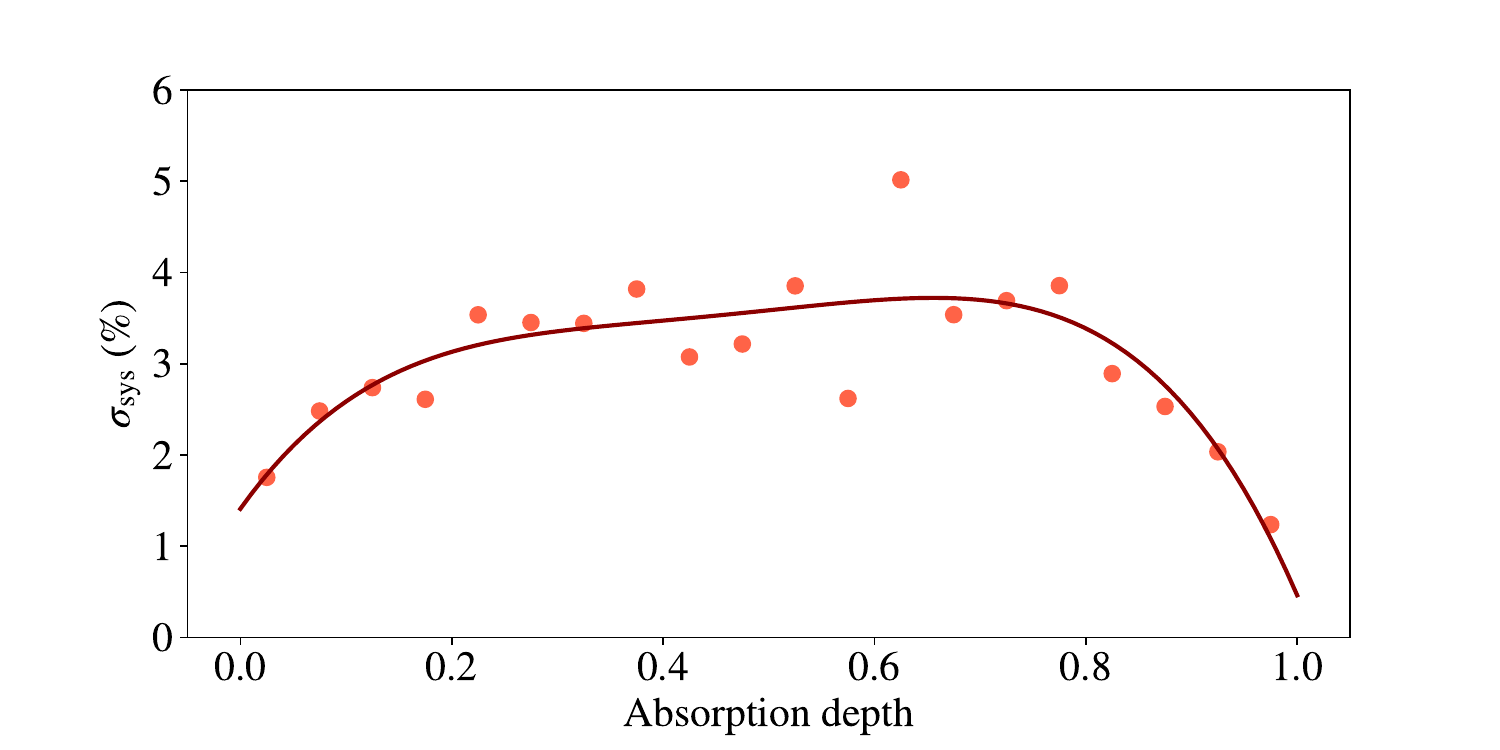}        \caption{Estimation of the systematic uncertainty of our model (expressed in \% of the reference spectrum). Dots show the measured systematic uncertainties $\sigma_{\rm sys}$ for pixels grouped by absorption depth. The solid line is a fit with a 4$^{\rm th}$ order polynomial, used to estimate the systematic uncertainty $\sigma_{k,\, \rm sys}$ of each pixel based on its absorption depth.}
    \label{Fig. systematic errors}
\end{figure}

These sources of uncertainties are difficult to model directly. To account for them, we introduced, for each pixel $k$, a systematic error $\sigma_{k,\, \rm sys}$, to be added quadratically to the tabulated noise. We assumed that, once the data are normalized by the reference spectrum, the systematic uncertainty depends mainly on the absorption depth. To quantify this dependence, the fitted pixels were grouped into categories according to their absorption depths: 0–5\%, 5–10\%, 10–15\%, and so on, up to 95–100\%. For each category, we determined the systematic error $\sigma_{\rm sys}$ that, when added in quadrature to the tabulated noise, brings the reduced $\rchi^2$ of the corresponding pixels to unity. The result of this calculation is shown on Fig.~\ref{Fig. systematic errors}: the systematic uncertainty reaches $\sim$4 \% of the continuum for pixels at $\sim$50 \% absorption depth (which is about 1--2 times larger than the tabulated noise), and $\sim$1\% when the absorption depth is close to 0 or 1. The systematic errors shown on Fig. \ref{Fig. systematic errors} were then fitted with a 4$^{\rm th}$ order polynomial, allowing us to compute the systematic error $\sigma_{k,\, \rm sys}$ of any pixel $k$, based on its absorption depth. The $\rchi^2$ of our model is then computed as: 
$$
\rchi^2 = \displaystyle \sum_{k} \frac{\left(  f_{k,\, \rm obs} - f_{k,\, \rm model} \right)^2}{\sigma_{k,\, \rm tab}^2 + \sigma_{k,\, \rm sys}^2},
$$
where $f_{k,\, \rm obs}$, $f_{k,\, \rm model}$  and $\sigma_{k,\, \rm tab}$ are respectively the observed flux, the modelled flux, and the instrument pipeline tabulated uncertainties, normalised by the reference spectrum.

It should be noted that this method does not account for possible correlations between the systematic uncertainties of adjacent pixels. Nevertheless, it allows us to obtain more realistic estimates of the uncertainties on the fitted parameters. Running again the fit with the systematic uncertainties included yielded a reduced $\rchi^2$ of 0.97, close to unity. The mean value and error bars of the fitted parameters were then estimated from the posterior distribution of the MCMC chains, and are provided in Table \ref{Tab. Fitted parameters}.

\section{Overview of the fitted model}
\label{Sect. Discussion}

\begin{figure*}[h!]
\centering
    \includegraphics[scale = 0.385,     trim = 20 25 55 15,clip]{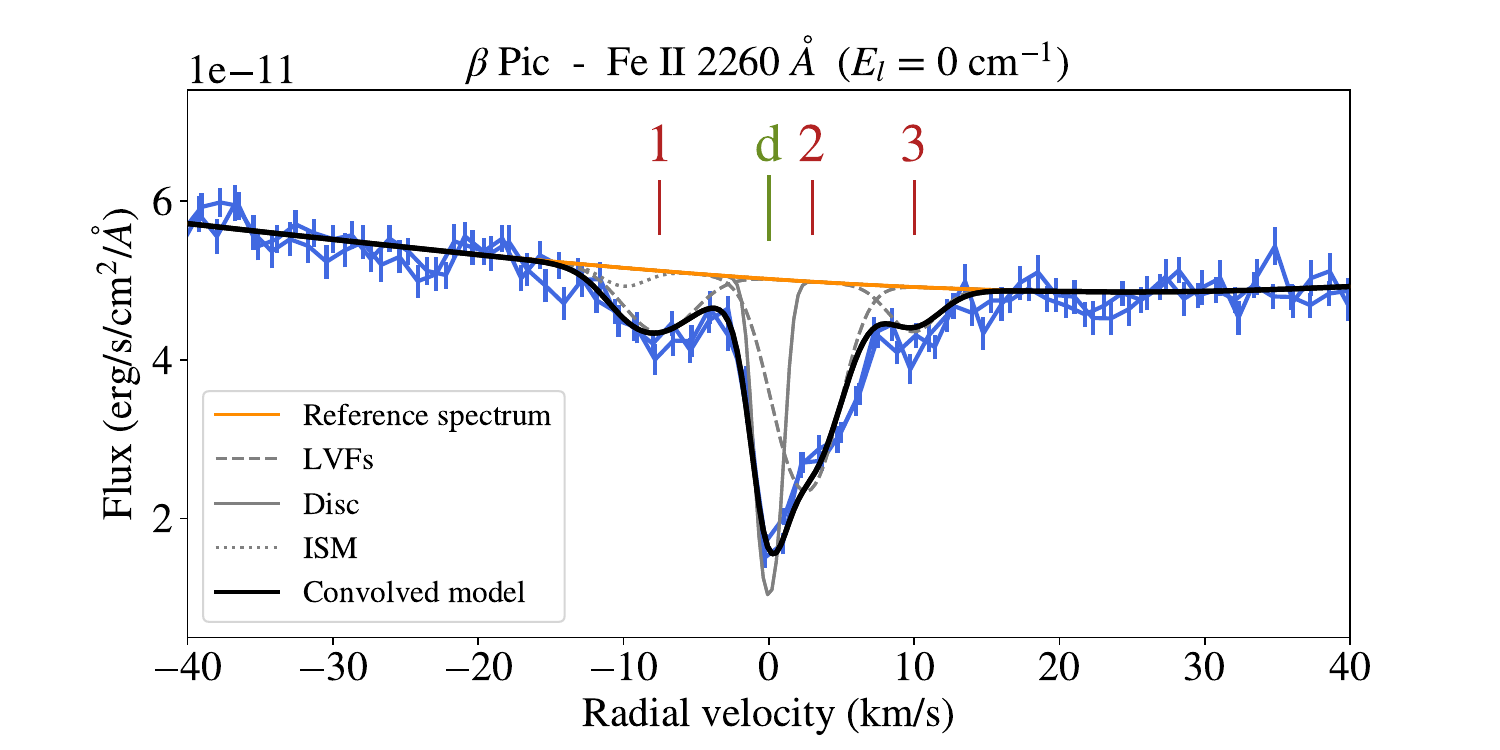}    
    \includegraphics[scale = 0.385,     trim = 20 25 55 15,clip]{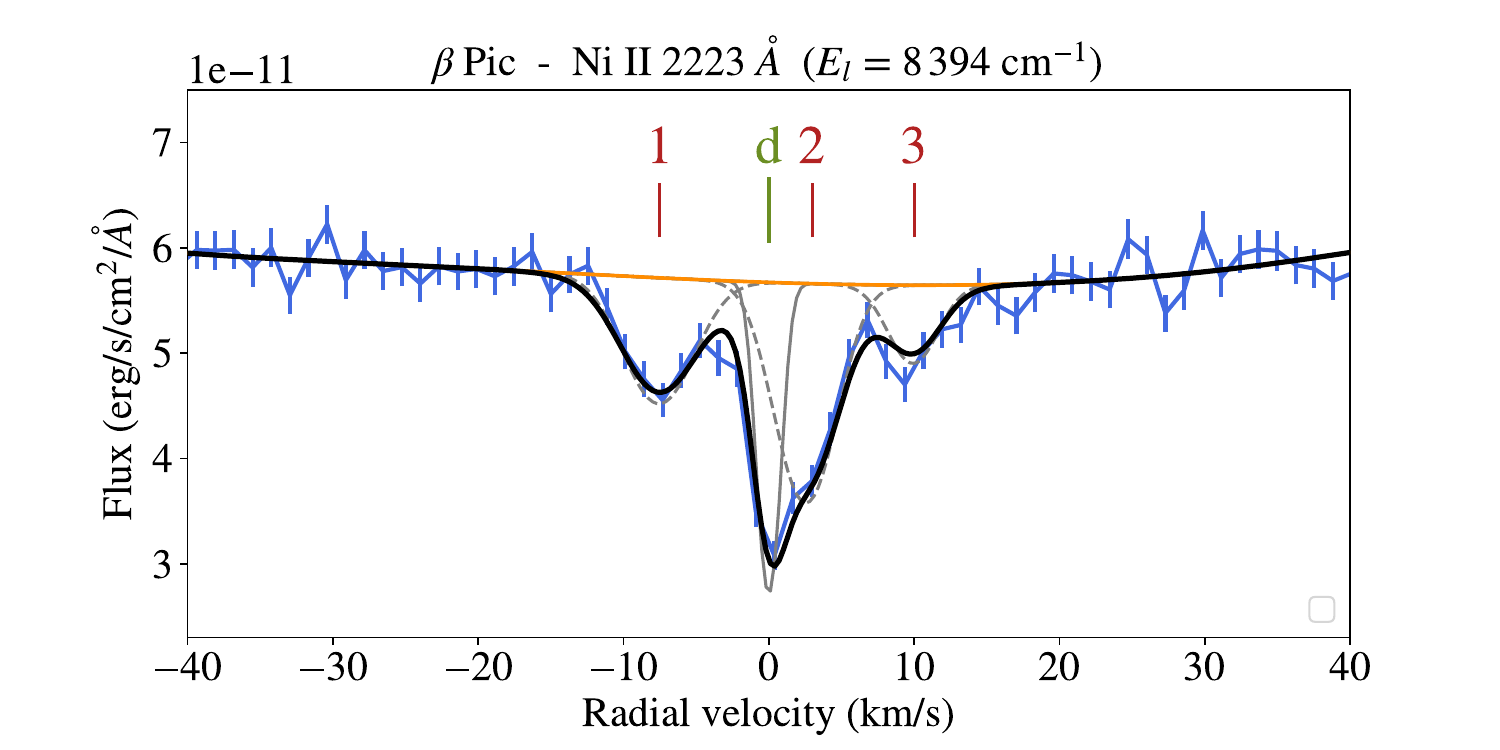}    
    
    \includegraphics[scale = 0.385,     trim = 20 25 55 0,clip]{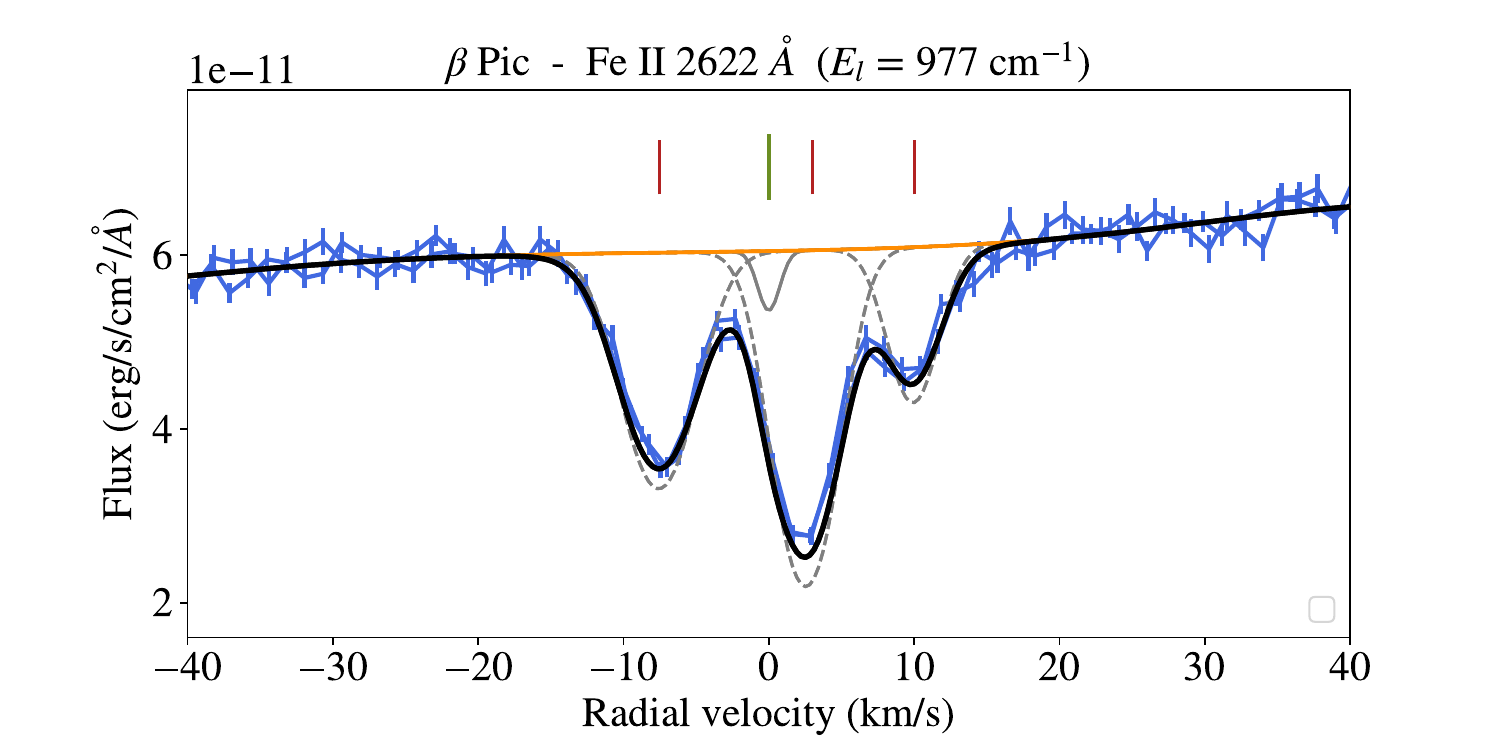}   
    \includegraphics[scale = 0.385,     trim = 20 25 55 0,clip]{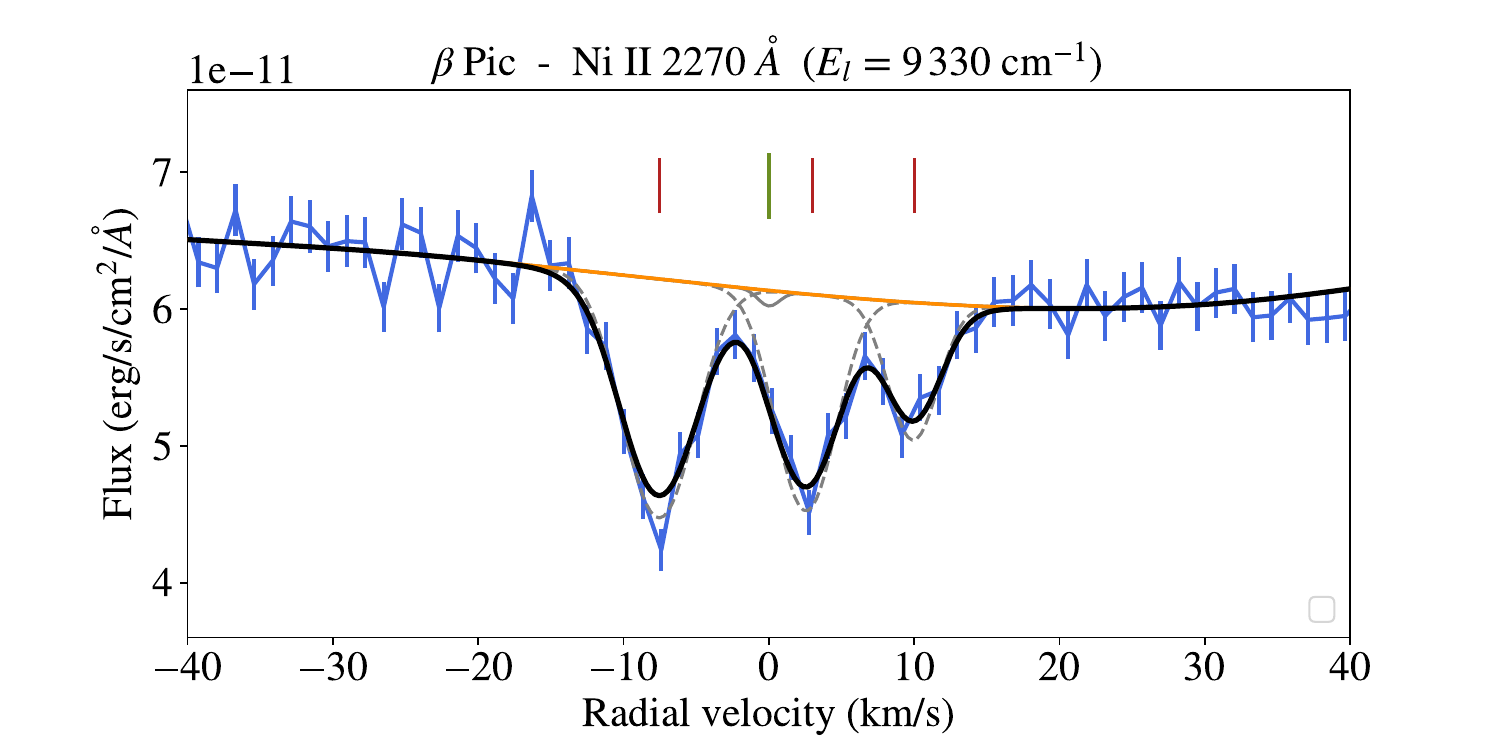}  
    
    \includegraphics[scale = 0.385,     trim = 20 25 55 0,clip]{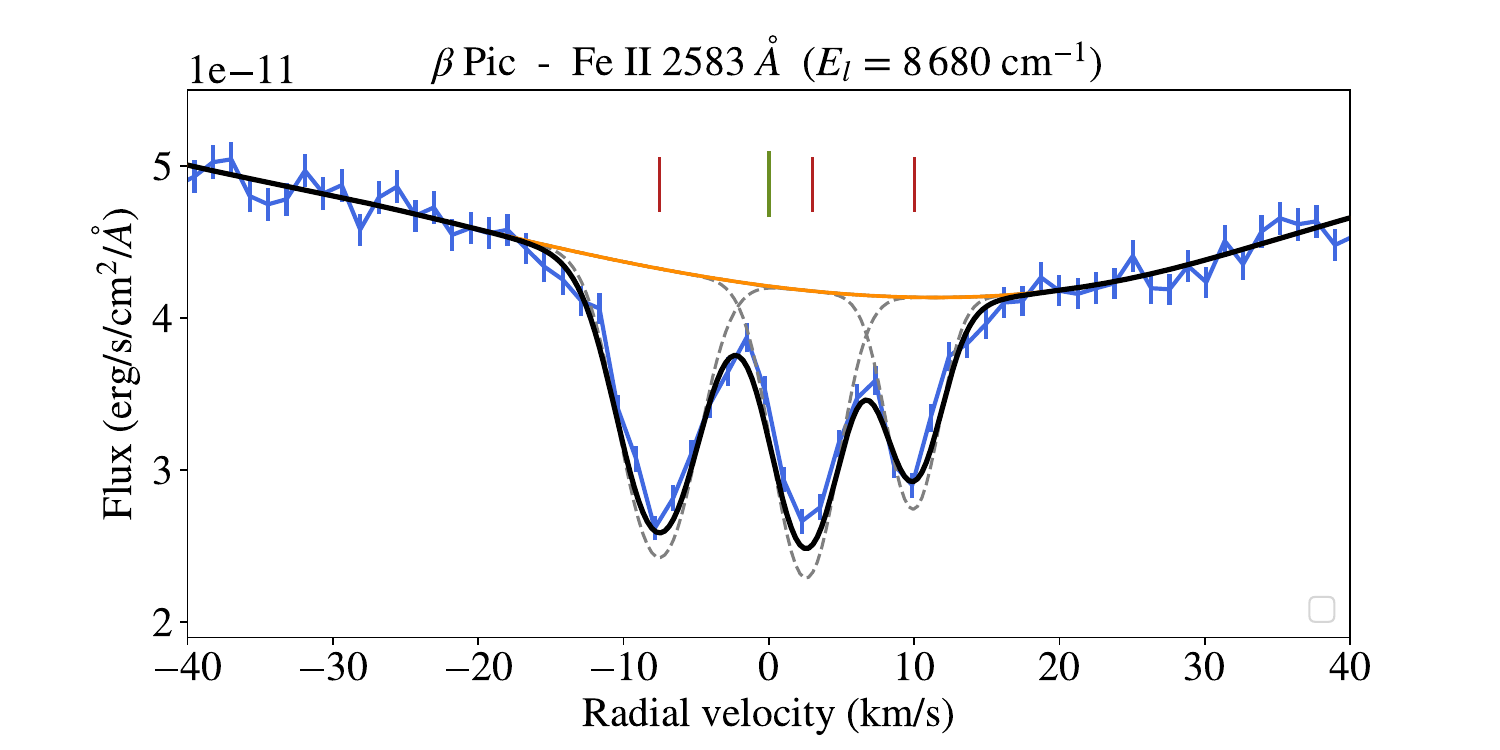}    
    \includegraphics[scale = 0.385,     trim = 20 25 55 0,clip]{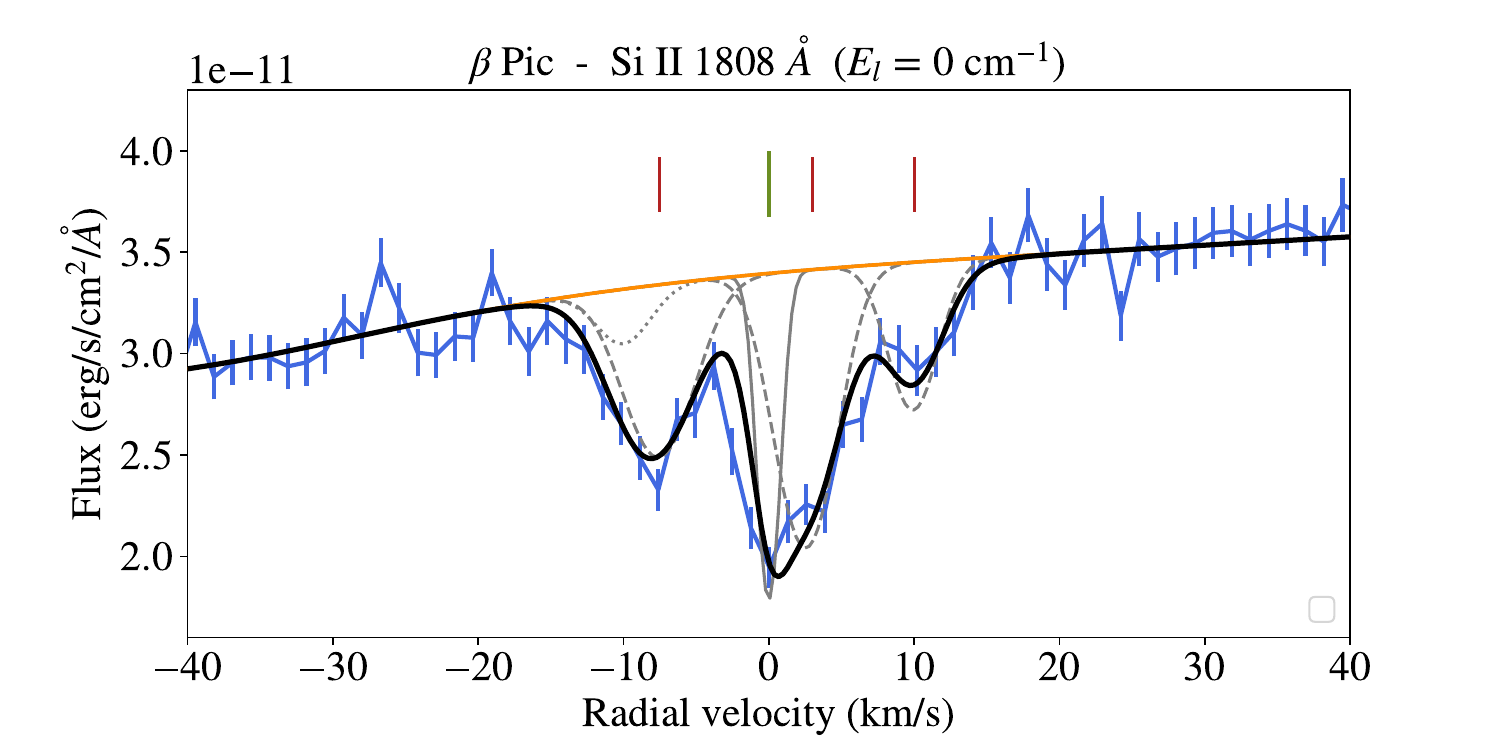}  
    
    \includegraphics[scale = 0.385,     trim = 20 0 55 0,clip]{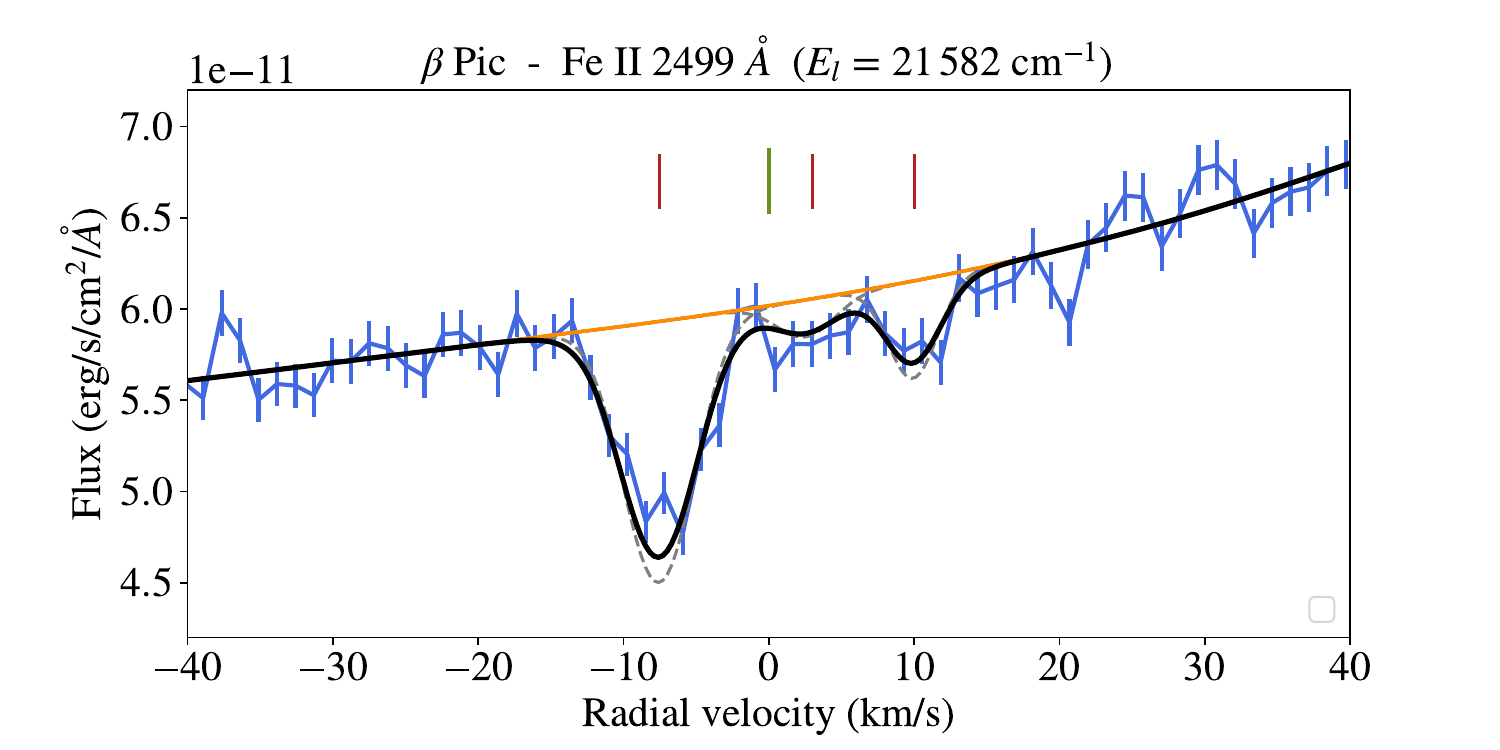}    
    \includegraphics[scale = 0.385,     trim = 20 0 55 0,clip]{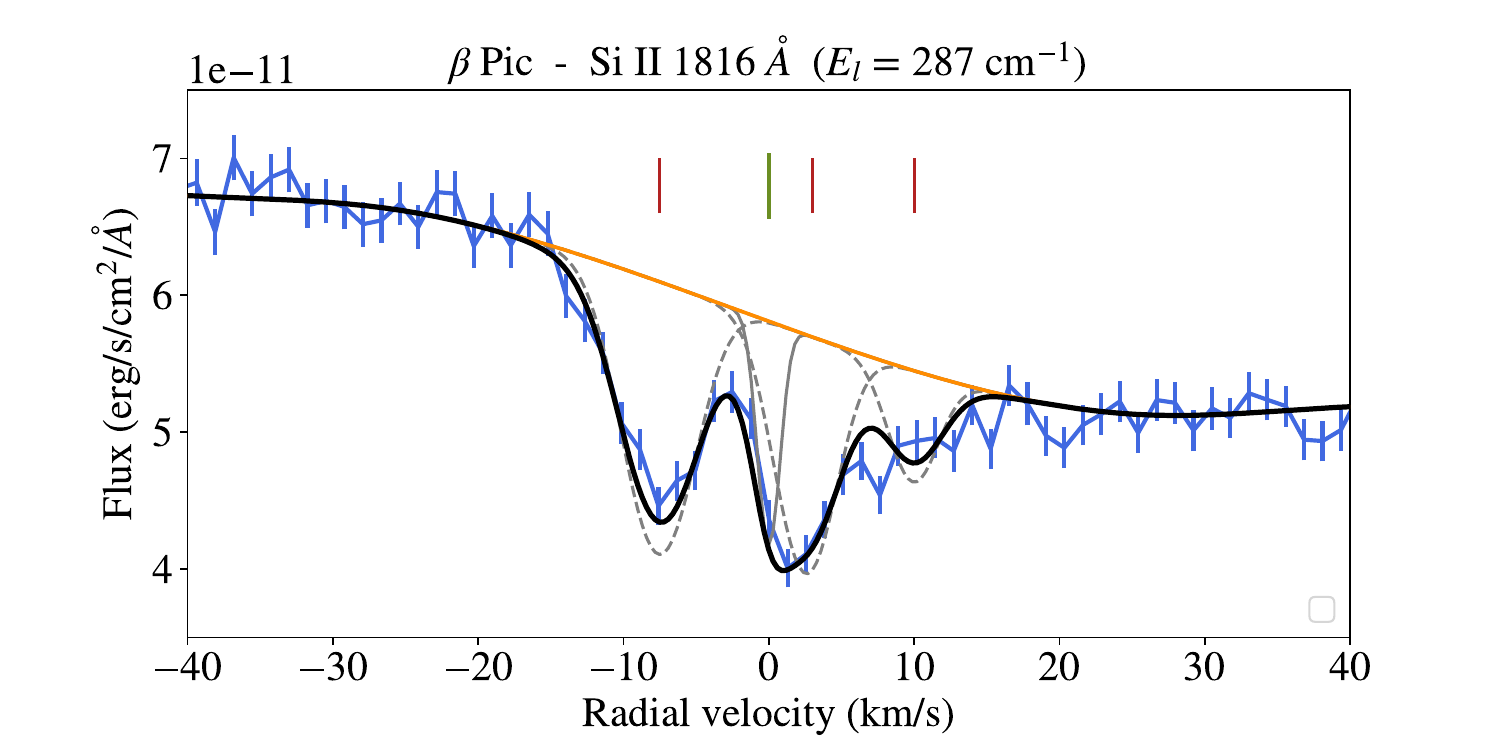}   
    \caption{Comparison between the observed spectrum and the fitted model, for \feii, \niii\ and \siii\ lines. The observed spectrum is shown with a blue line; errors bars reflect the uncertainties tabulated by the STIS pipeline (see Sect. \ref{Sect. systematic uncertainties}). The solid, orange line shows the reference spectrum. Absorptions from the LVCs, the disc and the ISM are shown in dashed, solid and dotted grey lines, respectively. The full model, convolved by the STIS line spread function, is shown in a solid, black line. Red ticks indicate the central velocities of the three LVCs, while the green tick at 0\,km/s indicates the RV of the disc.}
    \label{Fig. lines + model}
\end{figure*}

 Our best-fit model is shown in Figs.~\ref{Fig. lines + model} around several \feii, \niii, and \siii\ lines (among a total of 340~lines), together with the observed spectrum. Additional examples covering lines from all the studied species are provided in App.~\ref{App. Example of absorption lines}.

Despite the limitations detailed above (Sect. \ref{Sect. systematic uncertainties}), our model fits remarkably well the observed spectrum. The absorption signatures of the three LVCs and the circumstellar disc are well reproduced across all the studied species, all excitation levels, and over a large range of oscillator strengths, with no exceptions (see, e.g., Fig \ref{Fig. lines + model}). In particular, the excitation state of \feii\ is well captured, up to very high excitation energies ($E_l \sim 30\,000\,\text{cm}^{-1}$). This allows us to place tights constraints on the transit distance of each component (Sect. \ref{Sect. Distance to the star}) and on their compositions (Sect. \ref{Sect. ionisation state of the gas}). Besides \feii, the model also reproduces the complex excitation states of \niii\ and \crii\ in both the LVCs and the disc, further confirming the robustness of our model.

\begin{figure*}[h!]
\centering
    \includegraphics[scale = 0.44,     trim = 55 0 0 30,clip]{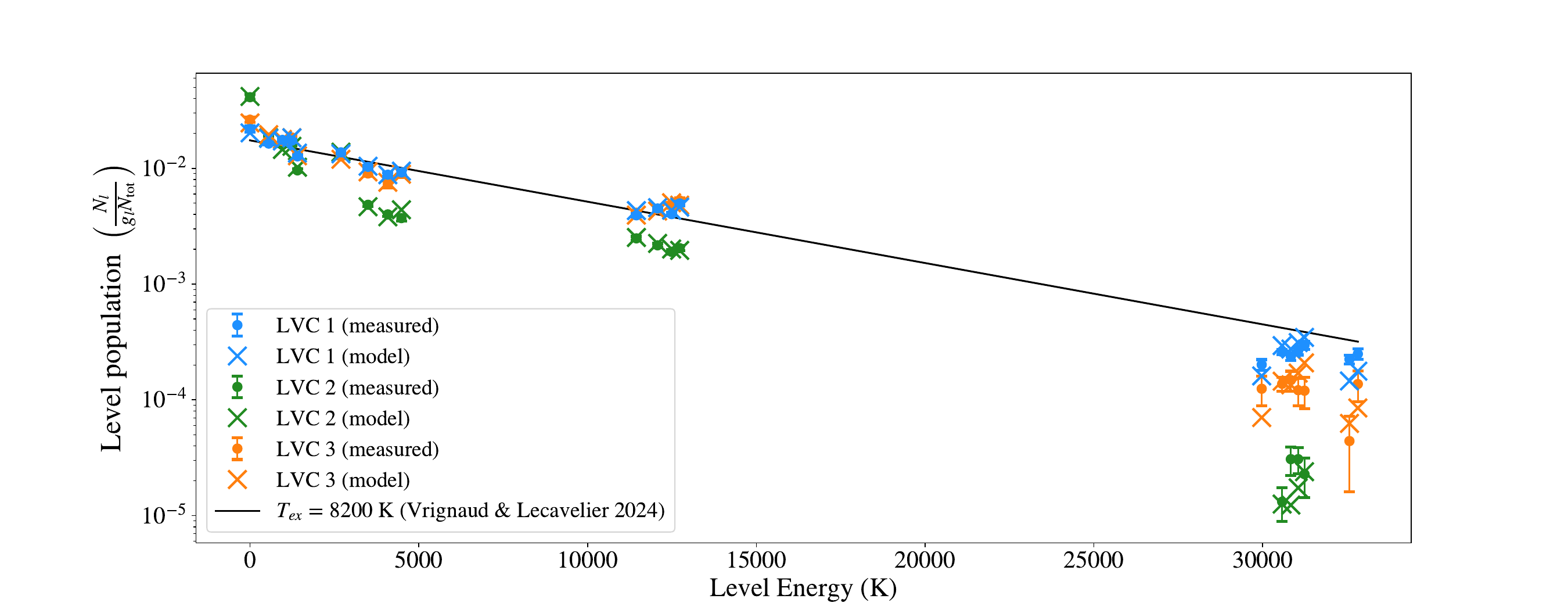}    
    \includegraphics[scale = 0.44,     trim = 55 0 0 10,clip]{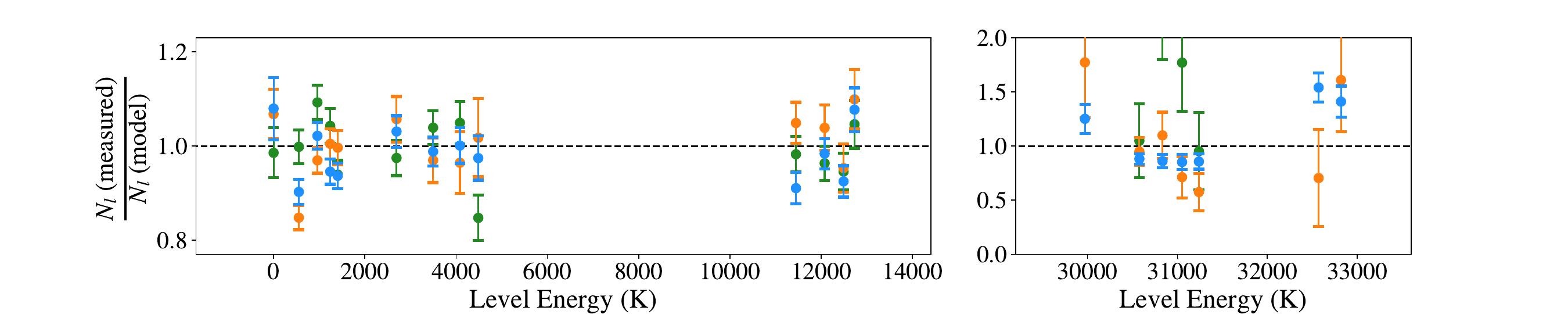}    
    \caption{Excitation \feii\ in the three LVCs. \textbf{Top}: Comparison between the excitation diagram of \feii\ measured in the three LVCs (dots with error bars) and the result from our excitation model (crosses). For clarity, we only show levels with significant detections. The solid, black line indicates the excitation temperature measured by \cite{Vrignaud24b} in a \bp\ exocomet observed on December 6, 1997 and located at < 0.4 au from the star. By comparison, LVCs \#1 and \#3 (blue and orange dots) are at located at $\sim1$\,au; and LVCs \#2 (green dots) is even much further away (4.7 au). 
    \textbf{Bottom}: Ratio of the measured and modelled abundances of the studied \feii\ levels. The horizontal, dotted line indicates the ratio of 1.}
    \label{Fig. Excitation diagram LVFs}
\end{figure*}

\subsection{Transit distances and excitations states}
\label{Sect. Distance to the star}

As anticipated in Sect. \ref{Sect. Closer look at the central components}, our model places the three LVCs at significantly different distances from the star: $0.88 \pm 0.08$ au for LVC~\#1, $4.7 \pm 0.3$ au for LVC~\#2, and $1.52 \pm 0.15$ au for LVC~\#3. These distances are primarily constrained by the many \feii\ lines detected in our spectrum, which rise from a wide range of stable and metastable states with energy between 0 and 30\,000 cm$^{-1}$. As pointed out in \cite{Vrignaud24b}, the relative population of these levels is determined by the amount of UV radiation reaching the gas, and thus by the distance to the star. Very close to \bp, exchanges between excitation states are dominated by the absorption and emission of UV photons, driving the excitation temperature toward $\sim$8000 K (the effective temperature of \bp). Conversely, at large distances from the star, UV pumping becomes negligible and metastable levels decay toward the ground state via forbidden infrared transitions. The three LVCs studied here are in a intermediate regime, where the excitation state is set by the balance between between UV radiative pumping and the radiative decay of metastable states.

Figure \ref{Fig. Excitation diagram LVFs} provides the excitation diagrams of \feii\ in the three LVCs for the most easily detected states. Crosses show the level populations calculated using our best-fit model, while dots with error bars show direct measurements of the level populations in the LVCs. These estimates were obtained by varying the column density $N_{\text{\feii},\, i}$ of each excitation state $i$ individually, keeping all other levels fixed, and finding the value that best reproduces the data. The modelled and measured excitation diagrams agree very well for all three LVCs, with most level populations being accurately predicted to within 10\%. Owing to its larger distance to \bp, LVC\,\#2 is, by far, the least excited component: all metastable states above 2400 cm$^{-1}$ are depleted by 0.3 to 1 dex compared to the two others LVCs, while the ground state is enhanced by 0.3 dex. In addition, we note that the three LVCs exhibit lower populations of very excited \feii\ states ($\geq 30\,000$\,K) compared to the exocomet observed on December 6, 1997 and analysed in \cite{Vrignaud24b}, which was found to transit at less than 0.4 au. This illustrates the tight dependency between the transit distance of an exocometary tail and the excitation state of the gas.

\begin{figure}[h!]
\centering
    \includegraphics[scale = 0.385,     trim = 25 0 0 35,clip]{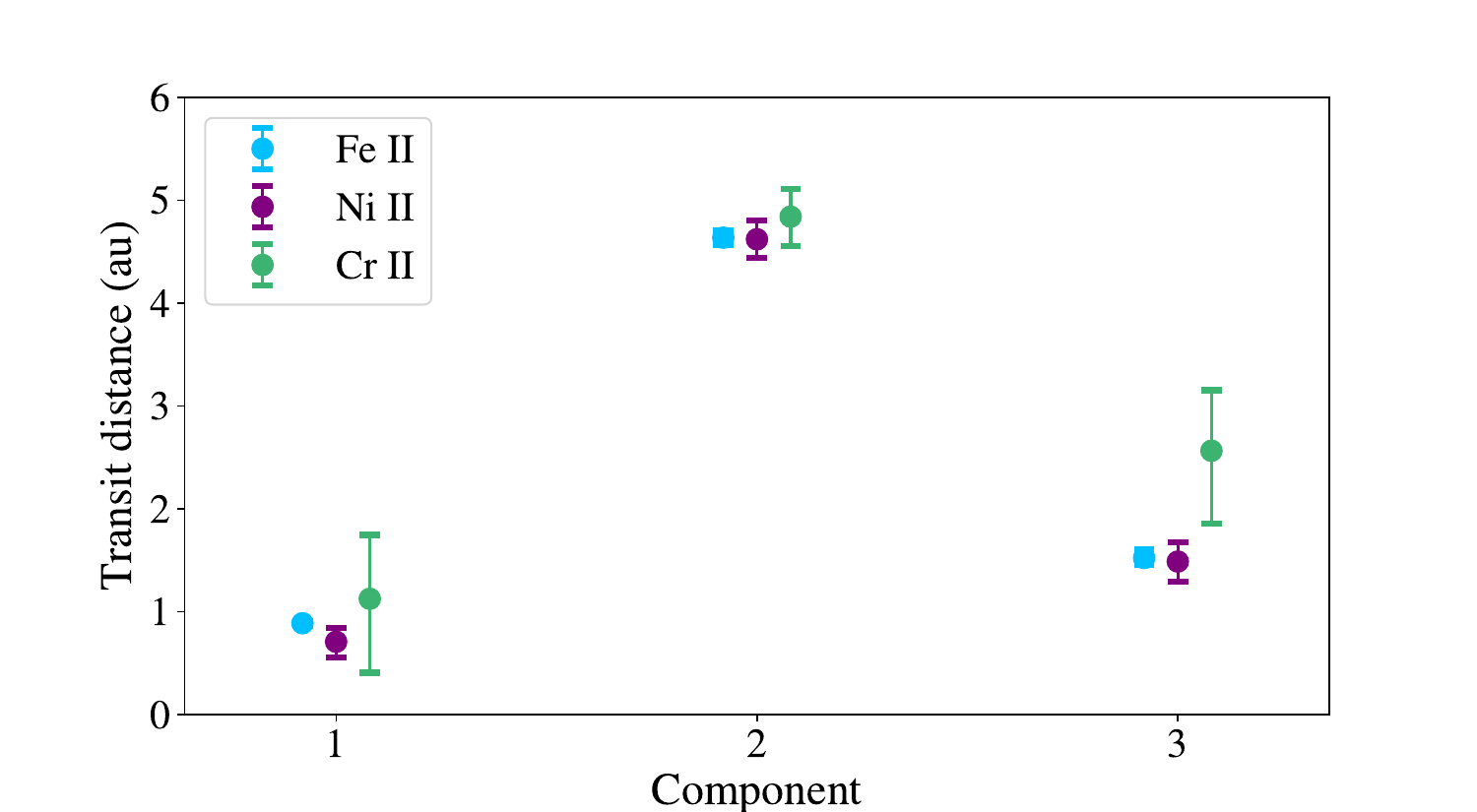}        
    \caption{Transit distances of the three LVCs derived from independant fits to the \feii, \niii, and \crii\ lines. The three species provide similar constraints on the transit distances of the LVCs.
    }
    \label{Fig. Distances Fe2 vs Ni2}
\end{figure}

In addition to \feii, the two species that provide the strongest constraints on the transit distance are \crii\ and \niii, as they are detected in multiple electronic configurations. To verify that \feii, \niii, and \crii\ yield consistent estimates of $d$, we performed three additional fits in which the lines of each species were fitted independently. We find that these fits constrain the transit distances of the LVCs to similar values (Fig.~\ref{Fig. Distances Fe2 vs Ni2}), demonstrating the robustness of our transit distance estimates.

Our fit places the circumstellar disc at $38_{-11}^{+15}$ au from \bp. Unlike the three LVCs, this distance is primarily constrained by the abundance of the \niii\ level at 8393 cm$^{-1}$, which remains substantially populated even at large distance from \bp. In practice, the inferred distance corresponds to the average radius of the ionised disc, which has a significant radial extent \citep[e.g.,][]{Brandeker2004}. Probing this radial extent with the method developed here is however difficult, because absorption spectroscopy only provides access to the global level populations in the disc, integrated along the line of sight. Allowing the disc to have a non-zero radial extent does not improve the fit; for instance, placing the disc at a single distance of~$\sim$~40 au, or distributing it uniformly from 20 to 60 au, provide fits which are equally good.

The typical radius of the disc is comparable, yet smaller, than values obtained from ground-based imagery in \fei, \nai\ and \caii\ \citep[][]{Olofsson2001, Brandeker2004, Nilsson2012}, which generally find the gas distribution to peak around 100 au. This difference could be due to the difference of spatial distribution between neutral and ionised species, as \fei\ and \nai\ are quickly photo-ionised close to \bp. In addition, the inner disc is difficult to probe in direct imaging due to its high opacity.

\subsection{The transverse size}

The parameter $f$, which characterises the ratio between the radial and transverse column densities of each component, is constrained to values between 0.09 and 0.20 for the three LVCs. These values suggest that the LVCs are about 5 times more extended in the radial direction than in the transverse direction. The $f$ parameter of the disc is not constrained by our fit.

The inferred distances of the three LVCs are not very sensitive to $f$: repeating the fit while imposing $f = 0$ for all components yields similar distance estimates (within 10 \%). However, introducing the parameter $f$ significantly improves the fit quality, with a reduced $\rchi^2$ of 9\,800 compared to 11\,000 when $f=0$.

\subsection{Temperatures and electronic densities}

\begin{figure}[h!]
\centering
    \includegraphics[scale = 0.385,     trim = 25 10 0 15,clip]{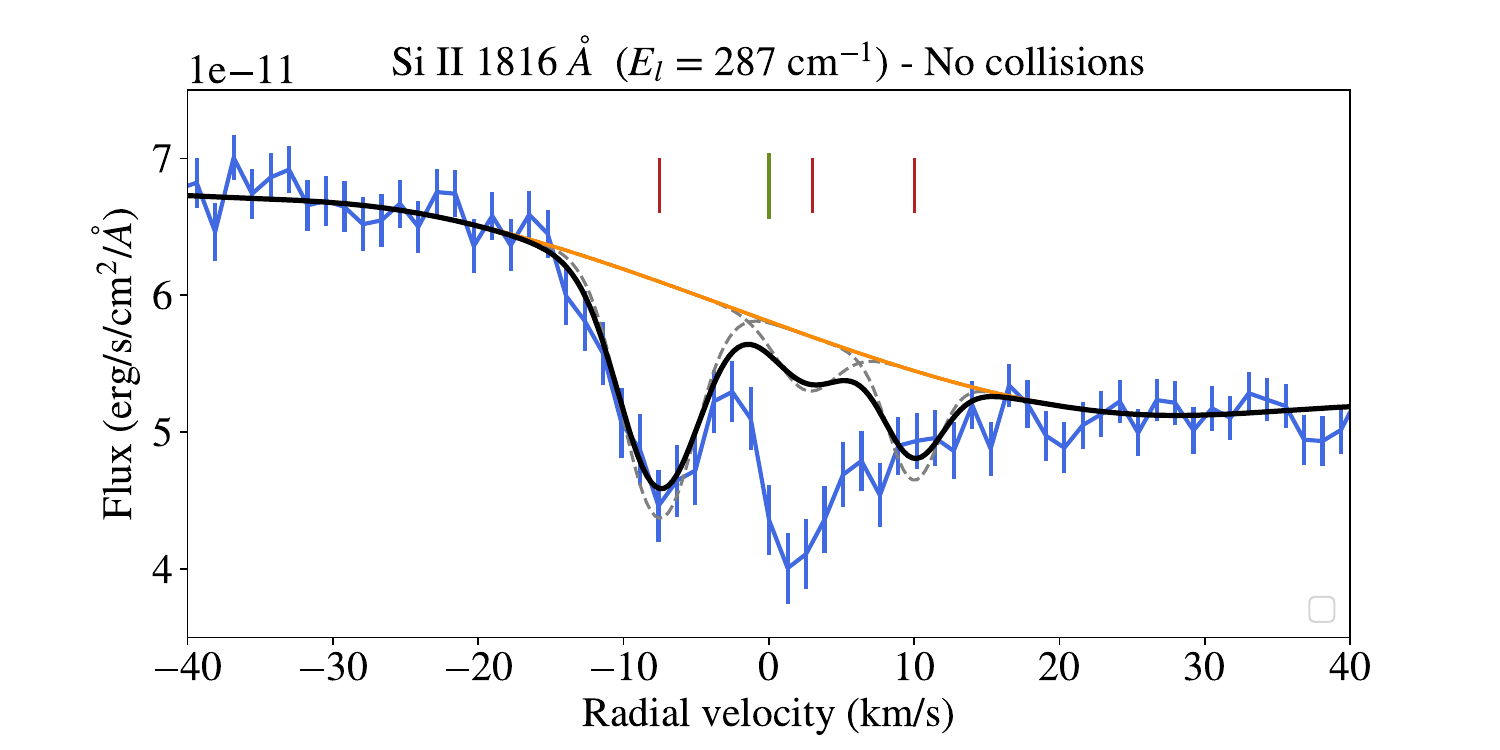}        
    \caption{Plot of our model in the 1816\,\A\ \siii\ line in the case where electronic collisions are disabled. The fit is poor, particularly for LVC~\#2 and for the disc, because far from \bp\ UV pumping is inefficient at populating the excited \siii\ level at 287\,cm$^{-1}$. Collisions must be taken into account in the model to properly fit the data 
    (see the bottom panel of Fig.~\ref{Fig. lines + model}).
    }
    \label{Fig. Si II 1816 no collisions}
\end{figure}

The excitation states of the studied species are primarily set by radiative excitation, in all four components. As a result, probing the electronic density and temperature is difficult. For LVC~\#1 and \#3, we only obtain upper limits on $n_e$, respectively $<10^6$ and $<10^5$\,cm$^{-3}$. For these components, including collisions does not significantly improve the fit. For LVC \#2, we find that collisions do improve the fit, particularly in \siii\ and \feii\ lines (see Fig.~\ref{Fig. Si II 1816 no collisions}). This is because LVC \#2 is located further out, and has a stronger self-opacity, leading to weaker radiative excitation. For this component, we constrain $\log\left(n_e\left[{\rm cm}^{-3}\right]\right) = 3.3 \pm 0.3$ and $\log\left(T_e \left[{\rm K}\right]\right) = 2.9 \pm 0.3$. Finally, for the disc, we obtain $\log\left(n_e\left[{\rm cm}^{-3}\right]\right) = 2.7 \pm 0.3$ and $\log\left(T_e \left[{\rm K}\right]\right) = 2.9 \pm 0.3$. These estimates are broadly consistent with the recent results of \cite{Wu2025}, which conducted a detailed modelling of the photochemistry along the \bp\ disc and found that the electronic density and temperature decrease from $\sim 10^4$\,cm$^{-3}$ and $10^3$\,K at 10 au, to $10$\,cm$^{-3}$ and $50$\,K at 200 au.

\vspace{-0.3 cm}
\section{Discussion}

\subsection{Large transit distances}

The transit distances of the three LVCs (0.88, 4.7 and 1.52 au) are much larger than the typical values obtained for high-velocity exocomets using acceleration measurements \citep[0.02–0.14 au;][]{Kennedy_2018}. This is not contradictory, however, as all the exocomets analysed by \cite{Kennedy_2018} belong to the S family, located closer to \bp\ than the D family (to which the LVCs belong). The absorption signatures of the three LVCs in \caii\ do not show any acceleration or temporal variation over the 2.25 hours of HARPS monitoring, supporting the fact that they are located far from the star. In addition, around \bp, the maximum RV for a bound object transiting at 1\,au is $\sim$55\,km\,s$^{-1}$ (25\,km\,s$^{-1}$ at 5\,au). The low RV of the three LVCs (-7.5, +2.5 and 10\,km/s) are thus consistent with objects transiting at 1--5\,au and having either small periastron longitudes, or small eccentricities and periastron distances similar to the measured transit distances.

Nonetheless, the transit distances of the three LVCs contradict the results of \cite{Kiefer_2014}, who inferred a typical transit distance of $19 \pm 4\,R_\star$ for the D family. Their estimates, based on converting the projected size of an exocomet into its distance to \bp\ using the hydrodynamical model of \cite{Beust_1993}, are however subject to important limitations. Several assumptions from the model of \cite{Beust_1993} now appear to be unrealistic, particularly that the ion dynamics can be treated using collisions with a neutral hydrogen cloud. Recent studies, \citep{Vrignaud24, Vrignaud2025} have shown that the tails of \bp\ exocomets are mostly ionised, and that drag forces are dominated by coulombian interactions between ions. This completely modifies the global dynamics of the tail, and therefore the link between the distance of a comet from \bp\ and the shape of its absorption profile.

Finally, the transit distances of the three LVCs ($0.88$, $4.7$ and $1.52$\,au) are surprisingly high as, at 1 au or more, the sublimation of refractory dust is very slow, if not impossible. Depending on the dust size and composition, the sublimation lines of silicates around \bp\ lie between 0.1 and 0.5\,au \citep[][]{Beust1996, Beust1998, Beust2001}. At larger distances, exocomets should not release significant amounts of refractory ions. Yet, this is not what is observed: the amount of \siii\ found in LVC\,\#2 ($\sim 5 \times 10^{14}$ cm$^{-2}$ covering the full stellar disc) corresponds to the full sublimation of $\sim 1$ km$^3$ of SiO$_2$, assuming a density of 2.3 g/cm$^3$. Such large quantities of \siii\ ions can only be produced by the complete sublimation of dense dust tails, if not whole cometary nuclei, at very short distance to the star. Thus, it seems that the large amounts of refractory ions found in the three LVCs were originally produced close to \bp\ (< 0.5 au), and subsequently transported to more distant regions. The mechanisms driving this migration are unclear; no study of the long-term evolution of cometary tails in the \bp\ system is available. In the models of \cite{Beust_1990} and  \cite{Beust_1993}, refractory ions (\caii, \mgii, \aliii...) are assumed to evolve in a surrounding cloud of neutral hydrogen; as a result, they are not efficiently retained and rapidly escape the tail due to radiation pressure. However, if the cometary tails are ionised \citep[which appears to be the case,][]{Vrignaud2025}, the gas may remain well-mixed for a much longer time, allowing it to expand and reach large stellocentric distances without being disrupted. Further modelling of the dynamics of exocometary tails in the \bp\ system are in preparation to better understand this behaviour, and in particular to explain why the studied LVCs are not observed to be strongly blue-shifted despite having migrated outward by one or several au.

\subsection{Ionisation state of the gas}
\label{Sect. ionisation state of the gas}
\begin{figure}[h!]
\centering
    \includegraphics[scale = 0.385,     trim = 5 0 0 30,clip]{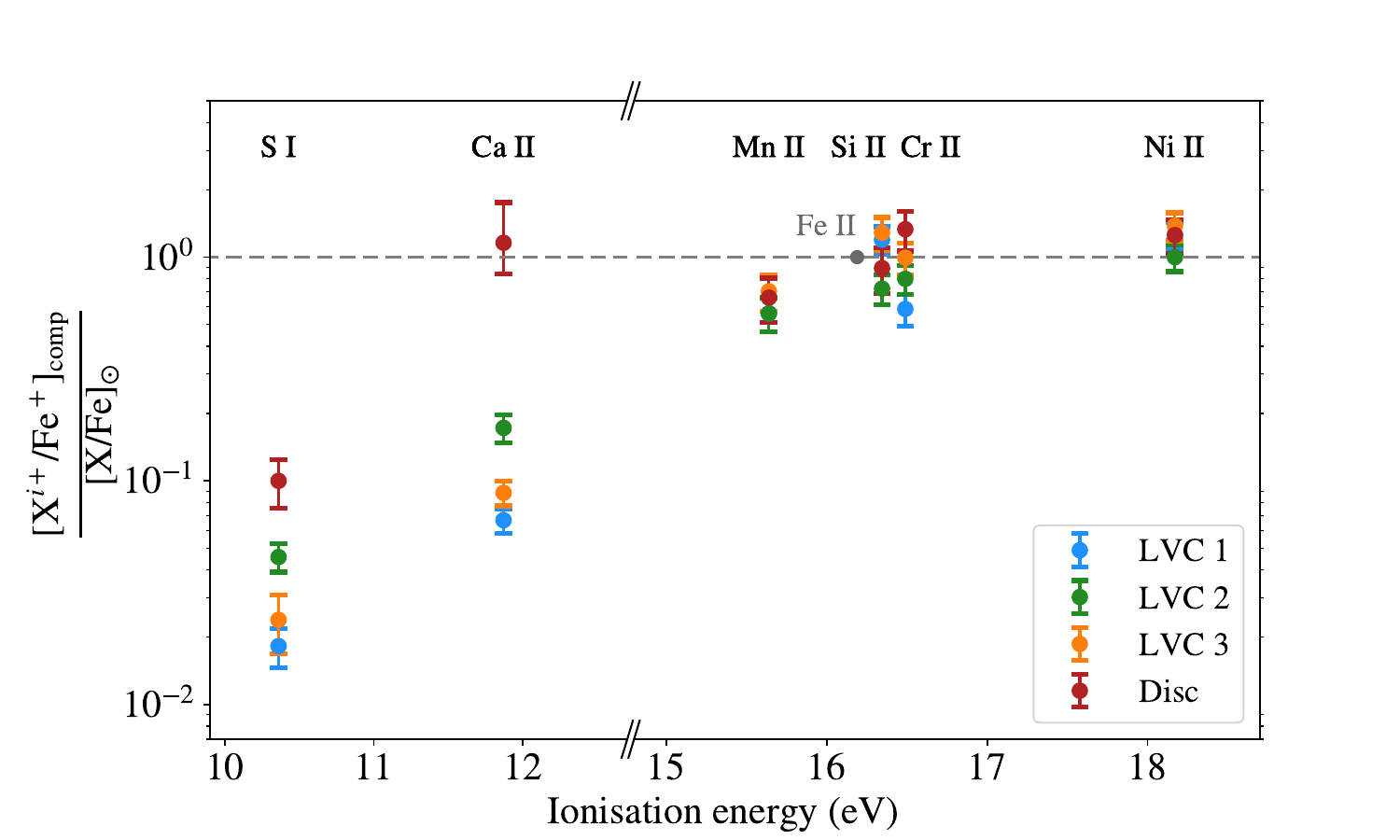}  
    \caption{Abundance of the studied species relative to \feii\ and normalised by the solar abundance ratios of the corresponding elements, plotted as a function of the ionisation potential. }
    \label{Fig. Composition LVF disc}
\end{figure}

The relative abundances of the various studied species in the three LVCs and in the disc are provided in Table~\ref{Tab. Composition}. The composition of the four components is also illustrated by Fig.~\ref{Fig. Composition LVF disc}, where the abundance ratios (e.g., \caii/\feii) are renormalised by the solar ratios of the corresponding elements (e.g., Ca/Fe). 

A clear dichotomy appears between species with ionisation energy in the 15-18 eV range (e.g., \mnii, \feii, \niii, ...), whose abundances are quasi-solar both in the disc and in the LVCs, and species with lower ionisation potentials (\Si\ and \caii, 10--12\,eV), which are depleted compared to the formers. This behaviour was already identified in \cite{Vrignaud2025} for exocomets at larger redshifts. In addition, easily ionised species appear to be more abundant in the disc and LVC~\#2 than in LVCs~\#1 and \#3, located closer to \bp. The \caii/\feii\ ratio of the disc is consistent with solar abundances, while it is 0.2$\times$ solar in LVC\#2, and 0.08$\times$ solar in LVCs\,\#1 and 3. Similarly, we find a \Si/\feii\ ratio $0.1 \times $ solar in the disc, 0.04$\times$ solar in LVC \#2, and only 0.02$\times$ solar in LVCs\,\#1 and \#3. The transiting components thus appear to have different ionisation states, in strong correlation with their transiting distances: components located further out in the system are also less ionised. In particular, Ca seems to be mostly present as \caii\ in the disc, while in the LVCs it is mostly ionised into \caiii\ (as shown by the low \caii/\feii\ ratio). Interestingly, the \caii/\feii\ ratios of LVCs\,\#1 and \#3 are very similar to the \caii/\feii\ measured in \cite{Vrignaud2025} for exocomets at shorter distances and larger velocities (\caii/\feii\ = $0.082 \pm 0.019\,\times$ solar), indicating that their ionisation state is somehow similar to that of HVCs.

\section{Conclusion}
\label{Sect. Conclusion}

We presented a new approach to measure the transit distance of exocometary tails, based on the excitation state of the gas. This method relies on the fact that, in diluted gaseous tails, the excitation state is primarily set by the absorption of photon coming from the star. The relative population of excited levels then only depends on the intensity of the incoming radiation and, therefore, on the distance to the star. 

We applied this approach to three low-velocity exocomets in the \bp\ system observed on April 29, 2025 with the HST and HARPS. We developed a detailed model of the absorption spectrum of these comets, where the excitation and the transmitted flux are computed iteratively along the line of sight within each gaseous cloud. Our model was fitted to the observations, allowing us to recover the transit distance and composition of the three exocomets. 

The measured transit distances of the three studied exocomets ($0.88 \pm 0.08$, $4.7 \pm 0.3$, and $1.52 \pm 0.15$ au) are much larger than previous estimates for low-velocities objects in the \bp\ system \citep[$19 \pm 4\, R_\star = 0.14 \pm 0.03\,$au,][]{Kiefer_2014}. These distances are also surprising as, at 1 au or more from \bp, dust sublimation is very inefficient, which is in contradiction with the large amounts of refractory ions found in the exocometary tails. This suggests that exocomet gaseous tails in the \bp\ system, most likely produced very close to the star ($<$0.5 au), can migrate over large distances without being disrupted, and remain detectable even several au from \bp.

\begin{acknowledgements}
We gratefully acknowledge funding from the Centre National d’\'Etudes Spatiales (CNES).
\end{acknowledgements}

\bibliographystyle{aa}
\bibliography{bibliography}

\newpage

\begin{appendix}

\onecolumn

\section{Fitted species}

\begin{table*}[h!]
    \centering 
    \begin{threeparttable}
    \caption{List of the species included in the model.}
    \renewcommand{\arraystretch}{1.25} \begin{tabular}{ c c c c c c}                    
    \cline{1-6}      
    \noalign{\smallskip}
    \cline{1-6}      
     
     Species & Ionisation potential    &  \ \ \ \ \ Main lines \ \ \ \ \ & Instrument & Radiative data & Collisional data  \\
             &  (eV) &       \\
                  
    \cline{1-6}      
    \noalign{\smallskip}

    \multirow{2}{*}{\Si} & \multirow{2}{*}{10.36} & 1200 - 1400 \A\ & COS  & \multirow{2}{*}{NIST} & \multirow{2}{*}{CHIANTI} \\
                         &                        & 1800 \A\      & STIS &                                                  \\

    \noalign{\medskip}
    \cdashline{1-6}      
    \noalign{\medskip}

    \caii & 11.87 & 3933, 3968 \A & HARPS  & NIST, [1] & CHIANTI \\

    \noalign{\medskip}
    \cdashline{1-6}      
    \noalign{\medskip}

    \mnii\tnote{a} & 15.64 & 2600 \A & STIS  & NIST, [2] & - \\

    \noalign{\medskip}
    \cdashline{1-6}      
    \noalign{\medskip}

    \multirow{2}{*}{\feii} & \multirow{2}{*}{16.19} & 1550 - 1700 \A & STIS  & \multirow{2}{*}{NIST, [3]} & \multirow{2}{*}{[3]} \\
                           &                        & 2200 - 2800 \A & STIS  &   &  \\
                           
    \noalign{\medskip}
    \cdashline{1-6}      
    \noalign{\medskip}
    
    \siii & 16.34 & 1800 \A & STIS  & NIST & CHIANTI \\
              
    \noalign{\medskip}
    \cdashline{1-6}      
    \noalign{\medskip}

    \multirow{2}{*}{\crii} & \multirow{2}{*}{16.50} & 2050 \A & STIS  & \multirow{2}{*}{NIST, [4], [5]} & \multirow{2}{*}{CHIANTI} \\
                           &                        & 2650 - 2850 \A & STIS  &   &  \\

    \noalign{\medskip}
    \cdashline{1-6}      
    \noalign{\medskip}

          &       & 1317, 1370 \A & C0S  &      &   \\
    \niii & 18.17 & 1700 - 1800 \A & STIS  & NIST, [7] & [8] \\
          &       & 2100 - 2400 \A & STIS  &      &   \\
                 
    \noalign{\smallskip}
    \cline{1-6}      
\end{tabular}

    \begin{tablenotes}
      \item[a] No set of collisional parameters was found for \mnii. 
      \item References: [1]~\cite{Melendez_2007}, [2]~\cite{Liggins2021}, [3]~\cite{Tayal2018}, [4] \cite{Nilsson2006}, [5] \cite{Tayal2020}, [6] \cite{Storey2016}, [7] \cite{Cassidy2016}. 
    \end{tablenotes}
    
\label{Tab. fitted species}

\end{threeparttable}
\end{table*}

\newpage

\section{Fitted parameters and compositions}

\begin{table*}[h!]
    \centering 
    \begin{threeparttable}
    \caption{List of the parameters fitted by the model.}
    \renewcommand{\arraystretch}{1.25} 
    \begin{tabular}{c c c c c c}                    
    \cline{1-6}      
    \noalign{\smallskip}
    \cline{1-6}      
    \noalign{\smallskip}

     &  & \hspace{1. cm} LVC~\#1 \hspace{1. cm}  & \hspace{1. cm} LVC~\#2 \hspace{1. cm}  & \hspace{1. cm}  LVC~\#3 \hspace{1. cm}  &  \hspace{1. cm}  Disc \hspace{1. cm}   \\

    \multicolumn{2}{r}{Nomenclature name\tnote{a}} & \bp\ C20250429a & \bp\ C20250429b & \bp\ C20250429c & \\
                  
    \noalign{\smallskip}
    \cline{1-6}   
    \noalign{\smallskip}
    \cline{1-6}      
    \noalign{\smallskip}

    Parameter & Unit & \\

    \noalign{\smallskip}
    \cline{1-6}      
    \noalign{\smallskip}
    
    $N_{\text{\Si}}$ & cm$^{-2}$ & $3.0 \pm 0.5\times 10^{12}$ & $1.29 \pm 0.10\times 10^{12}$ & $1.5 \pm 0.4\times 10^{12}$ & $1.0 \pm 0.2\times 10^{13}$ \\

    $N_{\text{\caii}}$ & cm$^{-2}$ & $1.64 \pm 0.05 \times 10^{12}$ & $7.4 \pm 0.5 \times 10^{12}$ & $8.6 \pm 0.4 \times 10^{12}$ & $1.7^{+0.9}_{-0.4}\times 10^{13}$ \\
   
    $N_{\text{\mnii}}$ & cm$^{-2}$ & $2.02 \pm 0.11\times 10^{12}$ & $3.15 \pm 0.18\times 10^{12}$ & $9.0 \pm 0.8\times 10^{11}$ & $1.33 \pm 0.16\times 10^{12}$ \\
     
    $N_{\text{\feii}}$ & cm$^{-2}$ & $3.57 \pm 0.05 \times 10^{14}$ & $6.20 \pm 0.15 \times 10^{14}$ & $1.41 \pm 0.02 \times 10^{14}$ & $2.2 \pm 0.2 \times 10^{14}$ \\

    $N_{\text{\siii}}$ & cm$^{-2}$ & $4.8 \pm 0.4\times 10^{14}$ & $4.7 \pm 0.5\times 10^{14}$ & $1.8 \pm 0.2\times 10^{14}$ & $2.0 \pm 0.5\times 10^{14}$ \\
    
    $N_{\text{\crii}}$ & cm$^{-2}$ & $3.1 \pm 0.3 \times 10^{12}$ & $7.4 \pm 0.4 \times 10^{12}$ & $2.1 \pm 0.2\times 10^{12}$ & $4.4 \pm 0.4 \times 10^{12}$ \\

    $N_{\text{\niii}}$ & cm$^{-2}$ & $2.47 \pm 0.10\times 10^{13}$ & $3.40 \pm 0.12\times 10^{13}$ & $1.07 \pm 0.07\times 10^{13}$ & $1.51 \pm 0.10\times 10^{13}$ \\

    \noalign{\medskip}
    \cdashline{1-6}
    \noalign{\medskip}

    $d$ & au & $0.88 \pm 0.08$ & $4.7 \pm 0.3$ & $1.52 \pm 0.15$ & $38_{-11}^{+15}$ \\  
    $v$ & km\,s$^{-1}$ & $-7.57 \pm 0.03$ & $2.54 \pm 0.05$ & $9.92 \pm 0.04$ & 0 (fixed) \\
    $b$ & km\,s$^{-1}$ & $3.31 \pm 0.04$ & $2.79 \pm 0.04$ & $2.35 \pm 0.06$ & $1.03 \pm 0.09$ \\
    $f_{\text{esc}}$ &  & $0.20 \pm 0.02$ & $0.11 \pm 0.02$ & $0.09 \pm 0.03$ & -- \\
    $\log(n_e)$ & cm$^{-3}$ & $< 6$ & $3.3 \pm 0.3$ & $< 5$ & $2.7 \pm 0.3$ \\
    $\log(T_e)$ & K & -- & $2.9 \pm 0.3$ & -- & $2.9 \pm 0.3$ \\

    \noalign{\smallskip}
    \cline{1-6}      
\end{tabular}

\begin{tablenotes}
  \item[a] Following \cite{Lecavelier_2026}. 
\end{tablenotes}

\label{Tab. Fitted parameters}

\end{threeparttable}
\end{table*}

\vspace{0.5 cm}

\begin{table*}[h!]
    \centering 
    \begin{threeparttable}
    \caption{Abundances in the three LVCs and the circumstellar disc, normalized to \feii. The corresponding elemental ratios in the solar photosphere (e.g., S/Fe for \Si/\feii) are shown in the second column.}
    \renewcommand{\arraystretch}{1.3} 
    \begin{tabular}{c c c c c c}                    
    \cline{1-6}      
    \noalign{\smallskip}
    \cline{1-6}      
     
     Species & Sun (X/Fe)\tnote{a} & LVC~\#1 & LVC~\#2 & LVC~\#3 & Disc \\
                  
    \cline{1-6}      
    \noalign{\smallskip}

    \Si\,/\,\feii & $4.6 \pm 0.5 \times 10^{-1}$ & $8.4 \pm 1.4 \times 10^{-3}$ & $ 2.1 \pm 0.2 \times 10^{-2}$ & $1.1 \pm 0.3 \times 10^{-2}$  & $4.6 \pm 1.0 \times 10^{-2}$ \\
   
    \caii\,/\,\feii & $6.9 \pm 0.8 \times 10^{-2}$ & $4.6 \pm 0.2 \times 10^{-3}$ & $1.19 \pm 0.10 \times 10^{-2}$ & $6.1 \pm 0.3 \times 10^{-3}$ & $8_{-2}^{+4} \times 10^{-2}$ \\
        
    \mnii\,/\,\feii & $9.1 \pm 1.5 \times 10^{-3}$  & $5.1 \pm 0.3 \times 10^{-3}$ & $5.1 \pm 0.3 \times 10^{-3}$ & $6.4 \pm 0.5 \times 10^{-3}$ & $6.0 \pm 0.9 \times 10^{-3}$ \\
        
    \siii\,/\,\feii & $1.12 \pm 0.13$ & $1.34 \pm 0.11$ & $0.81 \pm 0.08$ & $1.44 \pm 0.18$ & $1.0 \pm 0.2$ \\
        
    \crii\,/\,\feii & $1.5 \pm 0.2 \times 10^{-2}$ & $8.8 \pm 0.8 \times 10^{-3}$ & $1.20 \pm 0.07 \times 10^{-2}$ & $1.49 \pm 0.14 \times 10^{-2}$ & $2.0 \pm 0.3 \times 10^{-2}$ \\
        
    \niii\,/\,\feii & $5.5 \pm 0.7 \times 10^{-2}$ & $6.9 \pm 0.3 \times 10^{-2}$ & $5.5 \pm 0.3 \times 10^{-2}$ & $7.6 \pm 0.5 \times 10^{-2}$ & $6.9 \pm 0.8 \times 10^{-2}$ \\

    \noalign{\smallskip}
    \cline{1-6}      
\end{tabular}   

\begin{tablenotes}
  \item[a] Obtained from \cite{Asplund2021}. 
\end{tablenotes}

\label{Tab. Composition}

\end{threeparttable}
\end{table*}

\newpage

\section{Examples of lines in the April 29, 2025 spectrum}
\label{App. Example of absorption lines}

\begin{figure}[h!]
\centering
    \includegraphics[scale = 0.38,     trim = 20 25 60 10,clip]{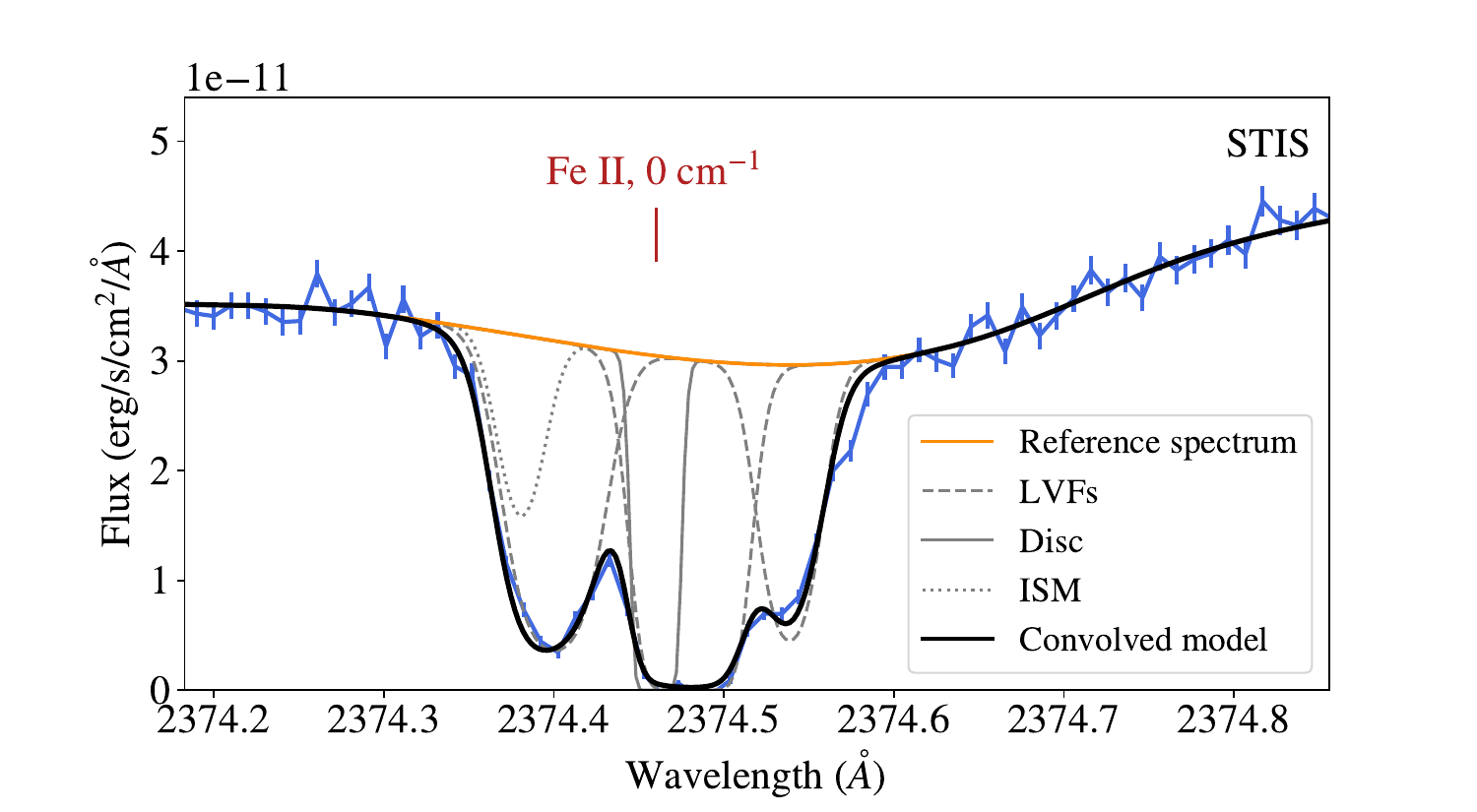}  
    \includegraphics[scale = 0.38,     trim = 20 25 60 10,clip]{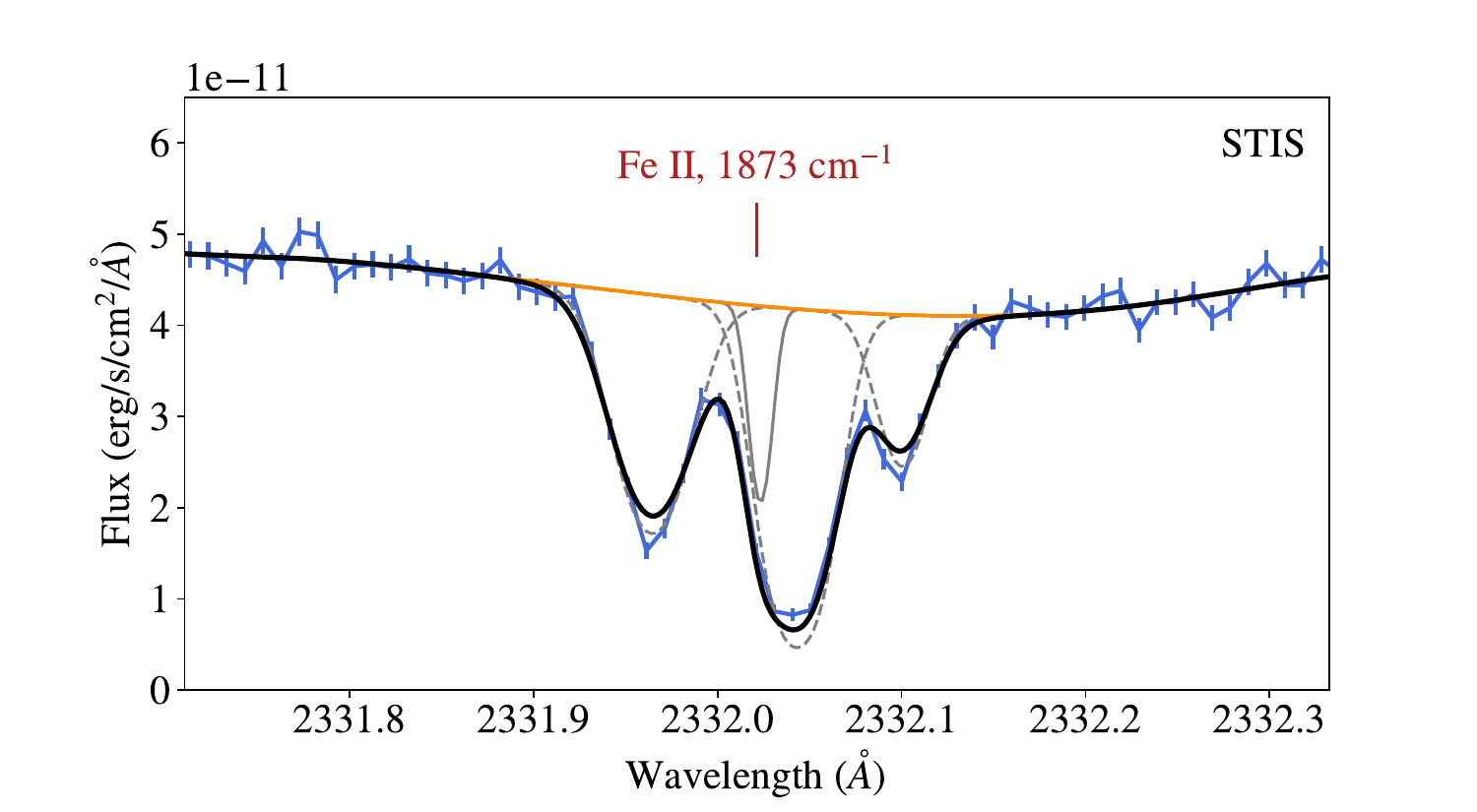} 

    \includegraphics[scale = 0.38,     trim = 20 25 60 10,clip]{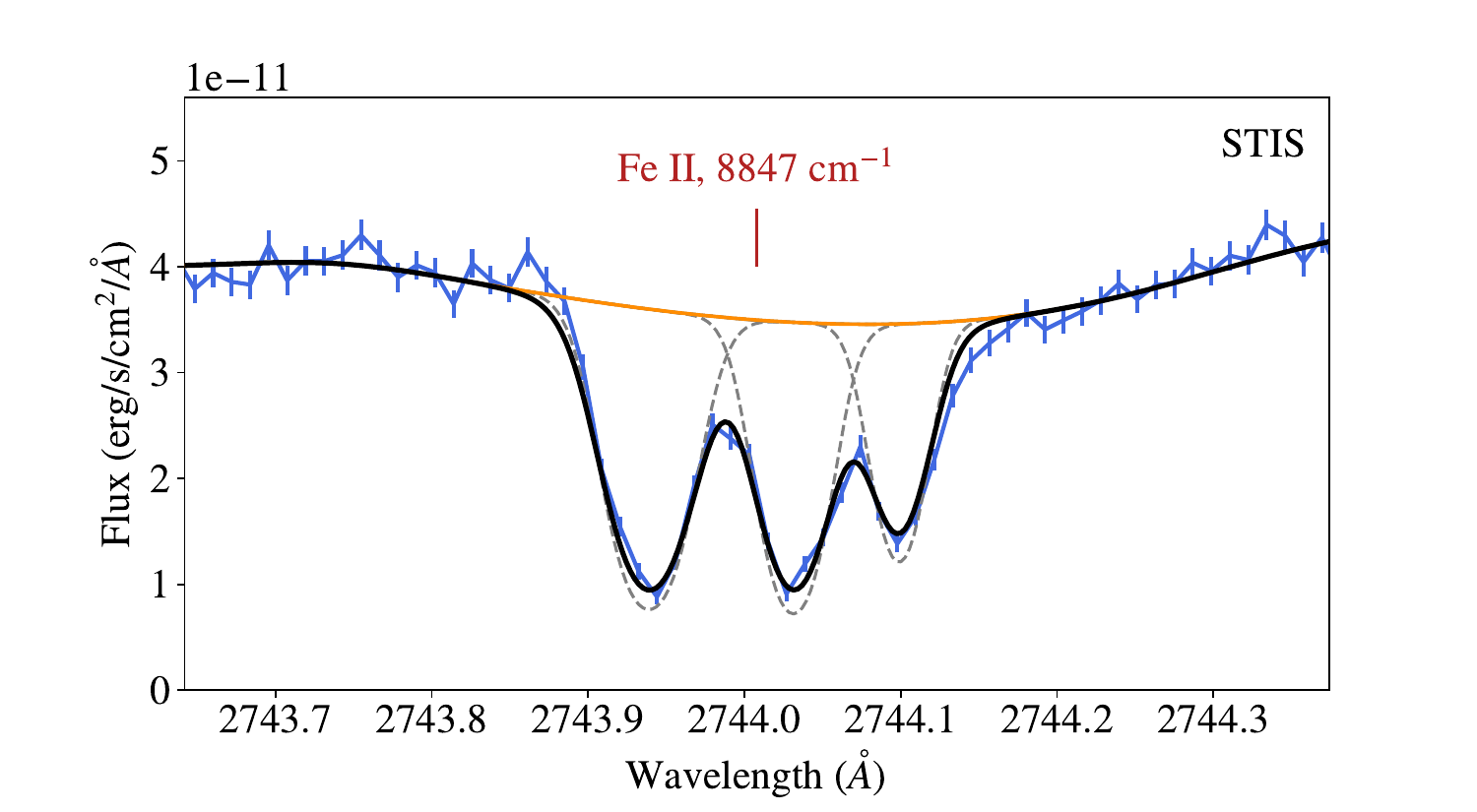} 
    \includegraphics[scale = 0.38,     trim = 20 25 60 10,clip]{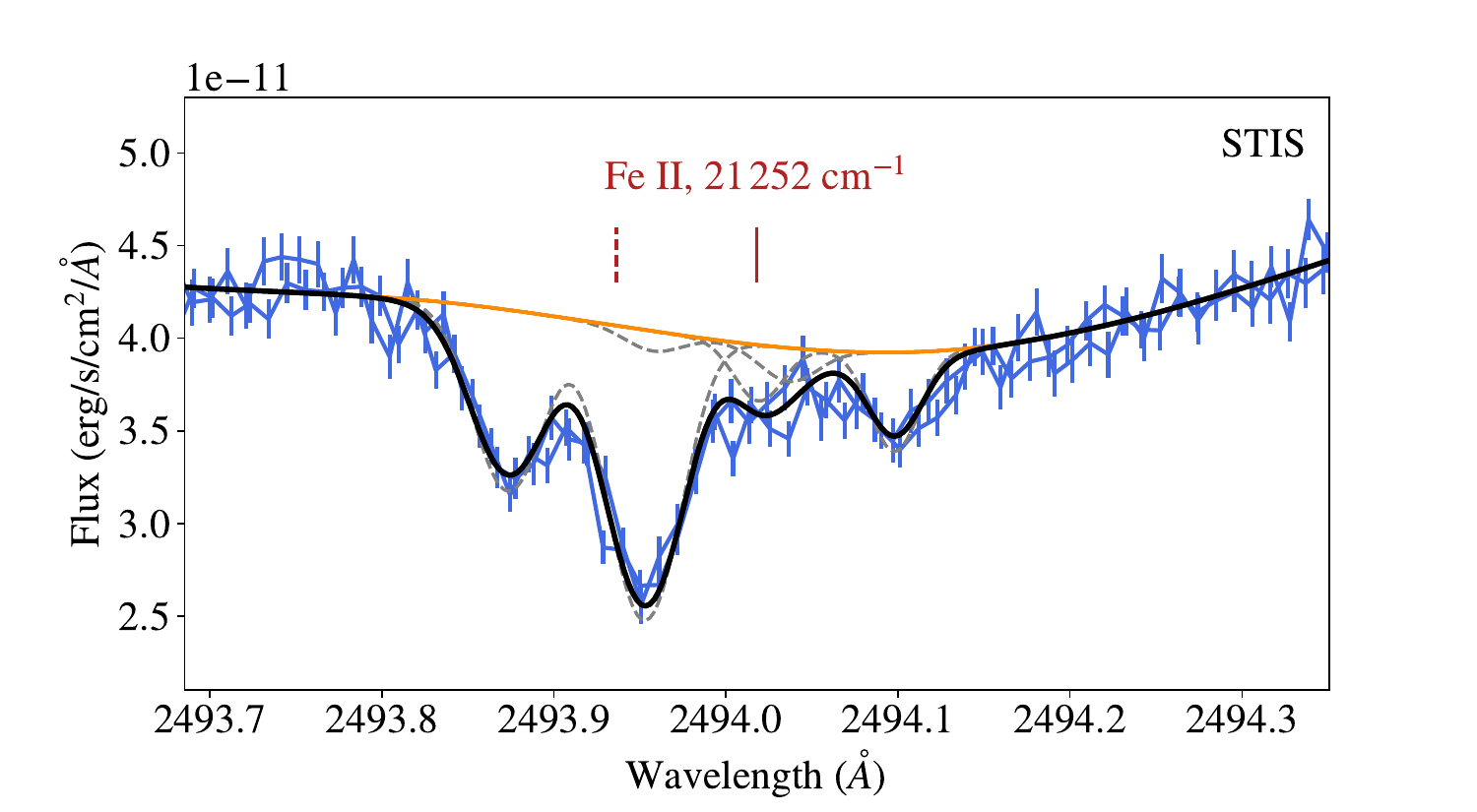}  
    
    \includegraphics[scale = 0.38,     trim = 20 25 60 10,clip]{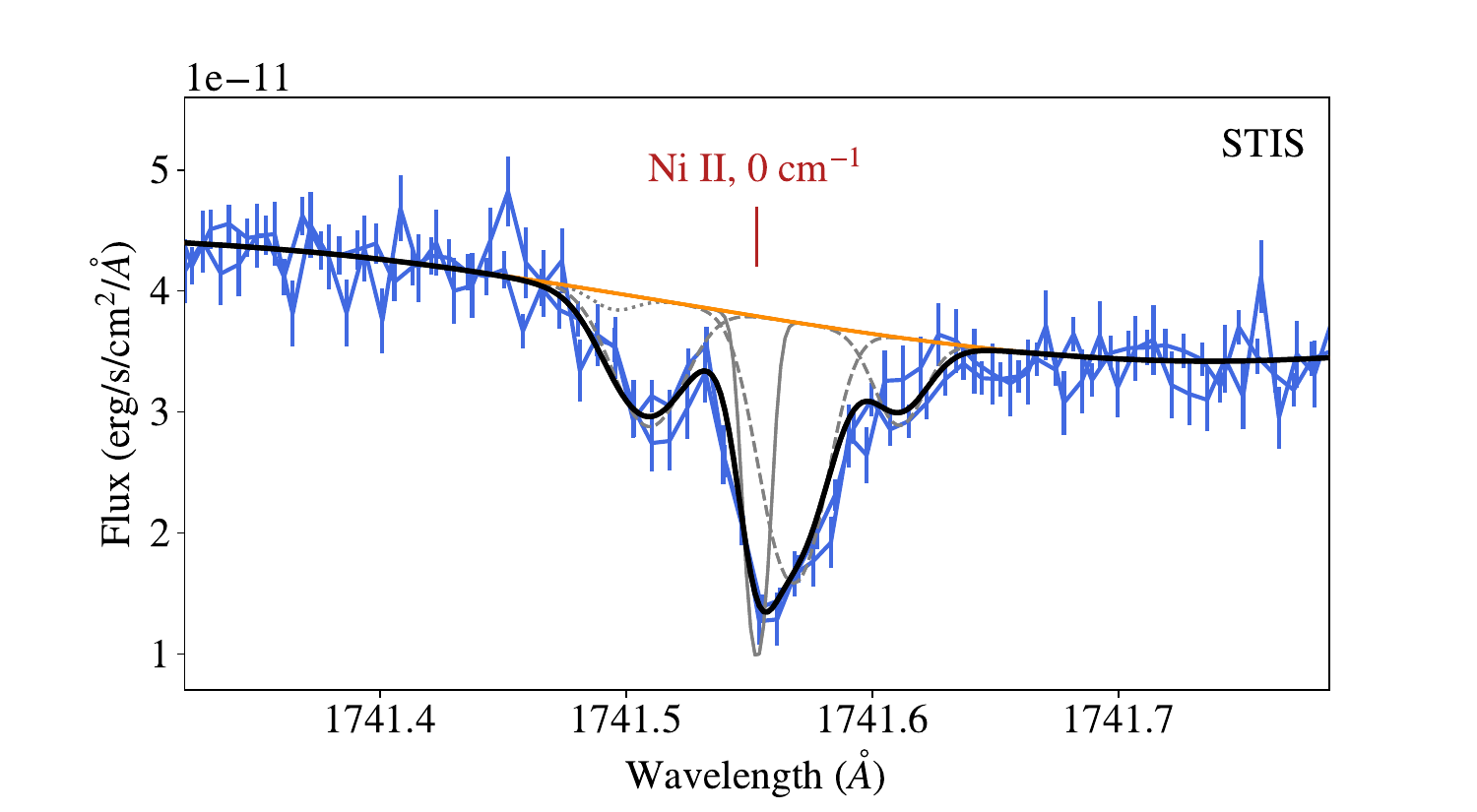}  
    \includegraphics[scale = 0.38,     trim = 20 25 60 10,clip]{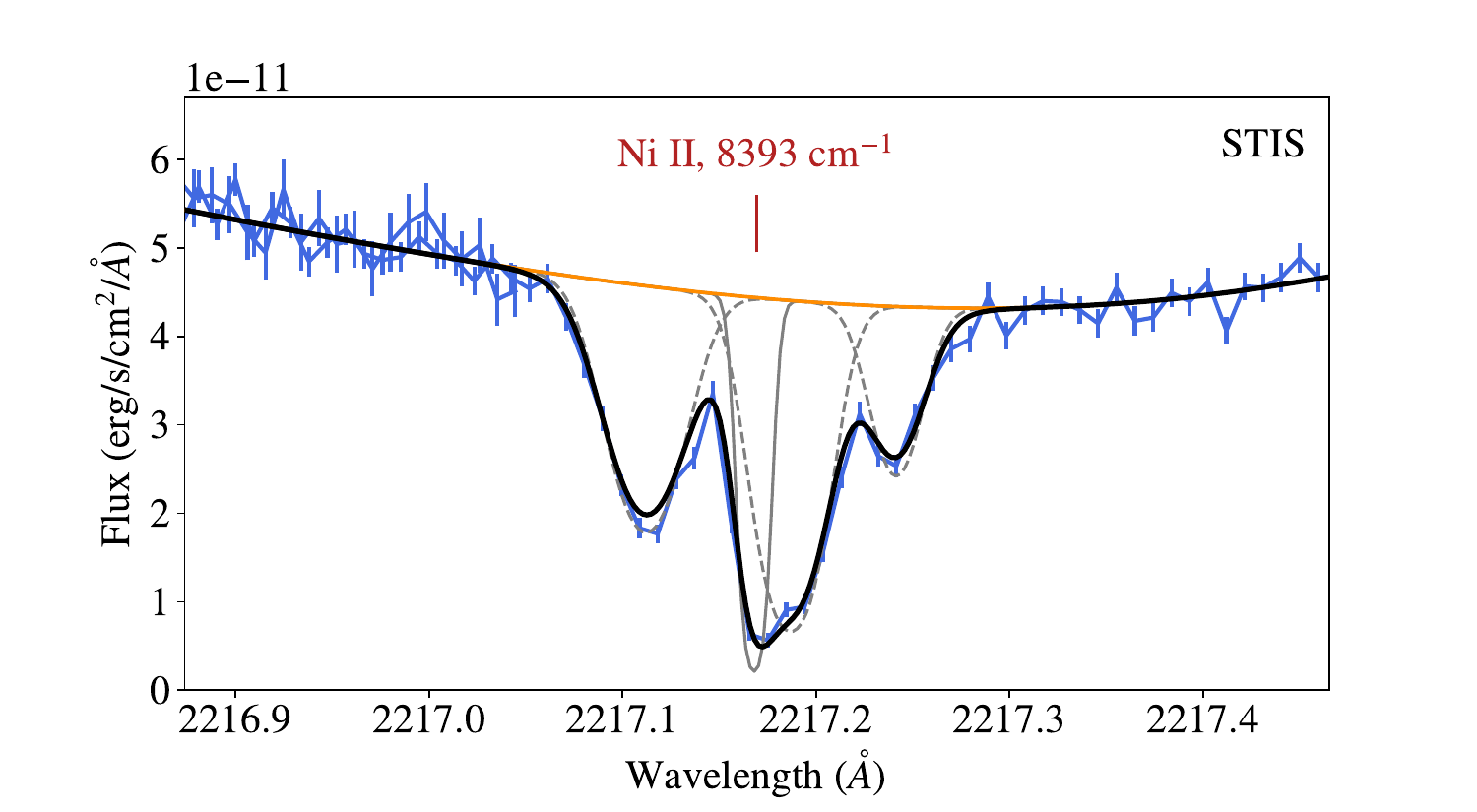} 

    \includegraphics[scale = 0.38,     trim = 20 0 60 10,clip]{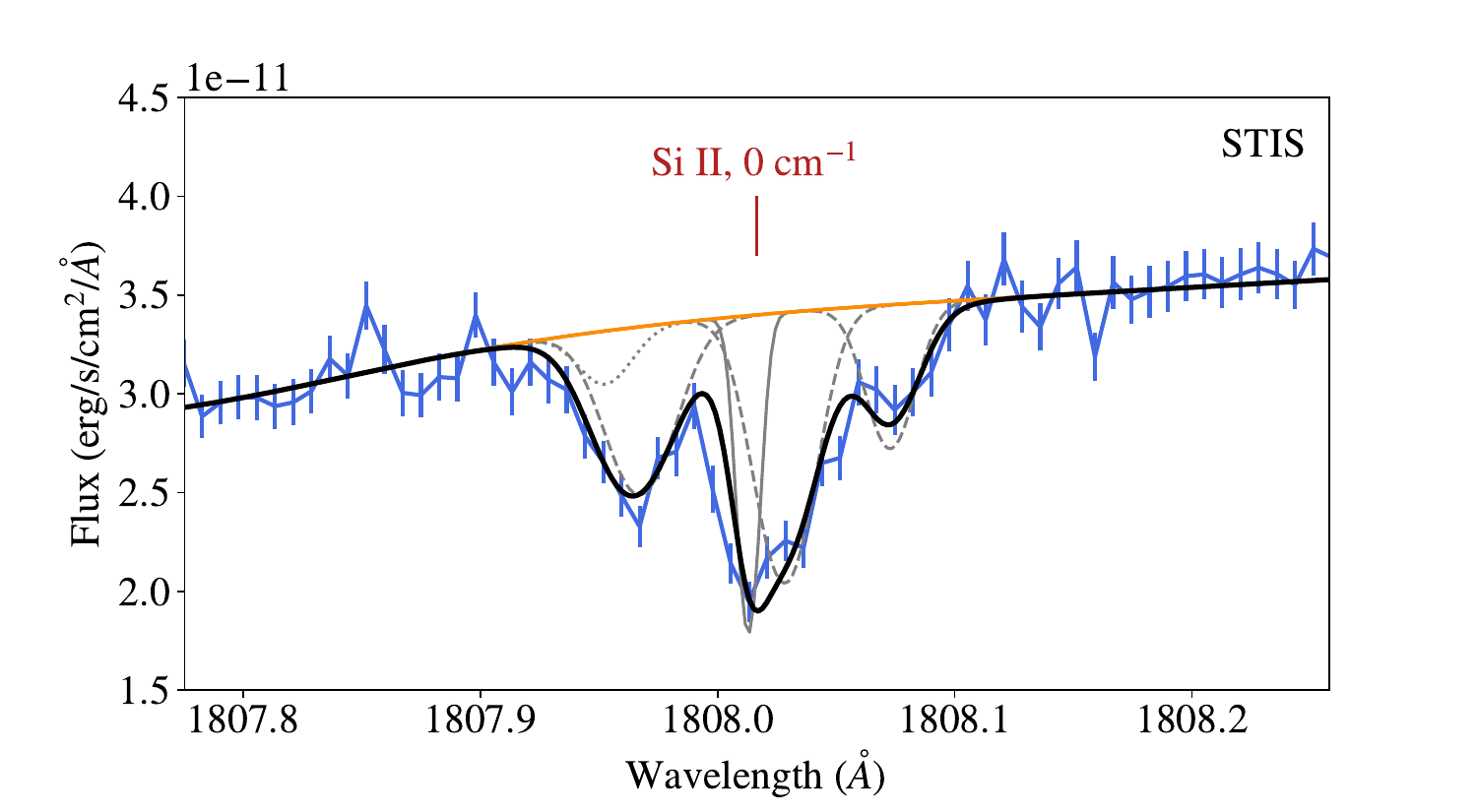}  
    \includegraphics[scale = 0.38,     trim = 20 0 60 10,clip]{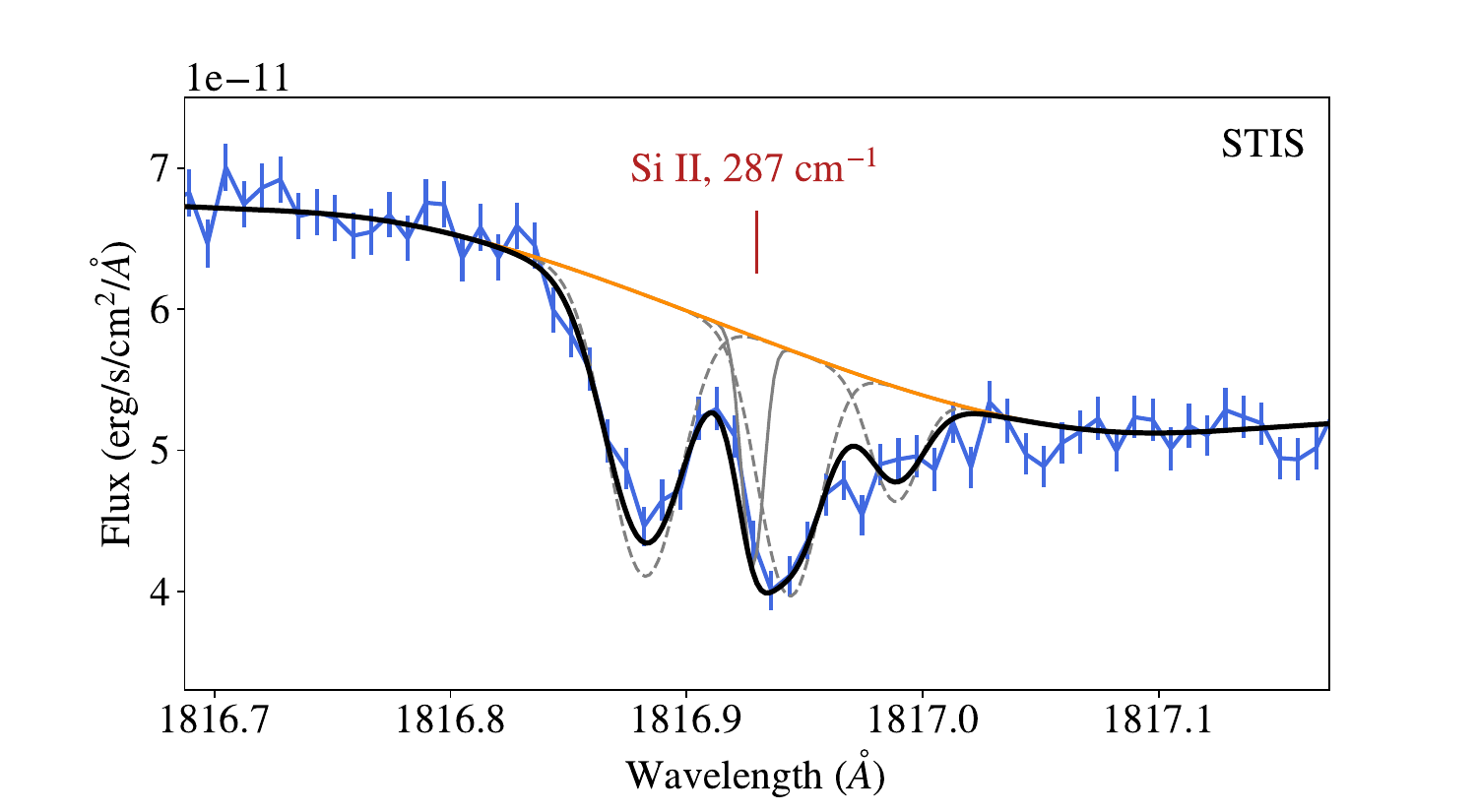} 

    \caption{Examples of lines studied in the present paper. The reference spectrum is indicated with a solid, orange line. Absorptions from the interstellar medium, from the three LVCs and from the circumstellar disc are shown with dotted, dashed and solid grey lines, respectively. The full model, convolved with the instrumental line spread function, is shown with a solid, black line. The observed spectrum is shown in blue; error bars only reflect the tabulated uncertainties and does not include the systematic uncertainties of our model (Sect \ref{Sect. systematic uncertainties}). For each line, the corresponding species and lower level energy are indicated in red. Note that the \feii\ line profile at 2494\,\AA\ results from a doublet of two lines separated by about 0.1\,\AA. }. 
  
    \label{Fig. Appendix lines}
\end{figure}

\begin{figure}[h!]
\centering

    \includegraphics[scale = 0.38,     trim = 20 25 60 10,clip]{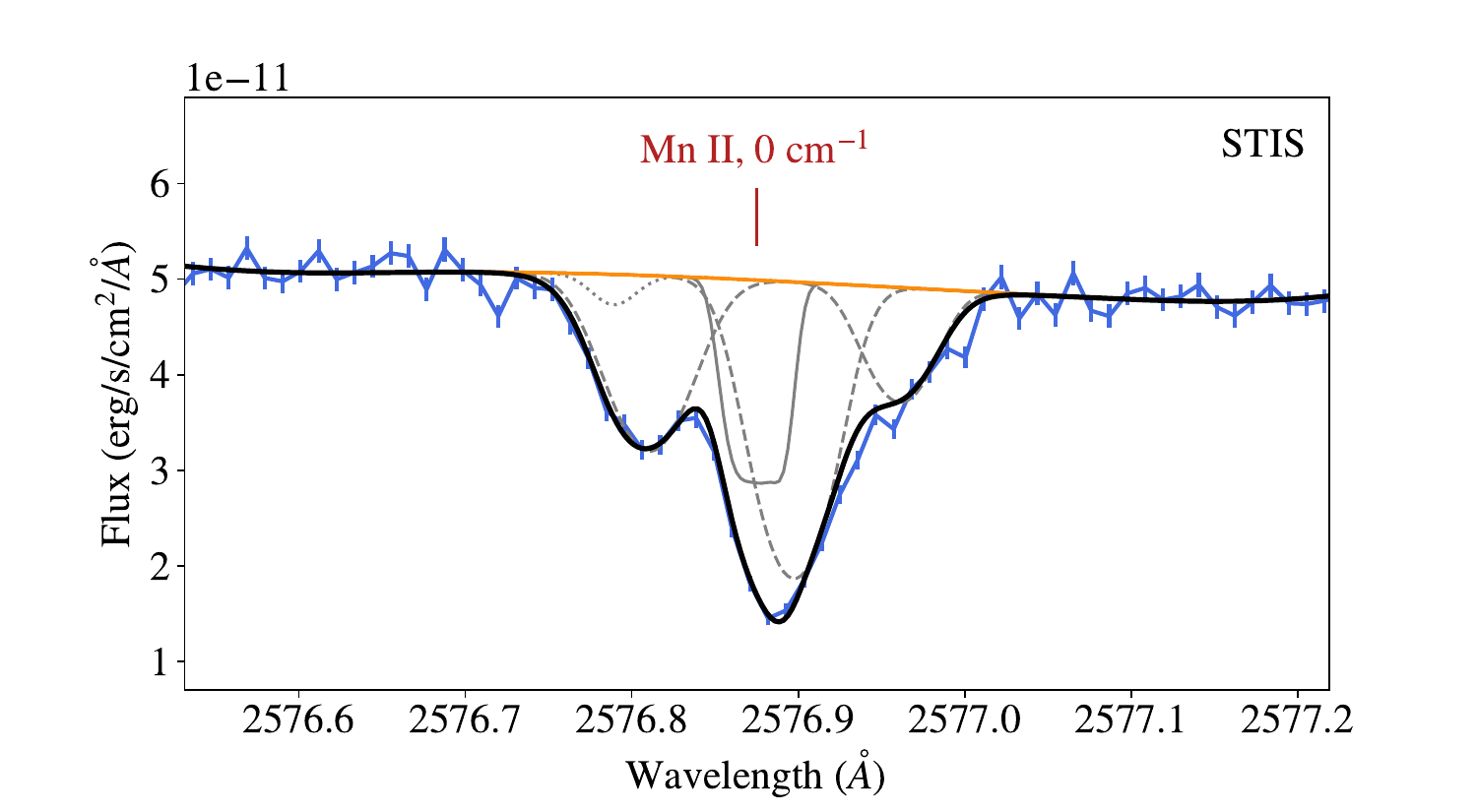}  
    \includegraphics[scale = 0.38,     trim = 20 25 60 10,clip]{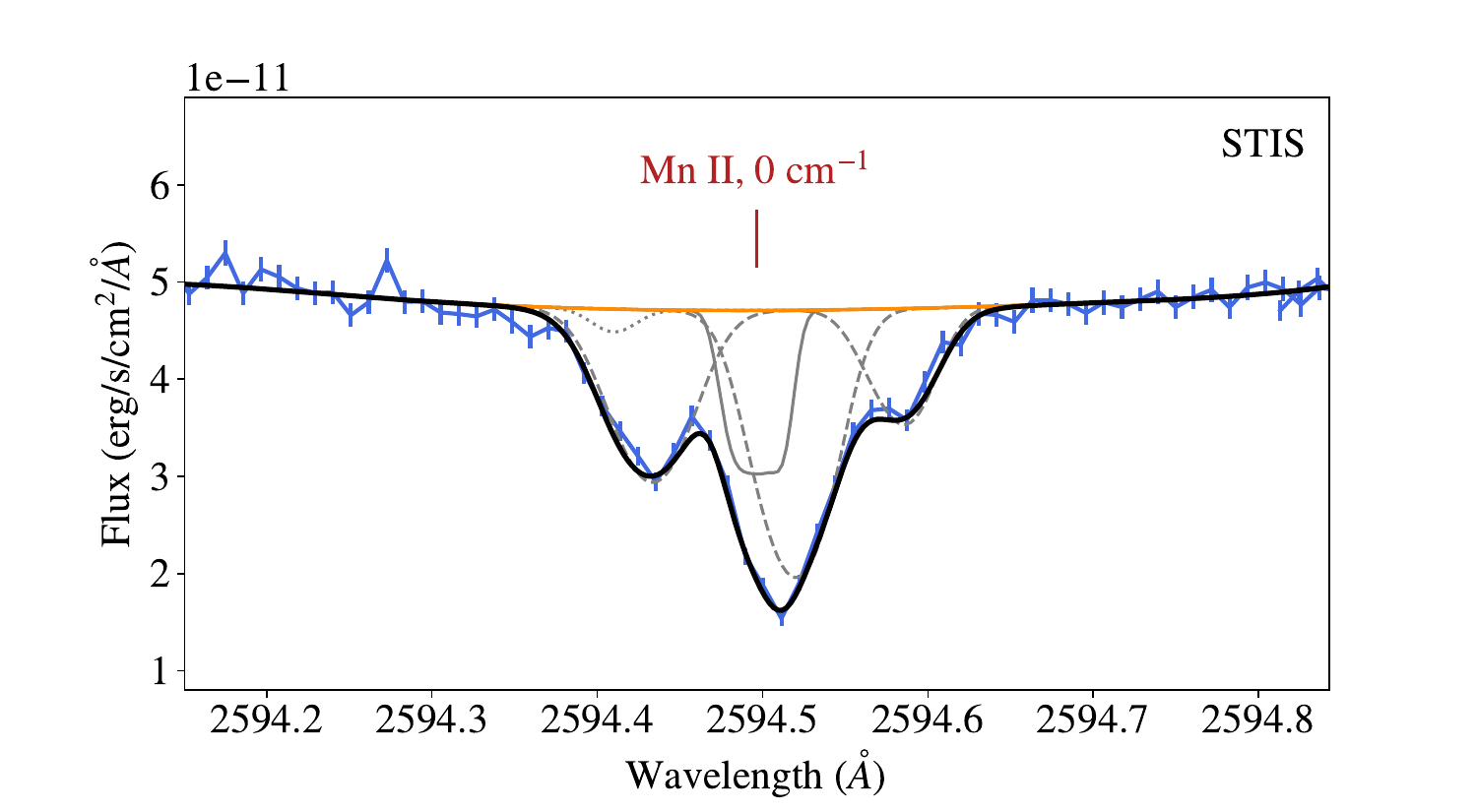} 

    \includegraphics[scale = 0.38,     trim = 20 25 60 10,clip]{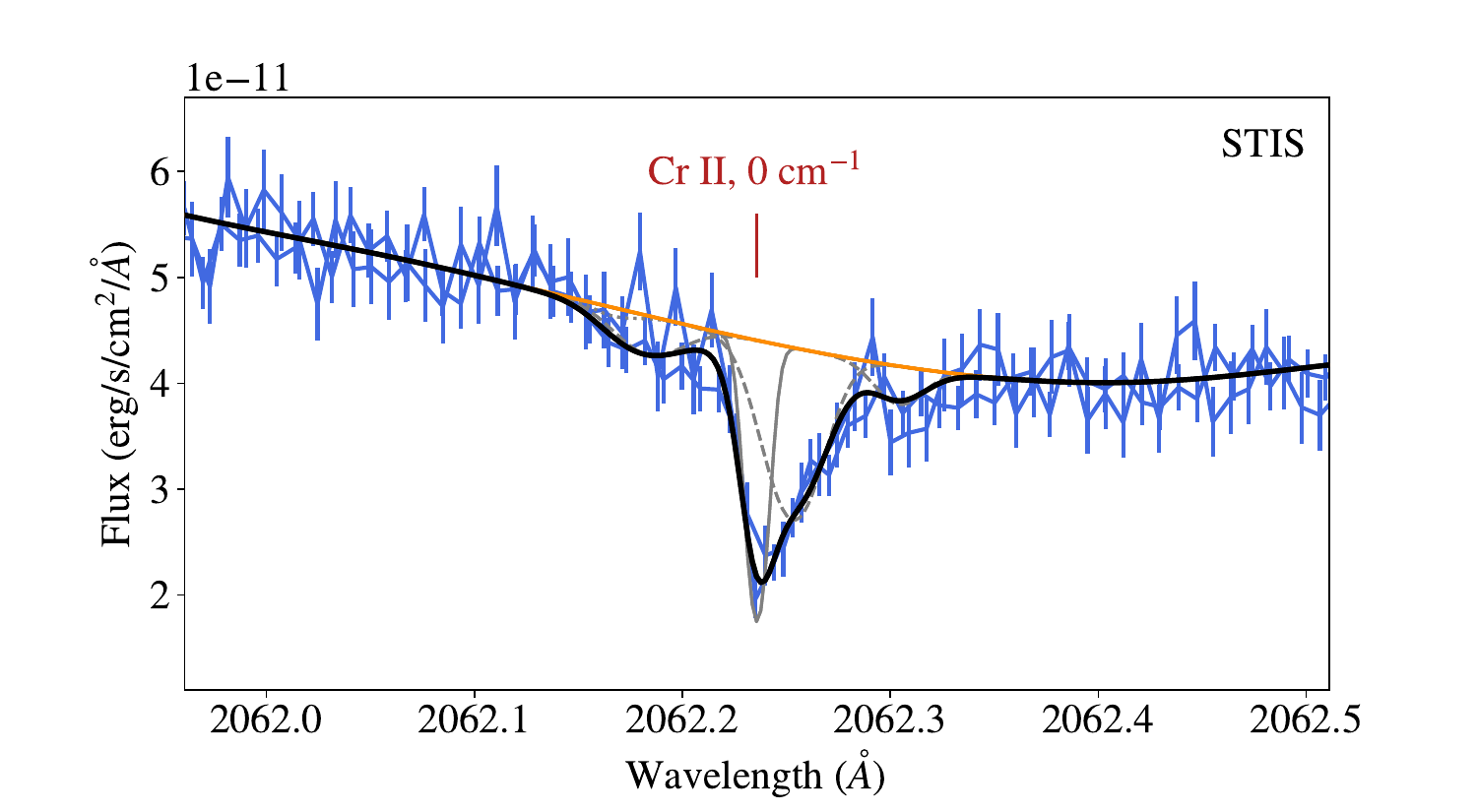}  
    \includegraphics[scale = 0.38,     trim = 20 25 60 10,clip]{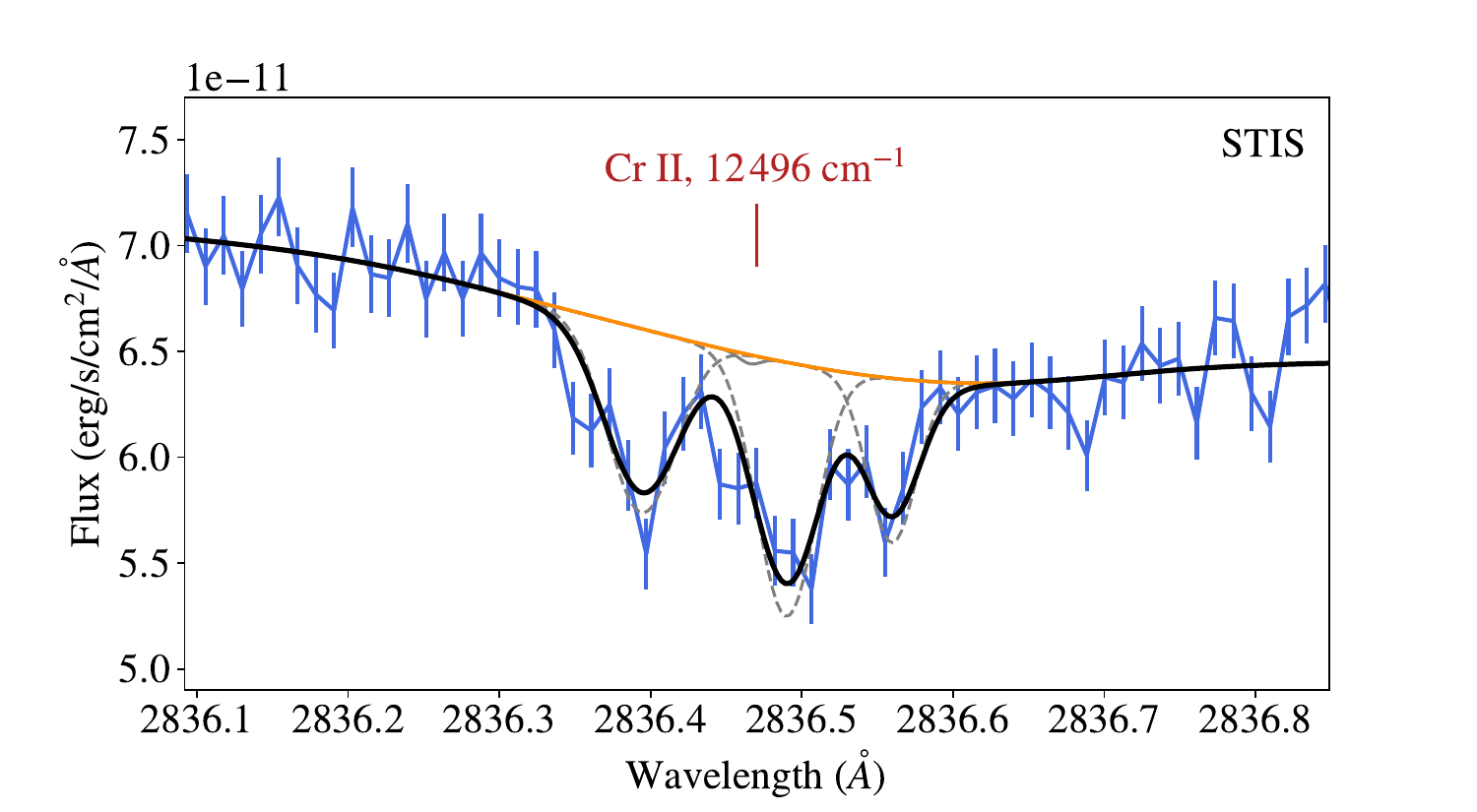} 

    \includegraphics[scale = 0.38,     trim = 20 25 60 10,clip]{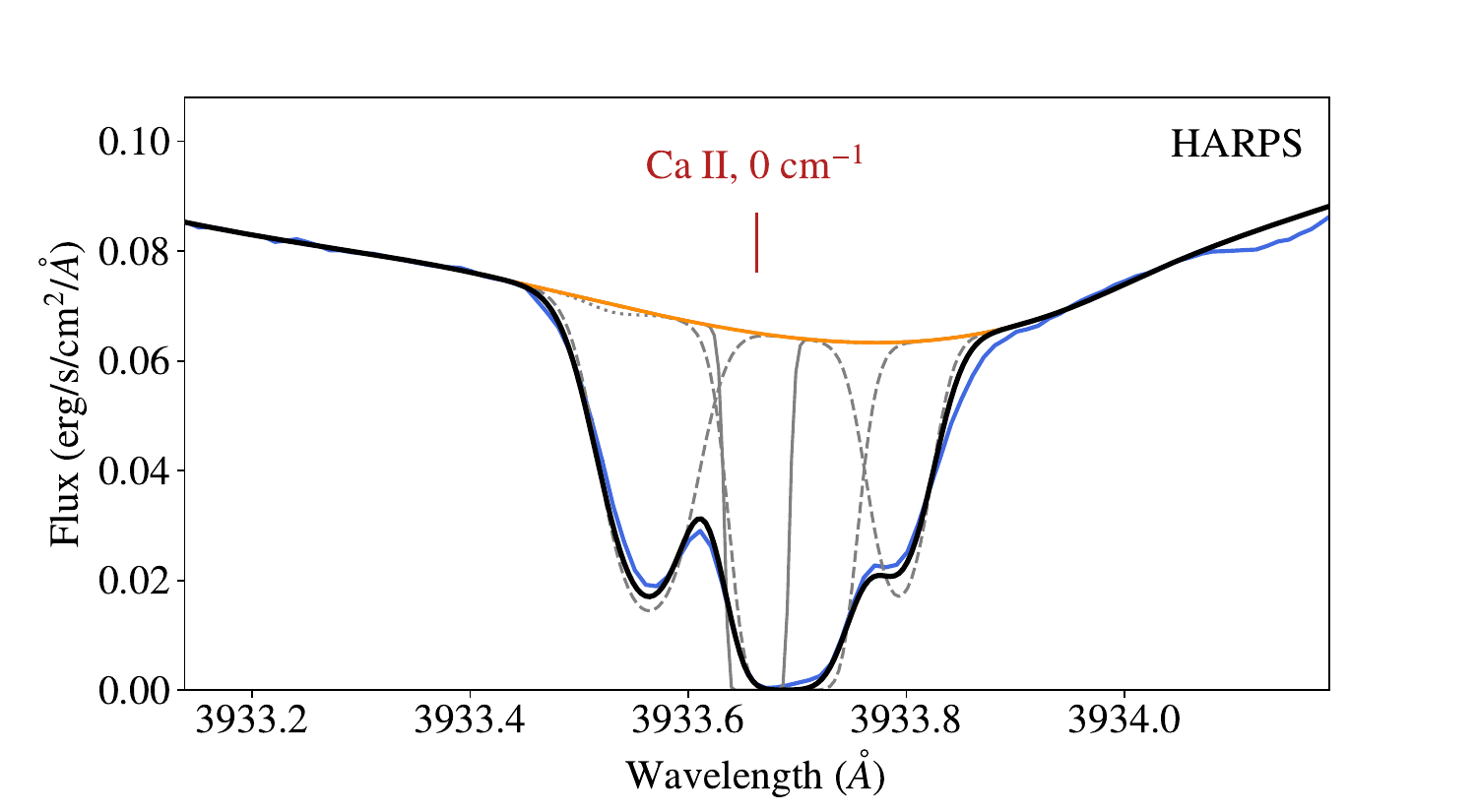}  
    \includegraphics[scale = 0.38,     trim = 20 25 60 10,clip]{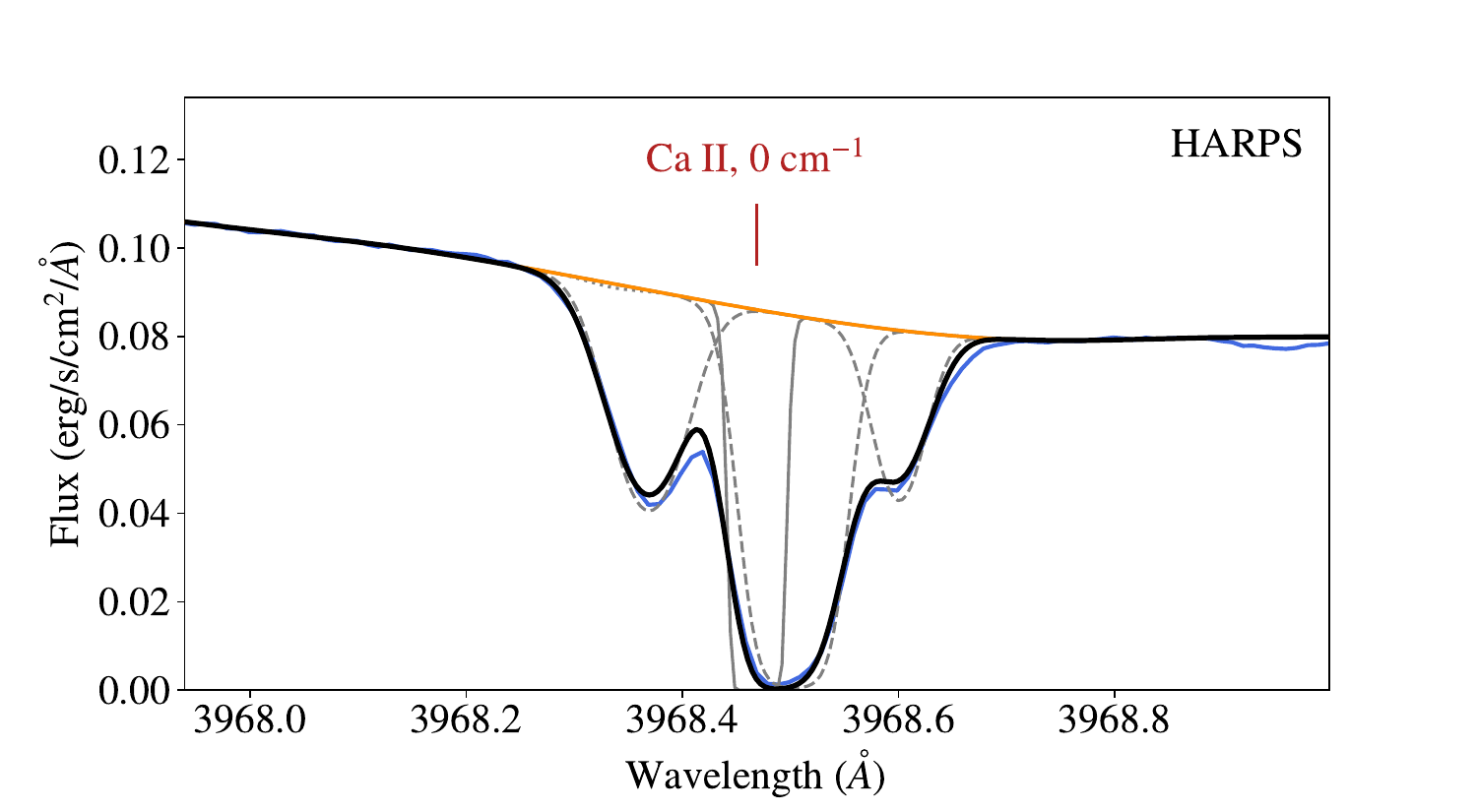} 
    
    \includegraphics[scale = 0.38,     trim = 20 0 60 10,clip]{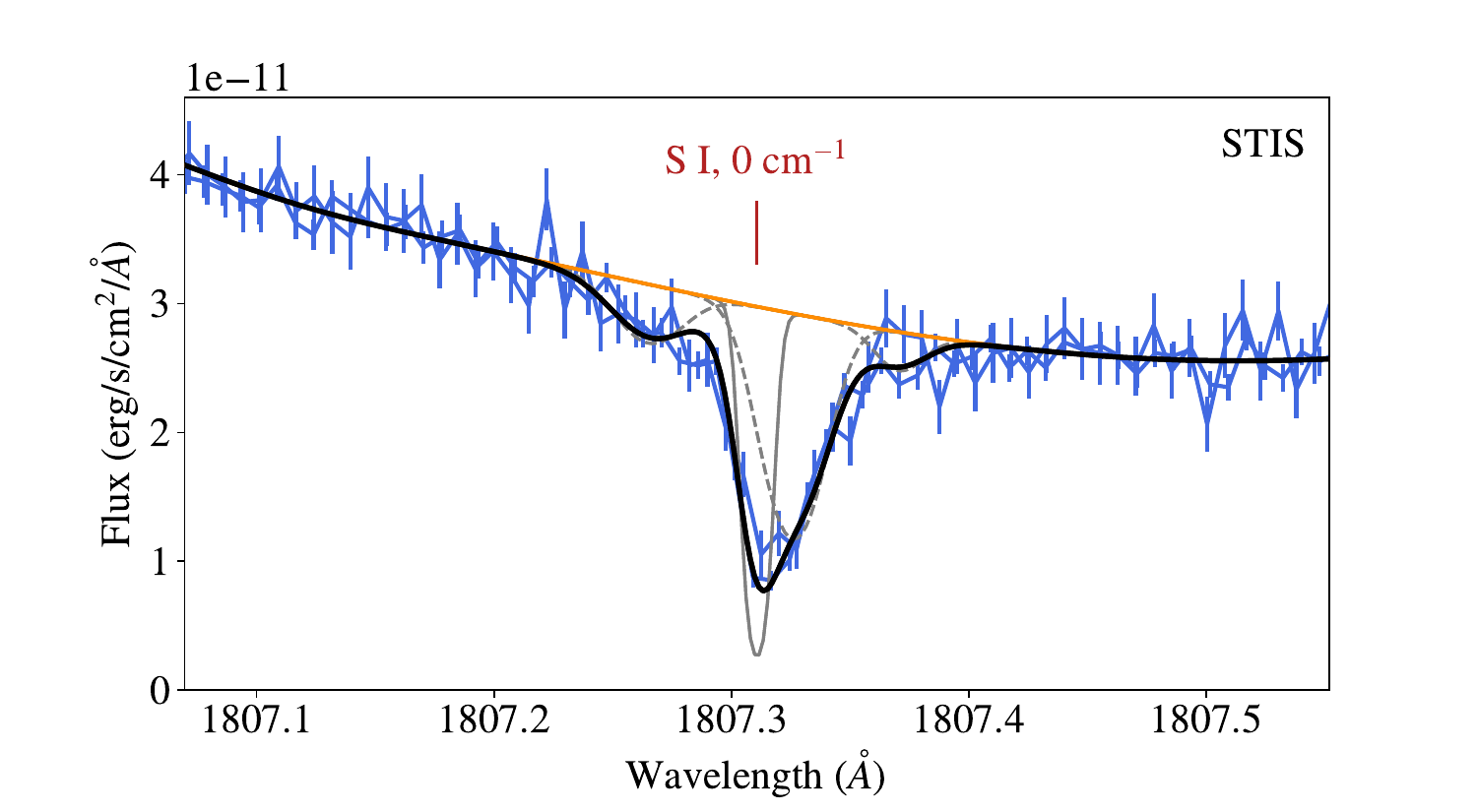} 
    \includegraphics[scale = 0.38,     trim = 20 0 60 10,clip]{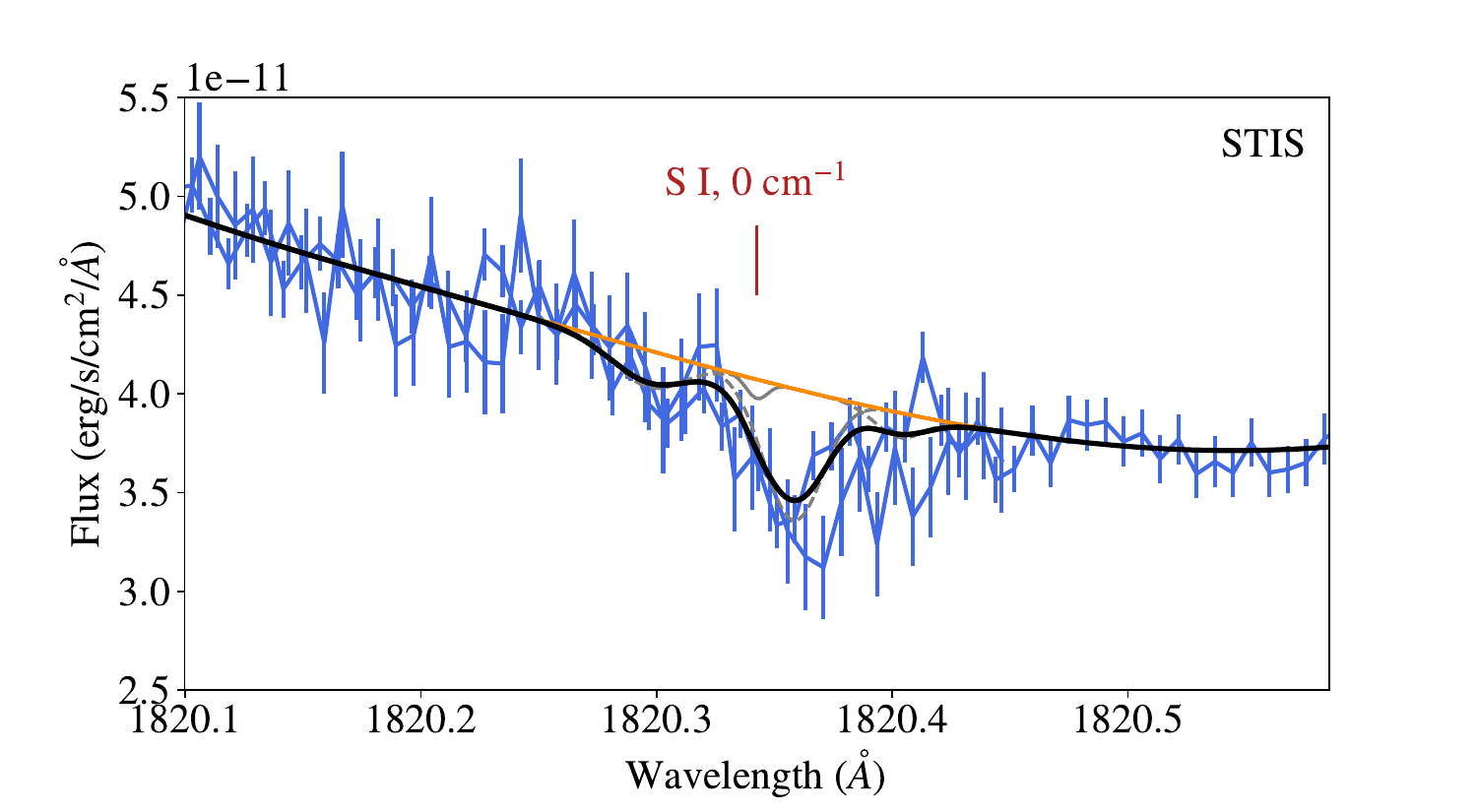} 
   
    \centering \small \textbf{Fig.~\ref{Fig. Appendix lines}.} Continued.
\end{figure}

\end{appendix}
\end{document}